\documentclass[usenatbib]{mn2e}

\usepackage{natbib,amssymb,color}
\usepackage{times}

\newcommand{\asz}{\ensuremath{{A_\mathrm{SZ}}}}
\newcommand{\bsz}{\ensuremath{{B_\mathrm{SZ}}}}
\newcommand{\csz}{\ensuremath{{C_\mathrm{SZ}}}}

\usepackage{sidecap}

\usepackage{url}
\usepackage{amsmath,graphics,graphicx}

\usepackage{epsfig}
\graphicspath{{figures/}}

\def\SPT2043{SPT-Cl\thinspace$J$2043$-$5035}

\newcommand{\ltsima}{$\; \buildrel < \over \sim \;$}
\newcommand{\ltsim}{\lower.5ex\hbox{\ltsima}}

\newcommand{\be}{\begin{equation}}
\newcommand{\ee}{\end{equation}}
\newcommand{\bea}{\begin{eqnarray}}
\newcommand{\eea}{\end{eqnarray}}

\usepackage{lmodern}
\DeclareMathAlphabet{\mathbfsf}{\encodingdefault}{\sfdefault}{bx}{sl}

\newcommand{\altaffilmark}[1]{$^{#1}$}
\usepackage{epsfig}
\graphicspath{{figures/}}

\def\Bonn{1}
\def\Munich{2}
\def\ExcellenceCluster{3}
\def\Bochum{4}
\def\Leiden{5}
\def\MPIA{6}
\def\KICPChicago{7}
\def\Cincinnati{8}
\def\FNAL{9}
\def\AAUChicago{10}
\def\PhysicsUChicago{11}
\def\ANL{12}
\def\Missouri{13}
\def\Montreal{14}
\def\MIT{15}
\def\Trieste{16}
\def\IFPU{17}
\def\INAF{18}
\def\INFN{19}
\def\CfA{20}

\defcitealias{schrabback10}{S10}
\defcitealias{massey2014}{M14}
\defcitealias{skelton14}{S14}
\defcitealias{becker11}{BK11}
\defcitealias{schrabback18}{S18}
\defcitealias{schrabback19}{S19}
\defcitealias{bocquet19}{B19}
\defcitealias{dietrich19}{D19}
\defcitealias{diemer19}{D19}
\defcitealias{raihan20}{R20}
\defcitealias{hernandez20}{H20}

\begin{document}
\title[Weak Lensing Study of 30 Distant SPT Clusters]{
  Mass calibration of distant SPT  galaxy clusters
 through expanded
weak lensing follow-up observations
with HST,  VLT \& Gemini-South
}

\author[T.~Schrabback et al.]
       {T.~Schrabback\altaffilmark{\Bonn}\thanks{E-mail: schrabba@astro.uni-bonn.de},
         S.~Bocquet\altaffilmark{\Munich,\ExcellenceCluster},
M.~Sommer\altaffilmark{\Bonn},
H.~Zohren\altaffilmark{\Bonn},
J.~L.~van den Busch\altaffilmark{\Bonn,\Bochum},\and
B.~Hern\'andez-Mart\'in\altaffilmark{\Bonn},
H.~Hoekstra\altaffilmark{\Leiden},
S.~F.~Raihan\altaffilmark{\Bonn},
M.~Schirmer\altaffilmark{\MPIA},
D.~Applegate\altaffilmark{\Bonn,\KICPChicago},\and
M.~Bayliss\altaffilmark{\Cincinnati},
 B.~A.~Benson\altaffilmark{\FNAL,\AAUChicago,\KICPChicago},
L.~E.~Bleem\altaffilmark{\KICPChicago,\PhysicsUChicago,\ANL},
J.~P.~Dietrich\altaffilmark{\Munich,\ExcellenceCluster},
B.~Floyd\altaffilmark{\Missouri}, \and
S.~Hilbert\altaffilmark{\Munich,\ExcellenceCluster},
J.~Hlavacek-Larrondo\altaffilmark{\Montreal},
M.~McDonald\altaffilmark{\MIT},
A.~Saro\altaffilmark{\Trieste,\IFPU,\INAF,\INFN}, \and
A.~A.~Stark\altaffilmark{\CfA} \&
N.~Weissgerber\altaffilmark{\Bonn}
\\
\vspace{0.4cm}\\
\parbox{\textwidth}{\large Author affiliations are listed after the reference list.\\}
}
\maketitle

\begin{abstract}
  Expanding from previous work we present weak lensing measurements for a total sample of 30 distant (\mbox{$z_\mathrm{median}=0.93$}) massive galaxy clusters from the South Pole Telescope
  Sunyaev-Zel'dovich (SPT-SZ) Survey, measuring galaxy shapes in
  {\it Hubble} Space Telescope (HST) Advanced Camera for Surveys
    images.
 We remove cluster members and preferentially select \mbox{$z\gtrsim 1.4$} background galaxies via $V-I$ colour,
 employing deep photometry from VLT/FORS2 and Gemini-South/GMOS.
 We apply revised calibrations for the weak lensing shape measurements and  the source redshift distribution
to estimate the cluster masses.
 In combination with earlier Magellan/Megacam  results for lower-redshifts clusters  we infer refined constraints on the
 scaling relation between the SZ detection significance and the cluster mass, in particular regarding
 its  redshift evolution.
 The mass scale inferred from the weak lensing data is lower by a factor $0.76^{+0.10}_{-0.14}$ (at our pivot redshift \mbox{$z=0.6$}) compared to what would be needed to reconcile a flat {\it Planck} $\nu\Lambda$CDM cosmology (in which the sum of the neutrino masses is a free parameter) with the observed SPT-SZ cluster counts.
 In order to sensitively test the level of (dis-)agreement between SPT clusters and
 {\it Planck}, further expanded weak lensing follow-up samples are needed.
\end{abstract}


\section{Introduction}
\label{sec:intro}
Massive galaxy clusters trace the densest regions of the cosmic large-scale structure.
Robust constraints on their number density as a function of mass and redshift
provide a powerful route to constrain the
growth of structure and
thereby cosmological parameters \citep[e.g.][]{allen11,mantz15,dodelson16,bocquet19}.
For this endeavour
to be successful we not only need large cluster samples that have a well-characterised selection function, but also accurate mass measurements.

Suitable cluster samples are now in place, where one particularly powerful technique
is provided by the  Sunyaev-Zel'dovich
\citep[SZ,][]{sunyaev70,sunyaev72}
effect.
This effect describes a characteristic spectral distortion  of the cosmic microwave background (CMB), caused by  inverse Compton scattering of CMB photons off the electrons in the hot intra-cluster plasma.
SZ surveys do not suffer from cosmic dimming, which is why
high-resolution wide-area surveys, such as the ones conducted by the South Pole Telescope \citep[SPT,][]{carlstrom11} and the Atacama Cosmology Telescope
\citep[ACT,][]{swetz11}, have delivered large samples of massive clusters
that extend out to the highest redshifts where these clusters exist
\citep{bleem15,bleem20,hilton18,hilton21,huang20}.
As a further benefit, the SZ signal provides a mass proxy with a comparably low intrinsic scatter
\citep[\mbox{$\sim 20\%$}, e.g.][]{angulo12},
      which reduces the impact  residual uncertainties regarding the selection function have on the  cosmological parameter estimation.

Accurate cluster cosmology constraints require a careful calibration
of mass-observable scaling relations. As a key ingredient,  weak lensing (WL) observations provide the
most direct route to obtain the absolute calibration of these relations \citep[e.g.][]{allen11}.
So far, the majority of constraints have been obtained for clusters at low and intermediate redshifts (\mbox{$z\lesssim 0.6$}) using ground-based WL data \citep[e.g.][]{vonderlinden14,hoekstra15,okabe16,mcclintock19,miyatake19,stern19,umetsu20,herbonnet20}.
However, cluster properties may evolve with redshift,  making it imperative to extend the empirical WL mass calibration to higher redshifts.
For higher-redshift clusters deeper imaging with higher resolution is required in order to resolve
the typically small and faint distant background galaxies for WL shape measurements.
Stacked analyses
of large samples  can still yield sensitive WL constraints for clusters out to \mbox{$z\sim 1$}
when using very deep optical images  obtained from the ground over  wide areas under excellent seeing conditions \citep[][]{murata19}.
However, in order to achieve tight measurements for rare high-mass, high-redshift clusters, even deeper data are needed, as provided  e.g.~by the {\it Hubble} Space Telescope
\citep[HST, see e.g.][]{jee11,jee17,thoelken18,kim19}.

In the context of SPT,
\citet[][\citetalias{schrabback18} henceforth]{schrabback18}
presented a
WL analysis of 13 distant  (\mbox{$0.57\le z \le 1.13$})
galaxy clusters from the SPT-SZ survey \citep{bleem15}, using  mosaic HST/ACS
imaging
for galaxy shape
measurements.
\citet[][\citetalias{dietrich19} henceforth]{dietrich19} combined the
resulting HST WL constraints
with Magellan WL measurements of
SPT-SZ clusters at lower redshifts
in order to constrain X-ray and SZ
mass-observable scaling relations.
The same combined WL sample has  been employed by
\citet[][\citetalias{bocquet19} henceforth]{bocquet19} to derive first directly WL-calibrated
constraints on cosmology
from the SPT-SZ cluster sample.

Here we
 update the
 \citetalias{schrabback18} analysis
 and
 present
results for an expanded sample.
For the clusters in the \citetalias{schrabback18} sample we report updated
constraints, employing updated
calibrations for WL shape estimates \citep[][\citetalias{hernandez20} henceforth]{hernandez20} and the source redshift distribution \citep[\citetalias{raihan20} henceforth]{raihan20},
and incorporating deeper VLT/FORS2 photometry for the source selection for six clusters.
To this we add new measurements for 16 intermediate-mass clusters  with single-pointing ACS F606W imaging and Gemini-South GMOS photometry plus
one relaxed cluster with mosaic HST/ACS F606W+F814W imaging.

As the primary goal, our measurements aim at
improving the mass calibration for high-redshift SPT clusters,
thereby tightening constraints on the redshift-evolution of the SZ-mass scaling relation.
This is particularly important in order to improve dark energy constraints based on the SPT-SZ cluster sample:
as demonstrated by \citetalias{bocquet19},
constraints
on the dark energy equation of state parameter $w$ show a strong degeneracy with the parameter $C_\mathrm{SZ}$, which describes the  redshift evolution of the  SZ-mass scaling relation.
In order to improve the $w$ constraints we therefore need to tighten the constraints on $C_\mathrm{SZ}$ by adding WL data over a broad cluster redshift range.

This paper is organised as follows:
We describe the data and image reduction in Sect.~\ref{sec:data}, followed by
the photometric analysis and weak lensing measurements in
Sect.~\ref{sec:analysis}.
After presenting the weak lensing results in Sect.~\ref{sec:results},
we use these to derive revised constraints on the SPT observable--mass scaling relation in
 Sect.~\ref{sec:scaling_relations}.
We summarise our findings and conclude in  Sect.~\ref{sec:conclusions}.

Unless noted differently we assume a standard flat $\Lambda$CDM
cosmology in this paper, characterised by $\Omega_\mathrm{m}=0.3$,  $\Omega_\Lambda=0.7$, and
\mbox{$H_0=70 \,h_{70}\mathrm{km/s/Mpc}$} with $h_{70}=1$,
as approximately consistent with
CMB
constraints \citep[e.g.][]{hinshaw13,planck2018cosmology}.
We additionally assume \mbox{$\sigma_8=0.8$}, \mbox{$\Omega_\mathrm{b}=0.046$}, and \mbox{$n_\mathrm{s}=0.96$}
when estimating the noise caused by
 large-scale structure projections for weak lensing
mass estimates, as well as the computation of the
concentration--mass relation according to \cite{diemer19}.
The term $\nu\Lambda$CDM denotes
a flat $\Lambda$CDM
cosmology in which the sum of the neutrino masses is treated as a free parameter.
Transverse separations listed in this paper are physical distances, not comoving ones.
All magnitudes are in the AB system and  corrected for extinction according to \citet{schlegel98}.
The (multivariate) normal distribution with mean $\mu$ and covariance matrix $\mathbf\Sigma$ is written as $\mathcal N(\mu, \mathbf\Sigma)$.

\section{Sample, data and data reduction}
\label{sec:data}

\begin{table*}
\caption{Basic properties of the clusters with mosaic ACS imaging.
\label{tab:clusters_mos}}
\begin{center}
\addtolength{\tabcolsep}{-2pt}
\begin{tabular}{clccccccc}
\hline
\hline
Cluster name & $z_\mathrm{l}$ & $\xi$ & \multicolumn{4}{c}{Centre coordinates [deg J2000]} & $M_\mathrm{500c,SZ}$ &Sample/Data\\
& & & SZ $\alpha$ & SZ $\delta$  & X-ray $\alpha$ & X-ray $\delta$ & [$10^{14} \mathrm{M}_\odot h_{70}^{-1}$] &\\
\hline
SPT-CL{\thinspace}$J$0000$-$5748 & 0.702 & 8.49 & 0.2499 & $-57.8064$ & 0.2518
                                                                & $-57.8094$ &  $4.33^{+0.65}_{-0.86}$
                                                                                                                              &
                                      \citetalias{schrabback18} + new VLT\\
  SPT-CL{\thinspace}$J$0102$-$4915 & 0.870 & 39.91 & 15.7294 & $-49.2611$ & 15.7350 & $-49.2667$ &  $13.15^{+2.08}_{-2.83}$
                                                                                                                              &  \citetalias{schrabback18}\\
SPT-CL{\thinspace}$J$0533$-$5005 & 0.881 & 7.08 & 83.4009 & $-50.0901$ &
                                                                         83.4018 & $-50.0969$ & $3.75^{+0.59}_{-0.82}$
                                                                                                                              &
                                                               \citetalias{schrabback18} + new VLT\\
SPT-CL{\thinspace}$J$0546$-$5345 & 1.066 & 10.76 & 86.6525 & $-53.7625$ &
                                                                          86.6532 & $-53.7604$ &   $4.85^{+0.74}_{-1.04}$
                                                                                                                              &  \citetalias{schrabback18}\\
SPT-CL{\thinspace}$J$0559$-$5249 & 0.609 & 10.64 & 89.9251 & $-52.8260$ &
                                                                          89.9357 & $-52.8253$ &  $5.33^{+0.80}_{-0.95}$
                                                                                                                              &  \citetalias{schrabback18}\\
SPT-CL{\thinspace}$J$0615$-$5746 & 0.972 & 26.42 & 93.9650 & $-57.7763$ &
                                                                          93.9652 & $-57.7788$ & $9.67^{+1.58}_{-2.16}$
                                                                                                                              &  \citetalias{schrabback18}\\
  SPT-CL{\thinspace}$J$2040$-$5725 & 0.930 & 6.24 & 310.0573 & $-57.4295$ & 310.0631$^*$ & $-57.4287^*$ & $3.35^{+0.60}_{-0.81}$
                                                                                                                              &  \citetalias{schrabback18} + new VLT\\
  SPT-CL{\thinspace}$J$2043$-$5035 & 0.723 & 7.18 & 310.8284 & $-50.5938$ & 310.8244 & $-50.5930$ &  $4.38_{-0.91}^{+0.72}$
 & new HST\\
SPT-CL{\thinspace}$J$2106$-$5844 & 1.132 & 22.22 & 316.5206 & $-58.7451$ &
                                                                           316.5174 & $-58.7426$ &  $7.76^{+1.19}_{-1.84}$
                                                                                                                              &  \citetalias{schrabback18}\\
SPT-CL{\thinspace}$J$2331$-$5051 & 0.576 & 10.47 & 352.9608 & $-50.8639$ &
                                                                           352.9610 & $-50.8631$ &  $5.17^{+0.75}_{-0.93}$
                                                                                                                              &  \citetalias{schrabback18}\\
SPT-CL{\thinspace}$J$2337$-$5942 & 0.775 & 20.35 & 354.3523 & $-59.7049$ &
                                                                           354.3516 & $-59.7061$ &  $7.67^{+1.14}_{-1.46}$
                                                                                                                              &  \citetalias{schrabback18} + new VLT\\
SPT-CL{\thinspace}$J$2341$-$5119 & 1.003 & 12.49 & 355.2991 & $-51.3281$ &
                                                                           355.3009 & $-51.3285$ &  $5.30^{+0.82}_{-1.09}$
                                                                                                                              &  \citetalias{schrabback18} + new VLT\\
SPT-CL{\thinspace}$J$2342$-$5411 & 1.075 & 8.18 & 355.6892 & $-54.1856$ &
355.6904 & $-54.1838$ & $3.86^{+0.64}_{-0.88}$
                                                                                                                              &
                                                                    \citetalias{schrabback18}\\
SPT-CL{\thinspace}$J$2359$-$5009 & 0.775 & 6.68 & 359.9230 & $-50.1649$ &
                                                                          359.9321 & $-50.1697$ &  $3.54^{+0.61}_{-0.76}$
                                                                                                                              &  \citetalias{schrabback18} + new VLT\\
                                                                          \hline
\end{tabular}
\addtolength{\tabcolsep}{+2pt}
\end{center}

\begin{justify}
Note. --- Basic data from \citet{mcdonald13}, \citet{bleem15}, \citet{chiu16}, and \citetalias{bocquet19} for the 14 clusters with mosaic HST imaging included in this
  weak lensing analysis.
{\it Column 1:} Cluster designation.
{\it Column 2:} Spectroscopic cluster redshift.
{\it Column 3:} Peak
signal-to-noise ratio
of the SZ
  detection.
{\it Columns 4--7:} Right ascension $\alpha$ and declination $\delta$ of the
SZ peak and X-ray centroid.
$^*$: X-ray centroid from XMM-Newton data,
otherwise {\it Chandra}.
{\it Column 8:} SZ-inferred mass from \citetalias{bocquet19}, fully marginalising over cosmology and scaling relation parameter uncertainties.
{\it Column 9:} Here we indicate the use of new HST or VLT data and whether the cluster was already
included in the  \citetalias{schrabback18} analysis.
\end{justify}
\end{table*}

\begin{table*}
\caption{Basic properties of the clusters with single-pointing ACS imaging.
\label{tab:clusters_snap}}
\begin{center}
\begin{tabular}{cccccc}
\hline
\hline
Cluster name & $z_\mathrm{l}$ & $\xi$ & \multicolumn{2}{c}{SZ peak position} & $M_\mathrm{500c,SZ}$\\
& & & $\alpha$ [deg J2000] &  $\delta$ [deg J2000]  &  [$10^{14} \mathrm{M}_\odot h_{70}^{-1}$] \\
  \hline
  SPT-CL{\thinspace}$J$0044$-$4037 & $1.02\pm0.09$ & 4.92 & 11.1232 & $-40.6282$ & $2.80_{-0.80}^{+0.58}$ \\
SPT-CL{\thinspace}$J$0058$-$6145 & $0.82\pm0.03$ & 7.52 & 14.5799 & $-61.7635$ & $4.27_{-0.91}^{+0.70}$ \\
SPT-CL{\thinspace}$J$0258$-$5355 & $0.99\pm0.09$ & 4.96 & 44.5227 & $-53.9233$ & $2.88_{-0.80}^{+0.54}$ \\
SPT-CL{\thinspace}$J$0339$-$4545 & $0.86\pm0.03$ & 5.34 & 54.8908 & $-45.7535$ & $3.01_{-0.78}^{+0.57}$ \\
SPT-CL{\thinspace}$J$0344$-$5452 & $1.05\pm0.09$ & 7.98 & 56.0922 & $-54.8794$ & $4.02_{-0.93}^{+0.67}$ \\
SPT-CL{\thinspace}$J$0345$-$6419 & $0.94\pm0.03$ & 5.54 & 56.2510 & $-64.3326$ & $3.08_{-0.79}^{+0.64}$ \\
SPT-CL{\thinspace}$J$0346$-$5839 & $0.70\pm0.04$ & 4.83 & 56.5733 & $-58.6531$ & $2.92_{-0.77}^{+0.56}$ \\
SPT-CL{\thinspace}$J$0356$-$5337 & $1.036$ & 6.02 & 59.0855 & $-53.6331$ & $3.21_{-0.81}^{+0.62}$ \\
SPT-CL{\thinspace}$J$0422$-$4608 & $0.66\pm0.04$ & 5.05 & 65.7490 & $-46.1436$ & $3.05_{-0.78}^{+0.59}$ \\
SPT-CL{\thinspace}$J$0444$-$5603 & $0.94\pm0.03$ & 5.18 & 71.1136 & $-56.0576$ & $2.91_{-0.77}^{+0.55}$ \\
SPT-CL{\thinspace}$J$0516$-$5755 & $0.97\pm0.03$ & 5.73 & 79.2398 & $-57.9167$ & $3.05_{-0.77}^{+0.58}$ \\
SPT-CL{\thinspace}$J$0530$-$4139 & $0.78\pm0.05$ & 6.19 & 82.6754 & $-41.6502$ & $3.92_{-0.89}^{+0.68}$ \\
SPT-CL{\thinspace}$J$0540$-$5744 & $0.761$ & 6.74 & 85.0043 & $-57.7405$ & $3.67_{-0.78}^{+0.62}$ \\
SPT-CL{\thinspace}$J$0617$-$5507 & $0.95\pm0.09$ & 5.53 & 94.2808 & $-55.1321$ & $3.23_{-0.85}^{+0.63}$ \\
SPT-CL{\thinspace}$J$2228$-$5828 & $0.73\pm0.05$ & 5.15 & 337.2153 & $-58.4686$ & $3.27_{-0.83}^{+0.63}$ \\
SPT-CL{\thinspace}$J$2311$-$5820 & $0.93\pm0.09$ & 5.72 & 347.9924 & $-58.3452$ & $2.97_{-0.74}^{+0.60}$ \\
\hline
\end{tabular}
\end{center}
      \begin{justify}
  Note. --- Basic data from \citetalias{bocquet19} for the SNAP clusters with single-pointing ACS imaging included in this
  weak lensing analysis.
 {\it Column 1:} Cluster designation.
{\it Column 2:} Cluster redshift. Photometric (spectroscopic) redshifts are indicated with (without) error-bars.
{\it Column 3:} Peak
signal-to-noise ratio
of the SZ
  detection.
{\it Columns 4--5:} Right ascension $\alpha$ and declination $\delta$ of the
SZ peak
location.
{\it Column 6:} SZ-inferred mass from \citetalias{bocquet19}, fully marginalising over cosmology and scaling relation parameter uncertainties.
\end{justify}
\end{table*}

All targets of our weak lensing analysis originate from the
2,500 deg$^2$ SPT-SZ galaxy cluster survey
\citep{bleem15}.
Here we employ updated cluster redshift estimates (see Tables \ref{tab:clusters_mos} and \ref{tab:clusters_snap} for a summary of basic
properties) from \citet{bayliss16} and
\citetalias[][]{bocquet19}.

\subsection{HST/ACS observations}

\subsubsection{High-mass clusters with ACS mosaics}
\label{se:data:acs:highmass}
\citetalias{schrabback18} presented a weak lensing analysis for 13
 high-redshift SPT-SZ
clusters. They measured galaxy shapes in \mbox{$2\times 2$} HST/ACS F606W mosaic
images (1.92ks per pointing) and incorporated HST/ACS F814W imaging for the source selection (a single central F814W pointing for all clusters plus a \mbox{$2\times 2$} mosaic for SPT-CL{\thinspace}$J$0615$-$5746).
We include these clusters in our analysis, where we apply updated shape and
redshift calibrations for the source galaxies for all clusters (see
Sect.\thinspace\ref{sec:analysis}),
and additionally incorporate  deeper VLT/FORS2 $I_\mathrm{FORS2}$ band imaging for the source
selection for six of the clusters (see Sect.\thinspace\ref{se:data_vlt}).
We refer readers to \citetalias{schrabback18} for details on the original
data sets and analysis for these clusters, and primarily describe
 changes compared to this earlier analysis in the current work.

With {\SPT2043} we include a further cluster with \mbox{$2\times 2$} HST/ACS
mosaics in our analysis. This target was observed as part of a joint {\it
  Chandra}+HST programme \citep[HST programme ID
14352, PI: J.~Hlavacek-Larrondo, see also][]{mcdonald19}, which has obtained imaging
in both F606W (1.93ks per pointing) and F814W (1.96ks per pointing).
For this cluster we also incorporate central single pointing HST/ACS
F606W imaging (1.44ks) obtained as part of the SPT ACS  Snapshot Survey (SNAP
13412, PI:
T.~Schrabback).

\subsubsection{Intermediate-mass clusters with single-pointing ACS imaging}

From the
SPT ACS  Snapshot Survey (see Sect.\thinspace\ref{se:data:acs:highmass})
 we additionally
 incorporate single pointing
ACS F606W imaging for an additional 16 SPT-SZ clusters\footnote{The SPT ACS
  Snapshot Survey observed a total of 46 SPT-SZ clusters between Oct 23,
  2013 and Sep 7, 2015. We limit the
  current analysis to targets for which adequate $I$-band imaging is
  available for the source colour selection.}.
These observations have total integration times between 1.44ks and 2.32ks
(see Table \ref{tab:snapdata}), depending on cluster redshift and orbital visibility.
These clusters have lower SZ detection significances and are
therefore expected to be less massive
compared to most of the clusters with mosaic ACS data (compare Tables \ref{tab:clusters_mos} and \ref{tab:clusters_snap}), leading to a smaller physical extent (e.g.~in terms of the radius $r_\mathrm{500c}$,
within which
the average density is 500 times the critical density of the Universe at the cluster redshift).
While not ideal, the limited radial coverage provided by single-pointing ACS data is
therefore still acceptable for these lower mass systems.

\subsubsection{HST data reduction}
For all data sets the
observations were split into four exposures per pointing and filter, in
order to
facilitate good cosmic ray removal.
We employ \texttt{CALACS} for basic image
reductions, except for the correction for charge-transfer inefficiency
(CTI), which is done using the method developed by \citet{massey2014}.
For further image reductions we employ scripts from
\citet{schrabback10} for the image registration and optimisation of masks
and weights,
as well as \texttt{MultiDrizzle} \citep{koekemoer2003} for the cosmic ray
removal and stacking (see \citetalias{schrabback18} for further details).

\subsection{VLT/FORS2 observations}
\label{se:data_vlt}

For six of the clusters
 initially
studied
by \citetalias{schrabback18} we incorporate new VLT/FORS2
imaging
obtained in the I\_BESS+77 filter (which we call $I_\mathrm{FORS2}$)
via programmes 0100.A-0217 (PI:  B.~Hern\'andez-Mart\'in),
0101.A-0694 (PI: H.~Zohren), and 0102.A-0189 (PI: H.~Zohren) into our analysis.
These new observations are significantly deeper and have a better image quality
(see Table \ref{tab:vltdata}) compared to the VLT data used by
\citetalias{schrabback18}, thereby allowing us to include fainter source
galaxies in the weak lensing analysis (see Sect.\thinspace\ref{sec:analysis}).
Following \citetalias{schrabback18} we reduce the new VLT images
using  \texttt{theli} \citep{erben05,schirmer13}, where we apply bias and flat-field corrections, relative photometric
calibration, and sky background subtraction employing \texttt{SExtractor} \citep{bertin1996}.
We do not include the earlier shallower observations in the stack for two reasons.
First, their inclusion would typically degrade the image quality in the
stack given their looser image quality requirements.
Additionally, they suffer from flat-field uncertainties \citep{moehler10}, which have been fixed prior to the new observations via an exchange of the FORS2 longitudinal atmospheric
dispersion corrector (LADC) prisms \citep{boffin16}.

\begin{table}
  \caption{The new VLT/FORS2 $I_\mathrm{FORS2}$ imaging data
     for clusters in the
    ``updated ACS+FORS2 sample''.
\label{tab:vltdata}}
\begin{center}
\begin{tabular}{cccc}
\hline
\hline
Cluster name & $t_\mathrm{exp}$ & $I_\mathrm{lim}(0\farcs8)$ & $2r_\mathrm{f}^*$ \\
\hline
SPT-CL{\thinspace}$J$0000$-$5748 & 10.6ks  & 27.3 & 0\farcs70 \\
SPT-CL{\thinspace}$J$0533$-$5005 & 8.4ks  & 27.3 & 0\farcs59 \\
SPT-CL{\thinspace}$J$2040$-$5726 & 7.3ks  & 27.1 & 0\farcs62 \\
SPT-CL{\thinspace}$J$2337$-$5942 & 7.1ks  & 27.3 & 0\farcs64 \\
SPT-CL{\thinspace}$J$2341$-$5119 & 6.6ks  & 27.4 & 0\farcs63 \\
SPT-CL{\thinspace}$J$2359$-$5009 & 6.8ks  & 27.4 & 0\farcs69 \\
\hline
\end{tabular}
\end{center}
\begin{justify}
Note. --- Details of the analysed VLT/FORS2 imaging data.
{\it Column 1:} Cluster designation.
{\it Column 2:} Total co-added exposure time.
{\it Column 3:} $5\sigma$-limiting magnitude using 0\farcs8 apertures,
computed by placing apertures at random field locations that do not overlap with detected objects.
{\it Column 4:} Image Quality defined as $2\times$ the \texttt{FLUX\_RADIUS} estimate of stellar sources
 from \texttt{SExtractor}.
\end{justify}
\end{table}

\begin{table}
  \caption{Properties of HST/ACS SNAP and Gemini-South GMOS $i_\mathrm{GMOS}$ imaging data for clusters in the ``ACS+GMOS sample''.
\label{tab:snapdata}}
\begin{center}
\begin{tabular}{ccccc}
\hline
\hline

Cluster name & $t_\mathrm{exp}^\mathrm{ACS}$ & $t_\mathrm{exp}^\mathrm{GMOS}$ & $i_\mathrm{lim}(1\farcs5)$ &  $2r_\mathrm{f}^*$\\
\hline
SPT-CL{\thinspace}$J$0044$-$4037 & 2.1ks &  6.2ks & 26.2 & 0\farcs93 \\
SPT-CL{\thinspace}$J$0058$-$6145 & 2.3ks &  6.7ks & 25.8 & 0\farcs92 \\
SPT-CL{\thinspace}$J$0258$-$5355 & 2.3ks &  6.2ks & 26.0 & 0\farcs70 \\
SPT-CL{\thinspace}$J$0339$-$4545 & 2.1ks &  4.8ks & 26.0 & 0\farcs88 \\
SPT-CL{\thinspace}$J$0344$-$5452 & 2.3ks &  5.6ks & 25.4 & 0\farcs92 \\
SPT-CL{\thinspace}$J$0345$-$6419 & 2.3ks &  5.6ks & 26.1 & 0\farcs69 \\
SPT-CL{\thinspace}$J$0346$-$5839 & 1.4ks &  5.4ks & 25.9 & 0\farcs82 \\
SPT-CL{\thinspace}$J$0356$-$5337 & 2.3ks &  5.2ks & 26.0 & 0\farcs77 \\
SPT-CL{\thinspace}$J$0422$-$4608 & 1.4ks &  5.2ks & 25.9 & 0\farcs66 \\
SPT-CL{\thinspace}$J$0444$-$5603 & 2.3ks &  7.9ks & 25.9 & 0\farcs72 \\
SPT-CL{\thinspace}$J$0516$-$5755 & 2.3ks &  5.2ks & 25.8 & 0\farcs85 \\
SPT-CL{\thinspace}$J$0530$-$4139 & 1.4ks &  5.0ks & 26.1 & 0\farcs77 \\
SPT-CL{\thinspace}$J$0540$-$5744 & 1.4ks &  5.9ks & 25.8 & 0\farcs72 \\
SPT-CL{\thinspace}$J$0617$-$5507 & 2.3ks &  5.2ks & 26.0 & 0\farcs91 \\
SPT-CL{\thinspace}$J$2228$-$5828 & 2.3ks &  5.4ks & 25.8 & 0\farcs75 \\
SPT-CL{\thinspace}$J$2311$-$5820 & 1.4ks &  5.6ks & 25.9 & 0\farcs99 \\
\hline
\end{tabular}
\end{center}
\begin{justify}
Note. --- Details of the analysed ACS and Gemini-South GMOS imaging data.
{\it Column 1:} Cluster designation.
{\it Column 2:} Total co-added exposure time with ACS in F606W.
{\it Column 3:} Total co-added exposure time with GMOS in $i_\mathrm{GMOS}$.
{\it Column 4:} $5\sigma$-limiting magnitude using 1\farcs5 apertures,
computed by placing apertures at random field locations that do not overlap with detected objects.
{\it Column 5:}  Image Quality defined as $2\times$ the \texttt{FLUX\_RADIUS} estimate of stellar sources
from \texttt{SExtractor}.
\end{justify}
\end{table}

\subsection{Gemini-South observations}
We obtained Gemini-South GMOS $i$-band imaging
via NOAO programmes 2014B-0338
and 	2016B-0176 (PI: B.~Benson)
for a subset of the clusters observed by the SNAP programme.
In our analysis we include observations of 16 clusters, which have been observed to the full depth under good conditions (see Table \ref{tab:snapdata}).
Similarly to the VLT data we reduced the GMOS images using  \texttt{theli},
where we included only the central GMOS chip in the stack as it covers most of the ACS area.\footnote{This also avoids complications due to
  differences in the  quantum efficiency curves of the different GMOS-S
  CCD chips.}

\section{Analysis}
\label{sec:analysis}

\subsection{Shape measurements}
\label{se:shapes}
\citetalias{schrabback18} measured WL galaxy shapes for the clusters
with mosaic ACS plus FORS2 observations
(``ACS+FORS2 sample'') from the ACS F606W images, employing \texttt{SExtractor}  \citep{bertin1996} for object detection and deblending, and the KSB+ formalism \citep{kaiser1995,luppino1997,hoekstra1998} for shape measurements as implemented by \citet{erben2001} and \citet{schrabback07}.
They modelled the spatial and temporal variations of the ACS point-spread function (PSF)  using  principal component analysis as done by \citet{schrabback10}.
Here we apply the same pipeline to also measure galaxy shapes for the remaining
clusters in our larger sample.

As a significant update we employ the revised calibration of our shape measurement pipeline
from
\citetalias{hernandez20}
for all of our targets.
This calibration was derived using custom \texttt{galsim} \citep{rowe15} image simulations that closely resemble our ACS data.
\citetalias{hernandez20}  mimic our observations  in terms of depth, detector characteristics and point-spread function, and, importantly,
adjust the galaxy sample such that its measured distributions in
magnitude, size, and signal-to-noise ratio, as well as the ellipticity dispersion, closely match the corresponding observed quantities of our magnitude- and colour-selected source sample.
They also employ distributions of galaxy light profiles that approximately resemble our   colour-selected source population.
\citetalias{hernandez20} derive an updated correction  for noise bias, where they assume a power-law dependence on the
KSB signal-to-noise  ratio \mbox{$S/N_\mathrm{KSB}$} \citep[incorporating the KSB weight function, see][]{erben2001} similar to \citet{schrabback10}.
They also
obtain corrections to account for selection bias, the impact of neighbours and faint sources below the detection threshold
  \citep[see also][]{martinet19}, and the increased light contamination caused by cluster galaxies.
They demonstrate that our pipeline
does not suffer from significant non-linear multiplicative shear biases
in the regime of non-weak shears, which can occur in the inner cluster regions.
Furthermore, they show that galaxies with slightly lower signal-to-noise ratios \mbox{$S/N_\mathrm{flux}>7$},
defined via  \texttt{SExtractor} parameters \mbox{$S/N_\mathrm{flux}=\mathrm{FLUX\_AUTO/FLUXERR\_AUTO}$},
can be robustly included in the analysis when their revised noise-bias calibration is applied.
We therefore employ this updated cut to boost the source number density (for comparison, \citetalias{schrabback18} used galaxies with \mbox{$S/N_\mathrm{flux}>10$})\footnote{However, because of the
additional magnitude selection, which is applied to keep the photometric scatter small (see Sect.\thinspace\ref{se:photo_color}),
the average increase in the source density  compared to \citetalias{schrabback18} is quite small, amounting to $10\%$ for the ACS-only selection and 5\% for the ACS+FORS2 selection (for the clusters without new photometric data).}
and apply a bias correction
\begin{eqnarray}
m_{1,\text{corr}} & = &-0.358 \left( S/N_{\mathrm{KSB}}\right) ^{-1.145}\,-0.042 , \nonumber \\
m_{2,\text{corr}}& =& -0.357 \left( S/N_{\mathrm{KSB}}\right) ^{-1.298}\,-0.042\, , \label{eq:sn_corr_final}
\end{eqnarray}
based on the \citetalias{hernandez20} results\footnote{We adjust the $m_{2,\text{corr}}$ correction by $-0.003$ compared to Eq.{\thinspace}(14) of \citetalias{hernandez20} to compensate for their slight final residual $m_2$ bias after calibration.}
to the components of the KSB+ ellipticity estimates $\epsilon_\alpha^\mathrm{biased}$
on a galaxy-by-galaxy basis,
to obtain corrected ellipticity estimates
\begin{equation}
\epsilon_\alpha=\frac{\epsilon_\alpha^\mathrm{biased}}{1+m_{\alpha,\text{corr}}} \,,
\end{equation}
which act as unbiased estimates of the reduced shear $g$
\begin{equation}
  \langle\epsilon_\alpha \rangle = g_\alpha\,.
\end{equation}
  Varying various aspects of the simulations,  \citetalias{hernandez20}
  conclude that our fully calibrated KSB+ pipeline
  yields accurate estimates for the reduced shear $g$
  with an estimated relative systematic uncertainty of $1.5\%$,
  which we therefore include in our systematic error budget.

  When applying the same \mbox{$S/N_\mathrm{flux}>10$} selection as \citetalias{schrabback18} and considering
  the ACS-only colour selection,
  we find that the new calibration increases the reduced shear estimates for our galaxies on average by
3.5\%.
  Several effects contribute to this
  shift in the shear calibration, where the largest contributions come from the updated noise-bias correction, as well as the corrections for selection bias and the impact of faint sources below the detection threshold.
  The previously employed calibration from  \citetalias{schrabback10} did not account for the latter two effects, and its source samples did not adequately reflect our colour-selected sample of mostly background galaxies, leading to the shift in the noise bias correction.
We however stress that the shift in the shear calibration is still within the
the
4\% systematic shear calibration uncertainty, which was included in the \citetalias{schrabback18} analysis to
account for the limitations in the  \citetalias{schrabback10} shear calibration.

 Additional changes in the (noisy) reduced shear profiles for the previously studied clusters occur due to the inclusion of  galaxies with \mbox{$7<S/N_\mathrm{flux}<10$}, and the deeper photometric source selection in the case of clusters with new VLT data  (see Sect.\thinspace\ref{se:photo_color}).

Note that \citet{hoekstra15} apply a bias calibration for their KSB+ implementation which is a function of both the
galaxy signal-to-noise ratio and a resolution factor that depends on the half-light radii of the PSF and the galaxy.
Capturing such a size dependence is less important for space-based data as variations in PSF size are much smaller compared to typical seeing-limited ground-based data. In addition, the variation in galaxy sizes is smaller in our case given the selection of mostly high-redshift galaxies via colour (see Sect.\thinspace\ref{se:photo_color}).
\citetalias{hernandez20} show that the residual multiplicative shear bias of our KSB+ implementation
(after applying the
\mbox{$S/N_\mathrm{KSB}$}-dependent correction)
depends only weakly
on the  \texttt{FLUX\_RADIUS} parameter
$r_\mathrm{f}$
from \texttt{SExtractor} (within \mbox{$\sim \pm 5\%$} for most of the galaxies).
 Combined with
the weak dependence
of
 the average geometric lensing efficiency
 on $r_\mathrm{f}$   for our colour- and magnitude-selected source sample
(see Appendix \ref{se:beta_of_rf}),
we can therefore safely ignore second-order effects for the bias correction.

\begin{figure*}
  \includegraphics[width=0.69\columnwidth]{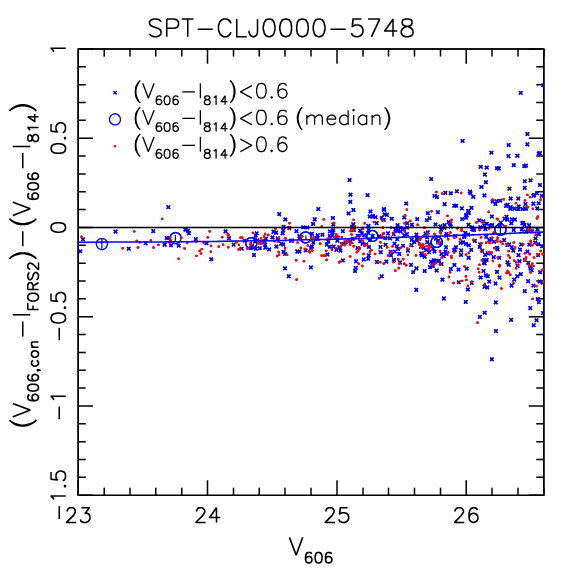}
  \includegraphics[width=0.69\columnwidth]{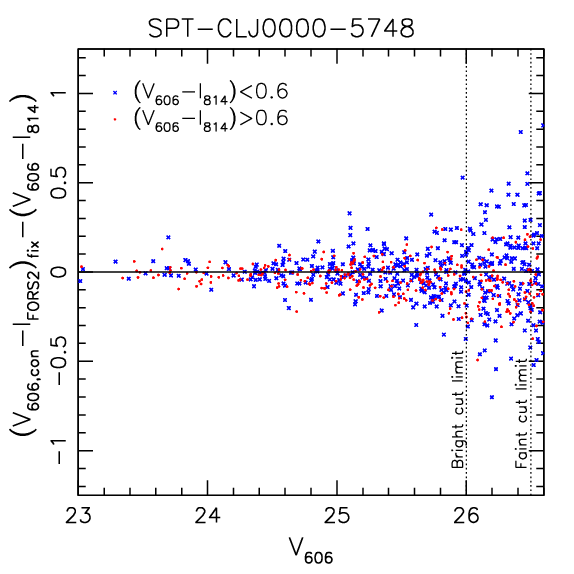}
  \includegraphics[width=0.68\columnwidth]{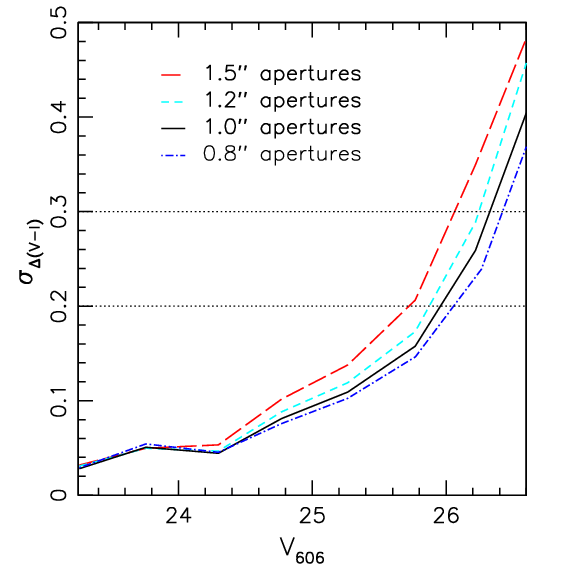}
\caption{{\it Left:}
Measured colour difference \mbox{$\Delta (V-I)=(V_{606,\mathrm{con}}-I_\mathrm{FORS2})-(V_{606}-I_{814})$}
between the PSF homogenised ACS+FORS2 colour estimate $V_{606,\mathrm{con}}-I_\mathrm{FORS2}$ (measured using 0\farcs8 apertures) and the ACS-only
colour estimate $(V_{606}-I_{814})$ in the inner region of SPT-CL{\thinspace}$J$0000$-$5748 as a function of \mbox{$V_{606}$}. Blue galaxies
  with \mbox{$(V_{606}-I_{814})<0.6$} are shown as small blue crosses, while red galaxies
  with \mbox{$(V_{606}-I_{814})>0.6$} are indicated as red points. The open circles
  show the median values for the blue galaxies
  in magnitude
  bins, where the (small) error-bars correspond to  the uncertainty on the mean for a
  Gaussian distribution and the curve shows their best-fit second-order
  polynomial interpolation. {\it Middle}: Here we show the same data after
  subtraction of this function. The photometric scatter
  distribution for the ACS+FORS2 selection is sampled from this distribution of offsets.
   The vertical
   lines separate  the magnitude ranges for the different colour cuts.
   {\it Right}:
Scatter in the model-subtracted \mbox{$\Delta (V-I)$} colour offsets
  as a function of $V_{606}$, averaged over all clusters listed in Table
  \ref{tab:vltdata}. The different curves correspond to different aperture
  diameters in the ACS+FORS2 analysis. The dotted horizontal lines indicate
  the scatter limits \citetalias{schrabback18} employed to define the bright cut and faint cut in their colour selection.
\label{fig:scatter}}
\end{figure*}

\subsection{Photometry and colour selection}
\label{se:photo_color}
As done by \citetalias{schrabback18} we select weak lensing source
galaxies via $V-I$ colour, allowing us to efficiently remove both red and blue cluster members (for clusters at redshifts \mbox{$0.6\lesssim z_\mathrm{l}\lesssim 1$}) as well as the majority of foreground galaxies, and keep most of the lensed background galaxies at \mbox{$z\gtrsim 1.4$}. For {\SPT2043} and the inner regions of the clusters with VLT observations (Table \ref{tab:vltdata}) we can directly employ \mbox{$V_{606}-I_{814}$} colours measured in the HST/ACS data (``ACS-only'' colours). Following \citetalias{schrabback18}   we here employ apertures with diameter 0\farcs7 to be consistent with the definitions of the photometric redshift catalogue from \citet[][see Sect.\thinspace\ref{se:redshiftdist}]{skelton14} and select  \mbox{$24<V_{606}<26$} galaxies with \mbox{$V_{606}-I_{814}<0.3$} plus \mbox{$26<V_{606}<26.5$} galaxies with \mbox{$V_{606}-I_{814}<0.2$}.

For the clusters in the ACS+GMOS sample (Table \ref{tab:snapdata}) as well as the outskirts of the clusters in the updated ACS+FORS2 sample (Table \ref{tab:vltdata}) we have to rely on PSF-homogenised colour measurements between the ACS F606W images and the ground-based $i_\mathrm{GMOS}$- or $I_\mathrm{FORS2}$-band images from Gemini-South/GMOS or VLT/FORS2, respectively.
After
homogenising the PSF\footnote{We convolve the ACS data with a Gaussian kernel in order to match  the \texttt{SExtractor} FLUX\_RADIUS of stars between the corresponding GMOS/FORS2 image and the convolved ACS image. For the clusters in the ACS+FORS2 sample we alternatively tested the use of a Moffat kernel, finding no significant improvement in the colour measurements when compared to the ACS-only colours.}
we measure convolved aperture colours \mbox{$V_{606,\mathrm{con}}-I_\mathrm{FORS2}$} and \mbox{$V_{606,\mathrm{con}}-i_\mathrm{GMOS}$}, respectively, using a range of aperture diameters.

For all data sets we employ conservative masks to remove regions near bright stars, very extended galaxies, and the image boundaries.

\subsubsection{ACS+FORS2 analysis}
\label{se:color:acs_fors2}
For the ACS+FORS2 sample the following steps of the colour measurements and colour selection closely follow  Appendix D of \citetalias{schrabback18}.
Here we only describe the updated analysis for the clusters with new VLT observations. For the other ACS+FORS2 clusters the colour measurements and selections were described in \citetalias{schrabback18} and have not been changed for this reanalysis.

In order to achieve a residual FORS2 zero-point calibration and a consistent colour selection between  the   \mbox{$V_{606,\mathrm{con}}-I_\mathrm{FORS2}$} and \mbox{$V_{606}-I_{814}$} colours
we compute colour offsets
\begin{equation}
  \Delta (V-I)=(V_{606,\mathrm{con}}-I_\mathrm{FORS2})-(V_{606}-I_{814})
\end{equation}
for blue galaxies in the overlap region of the $I_\mathrm{FORS2}$ images and the central ACS
F814W
images (see the left panel of Fig.\thinspace\ref{fig:scatter} for an example).
We then fit the median of these offsets as a function of $V_{606}$ aperture magnitude using a second-order polynomial  and subtract this model from the measured $V_{606,\mathrm{con}}-I_\mathrm{FORS2}$ colours,
providing corrected colour estimates \mbox{$(V_{606,\mathrm{con}}-I_\mathrm{FORS2})_\mathrm{fix}$} (see the middle panel of Fig.\thinspace\ref{fig:scatter}) not only in the inner cluster region, but also the full field covered by FORS2.

The right panel of Fig.\thinspace\ref{fig:scatter}
shows the measured scatter in $\Delta (V-I)$ as a function of  $V_{606}$ magnitude after the model subtraction
for different aperture diameters, averaged over the six clusters with new VLT data.
This clearly shows that the 1\farcs5 apertures employed by \citetalias{schrabback18} are not optimal for the new VLT data, which is a result of the excellent image quality of the new observations and the typically very small spatial extent of the faint blue galaxies constituting our source sample.
For the ACS+FORS2 analysis of the clusters with new VLT data we therefore employ smaller apertures with diameter 0\farcs8, which significantly reduces the scatter in the colour differences to the ACS-only colours.
Together with the longer FORS2 integration times this allows us to include fainter galaxies in the ACS+FORS2 colour selection compared to the \citetalias{schrabback18} analysis, where we now select  \mbox{$24<V_{606}<26$}  galaxies with  \mbox{$(V_{606,\mathrm{con}}-I_\mathrm{FORS2})_\mathrm{fix}<0.2$}  (``bright cut'' regime in the middle panel of Fig.\thinspace\ref{fig:scatter}) plus
\mbox{$26<V_{606}<26.5$} galaxies with \mbox{$(V_{606,\mathrm{con}}-I_\mathrm{FORS2})_\mathrm{fix}<0.0$} (``faint cut'' regime in the middle panel of Fig.\thinspace\ref{fig:scatter}).

When calibrating the source redshift distribution (see Sect.\thinspace\ref{se:redshiftdist})
we have to account for the impact of photometric scatter.
To model the scatter compared to the ACS-only colours
we
then
sample from the measured
scatter distribution in $\Delta (V-I)$ for each cluster in the ACS+FORS2 sample (see the middle panel of Fig.\thinspace\ref{fig:scatter} for an example), split into magnitude and colour bins as done by \citetalias{schrabback18}.

\subsubsection{ACS+GMOS analysis}
\label{se:color:acs_gmos}

For the ACS+GMOS sample ACS F814W imaging is not available, which is why we cannot directly apply the same colour calibration scheme.
Instead, we calibrate the colours via shallower Magellan/PISCO
$griz$ photometry, which itself has been calibrated using stellar locus regression to the SDSS photometric system
\citep[corrected for galactic extinction, see][]{bleem20}.

For the cluster  SPT-CL{\thinspace}$J$0615$-$5746
both  PISCO photometry and HST/ACS \mbox{$V_{606}-I_{814}$} colours (from \citetalias{schrabback18}) are available,
allowing us to calibrate the transformation
\begin{equation}
  \label{eq:pisco_trans}
  (V_{606}-I_{814})- (r-i) \simeq (0.222\pm 0.025) (g-i-1.0) + 0.096 \pm 0.014
\end{equation}
using stars with \mbox{$20<V_{606}<22$} and \mbox{$g-i<2$}.
Alternative choices to include fainter stars or galaxies
change the
fit coefficients in Eq.\thinspace\ref{eq:pisco_trans} slightly,
but affect the resulting transformed colour in the regime of our colour cuts
by \mbox{$\le 0.01$} mag only, providing sufficient accuracy for our study.

Employing Eq.\thinspace\ref{eq:pisco_trans} we compute
the transformed
\mbox{$V_{606}-I_{814}$} colours  for the PISCO objects
in the fields of the ACS+GMOS clusters.
Using overlapping bright
objects with \mbox{$20<V_{606}<23$}
from our ACS+GMOS photometry
we then derive the required transformation from \mbox{$V_{606,\mathrm{con}}-i_\mathrm{GMOS}$} to \mbox{$V_{606}-I_{814}$}.
Here we first compute a linear fit \mbox{$(V_{606}-I_{814})=a (V_{606,\mathrm{con}}-i_\mathrm{GMOS}) + b$} between these colours for each cluster field separately.
To reduce the sensitivity to outliers we then fix the slope to the median slope from all fields
\mbox{$a_\mathrm{med}=1.147\pm 0.013$} in a second step and redetermine  $b$ using a median estimate
for each cluster field, effectively providing the zero-point calibration for the GMOS data.
Here we exclude very red objects (\mbox{$V_{606}-I_{814}>1.2$}) to optimise the calibration close to the regime of our colour cut.

As the final ingredient for the ACS+GMOS photometric analysis we need to obtain a model for the photometric scatter. Different to the ACS+FORS2 analysis we cannot derive this from the comparison of in-field ACS \mbox{$V_{606}-I_{814}$} colour measurements.
Instead, we make use of GMOS $i$-band imaging that we obtained for cross-calibration in the centre of the GOODS-South field with similar characteristics to our cluster fields (exposure time 5.0ks).
For this field we can directly calibrate and compare to ACS  \mbox{$V_{606}-I_{814}$} colours similarly to the ACS+FORS2 analysis. We then apply the resulting magnitude- and colour-dependent photometric scatter distribution from this field as a scatter model in the redshift calibration of the ACS+GMOS clusters (see Sect.\thinspace\ref{se:redshiftdist}).

On average the image quality of our GMOS observations is significantly worse than for our new VLT observations  (compare Tables \ref{tab:vltdata} and \ref{tab:snapdata}).
Following  \citetalias{schrabback18} we therefore employ 1\farcs5 apertures for the ACS+GMOS photometry.
Thanks to the deep GMOS integration times we can still  include \mbox{$24<V_{606}<25.8$} galaxies in our analysis (selected via a cut \mbox{$V_{606}-I_{814}<0.2$} in transformed colour),
but we have to drop
\mbox{$V_{606}>25.8$} galaxies
given their increased photometric scatter.

\subsection{Number density checks}
\label{se:ngalprofile}
After accounting for masks, our colour and source selection  results in average galaxy number densities within the weak lensing fit range  (see Sect.\thinspace\ref{sec:results:shearprofile}) of
15.5/arcmin$^2$  for the ACS+FORS2 selection and 10.9/arcmin$^2$ for the ACS+GMOS selection
(values not corrected for magnification, see Table \ref{tab:beta} for the source densities of individual clusters).

An important consistency check for the source selection
is provided by the number density profile of the selected sources.
On average it should be consistent with flat  if cluster members have been accurately removed  and if the impact of masks and weak lensing
magnification  have been properly accounted for.
Sources appear brighter due to magnification, which increases the source counts.
However, at the depth of our data the change in solid angle
has a bigger impact,
leading to a net reduction in the measured source density  (\citetalias{schrabback18}).
To compensate for the  impact of magnification,
we follow \citetalias{schrabback18} and
employ the best-fit NFW reduced shear profile model for each cluster (see Sect.\thinspace\ref{sec:results:shearprofile}) to compute magnitude- and cluster redshift-dependent corrections
for the source density profile and the estimate of the mean geometric lensing efficiency (see Sect.\thinspace\ref{se:redshiftdist}). These corrections were derived by \citetalias{schrabback18} based on the magnitude-dependent source redshift distribution in CANDELS data.

\begin{figure}

  \includegraphics[width=0.99\columnwidth]{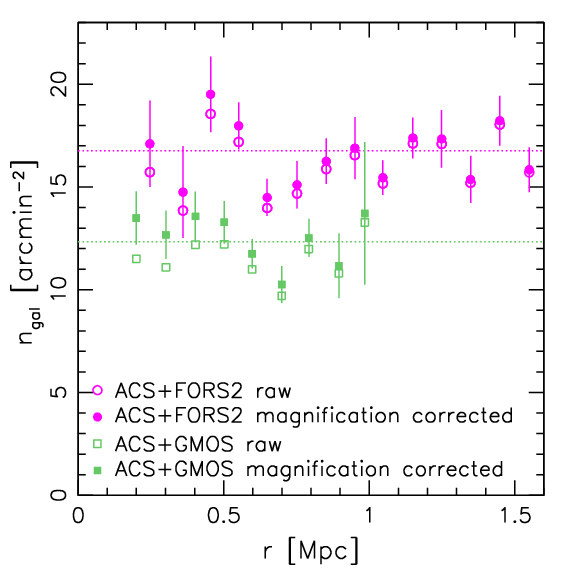}
  \caption{Weak lensing source density as a function of distance to the
    X-ray cluster centre for the ACS+FORS2 selection (magenta points) and the
    SZ cluster centre for  the ACS+GMOS selection (green squares).
    The points
    show the average number density from all available fields (including only clusters with new FORS2 data in case of the ACS+FORS2 selection), where open symbols correspond to raw (mask-corrected) values, while filled symbols have additionally been corrected for magnification assuming the best-fit NFW cluster models (Sect.\thinspace\ref{sec:results:shearprofile}).
    The error-bars indicate
    the uncertainty on the mean for the magnification-corrected values as estimated from the dispersion between the different fields. They are correlated due to large-scale structure variations.
    Error-bars for the raw values have a similar size but are not shown for clarity.
The horizontal lines correspond to the global average densities corrected for magnification.\label{fi:ngalprofile}}
\end{figure}
\begin{figure}
  \includegraphics[width=1\columnwidth]{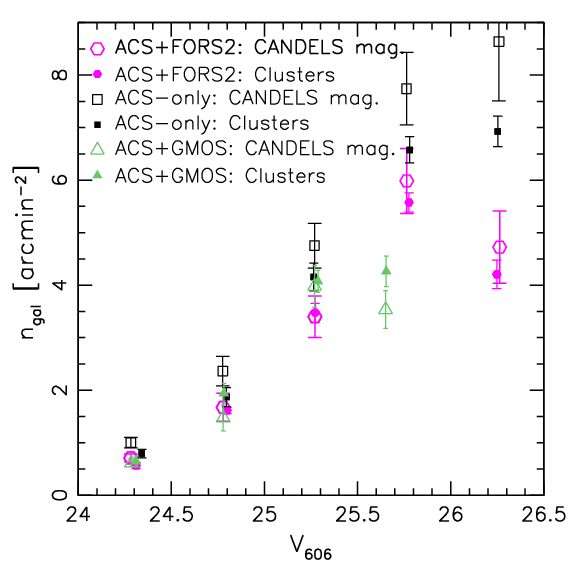}
 \caption{Number density of selected source galaxies $n_\mathrm{gal}$ as a function of $V_{606}$ magnitude, accounting for
   masks.
   Solid green triangles show the average source density in the ACS+GMOS data,
   while solid magenta hexagons and black squares correspond to the source densities for   the ACS+FORS2
   and ACS-only selections, respectively, averaged over the six cluster fields with new VLT/FORS2 imaging.
   The corresponding  source density
   estimates from
   the CANDELS fields
are shown with the large open symbols,
  applying a consistent selection,
  photometric scatter, and
  artificial magnification based on the best-fit cluster NFW models.
The error-bars indicate the uncertainty on the mean as estimated from the variation between
the contributing cluster fields or the five CANDELS fields, respectively,
assuming Gaussian scatter.
Errors are correlated between
magnitude bins due to  large-scale structure.
Especially at faint magnitudes source densities differ between the selections due to their differences in depth and applied colour limits.
\label{fig:ngal}}
\end{figure}

As visible in Fig.\thinspace\ref{fi:ngalprofile},
the corrected source density profile is consistent with flat for the ACS+FORS2
selection, as expected for an accurate cluster member removal.
Within the uncertainty this is also the case for the ACS+GMOS selection (error-bars
are correlated
due to large-scale structure variations in the source population, especially at small radii), but here the limited radial range limits the constraining power of the test.
As a further cross-check we therefore investigate
the measured number counts of the colour-selected sources
in the ACS+GMOS and ACS+FORS2 selected samples
(which apply consistent source selections at brighter magnitudes) in Fig.\thinspace\ref{fig:ngal}.
Their number counts  do not only agree well with each other, but also with the expected number counts from the CANDELS fields, which have been degraded to the same noise properties.
We therefore conclude that cluster members have  been removed accurately.
Note that our magnification correction does not account for miscentring of the cluster shear profile and mass distribution (see Sect.\thinspace\ref{sec:massmodelling}), likely  leading to a minor  over-correction at small radii.
This effect should be more pronounced
for the ACS+GMOS sample given the poorer SZ centre proxy.
This could be the cause for
the mild  increase that is tentatively visible (within the errors) for the magnification-corrected ACS+GMOS number density profile
in Fig.\thinspace\ref{fi:ngalprofile}
at small radii.

\begin{table*}
  \caption{Summary of geometric lensing efficiencies and source densities.
The three sets of rows correspond the ACS mosaic clusters with new observations, ACS mosaic clusters without new observations, and clusters from the ACS+GMOS sample, respectively.
    \label{tab:beta}}
\begin{center}
\begin{tabular}{lccccc}
  \hline
  \hline
Cluster & $\langle\beta\rangle$ & $\langle\beta^2\rangle$ & $\sigma_{\langle
  \beta \rangle_j}/\langle \beta\rangle$  & \multicolumn{2}{c}{$n_\mathrm{gal} [\mathrm{arcmin^{-2}}]$}\\
& & & & ACS-only & ACS+FORS2/GMOS\\
  \hline
  SPT-CL{\thinspace}$J$0000$-$5748 & 0.459 & 0.241 & 0.051 & 20.3 & 14.8 \\
SPT-CL{\thinspace}$J$0533$-$5005 & 0.372 & 0.163 & 0.061 & 20.7 & 16.9 \\
SPT-CL{\thinspace}$J$2040$-$5726 & 0.351 & 0.146 & 0.065 & 20.8 & 13.5 \\
SPT-CL{\thinspace}$J$2043$-$5035 & 0.441 & 0.226 & 0.073 & 20.2 & - \\
SPT-CL{\thinspace}$J$2337$-$5942 & 0.424 & 0.207 & 0.055 & 19.1 & 15.4 \\
SPT-CL{\thinspace}$J$2341$-$5119 & 0.323 & 0.124 & 0.069 & 21.3 & 14.8 \\
SPT-CL{\thinspace}$J$2359$-$5009 & 0.420 & 0.205 & 0.055 & 19.7 & 17.3 \\
\hline
  SPT-CL{\thinspace}$J$0102$-$4915 & 0.370 & 0.163 & 0.072 & 20.4 & 4.0 \\
SPT-CL{\thinspace}$J$0546$-$5345 & 0.299 & 0.108 & 0.095 & 13.8 & 3.3 \\
SPT-CL{\thinspace}$J$0559$-$5249 & 0.496 & 0.284 & 0.065 & 18.7 & 3.8 \\
SPT-CL{\thinspace}$J$0615$-$5746 & 0.331 & 0.132 & 0.084 & 19.9 & 2.9 \\
SPT-CL{\thinspace}$J$2106$-$5844 & 0.275 & 0.092 & 0.103 & 9.8 & 2.2 \\
SPT-CL{\thinspace}$J$2331$-$5051 & 0.514 & 0.304 & 0.066 & 19.8 & 8.1 \\
  SPT-CL{\thinspace}$J$2342$-$5411 & 0.294 & 0.104 & 0.097 & 15.2 & 2.6 \\
  \hline
SPT-CL{\thinspace}$J$0044$-$4037 & 0.309 & 0.116 & 0.115 & - & 13.2 \\
SPT-CL{\thinspace}$J$0058$-$6145 & 0.393 & 0.182 & 0.105 & - & 12.4 \\
SPT-CL{\thinspace}$J$0258$-$5355 & 0.322 & 0.125 & 0.109 & - & 12.2 \\
SPT-CL{\thinspace}$J$0339$-$4545 & 0.376 & 0.167 & 0.109 & - & 11.5 \\
SPT-CL{\thinspace}$J$0344$-$5452 & 0.299 & 0.109 & 0.103 & - & 7.8 \\
SPT-CL{\thinspace}$J$0345$-$6419 & 0.343 & 0.140 & 0.104 & - & 10.8 \\
SPT-CL{\thinspace}$J$0346$-$5839 & 0.453 & 0.238 & 0.098 & - & 9.0 \\
SPT-CL{\thinspace}$J$0356$-$5337 & 0.300 & 0.111 & 0.112 & - & 12.0 \\
SPT-CL{\thinspace}$J$0422$-$4608 & 0.476 & 0.259 & 0.084 & - & 7.8 \\
SPT-CL{\thinspace}$J$0444$-$5603 & 0.344 & 0.141 & 0.105 & - & 10.7 \\
SPT-CL{\thinspace}$J$0516$-$5755 & 0.331 & 0.131 & 0.096 & - & 9.8 \\
SPT-CL{\thinspace}$J$0530$-$4139 & 0.412 & 0.199 & 0.106 & - & 12.4 \\
SPT-CL{\thinspace}$J$0540$-$5744 & 0.422 & 0.208 & 0.090 & - & 11.0 \\
SPT-CL{\thinspace}$J$0617$-$5507 & 0.335 & 0.136 & 0.116 & - & 10.9 \\
SPT-CL{\thinspace}$J$2228$-$5828 & 0.441 & 0.224 & 0.082 & - & 10.6 \\
SPT-CL{\thinspace}$J$2311$-$5820 & 0.349 & 0.144 & 0.099 & - & 12.9 \\
  \hline
\end{tabular}
\end{center}
\begin{justify}
Note. ---
{\it Column 1:} Cluster designation.
{\it Columns 2--4:}  $\langle\beta\rangle$,  $\langle\beta^2\rangle$, and $\sigma_{\langle
  \beta \rangle_j}/\langle \beta\rangle$ averaged over both colour selection
schemes and all
magnitude bins that are included in the NFW fits according to their
corresponding shape weight sum.
{\it Columns 5--6:} Density of selected sources in the cluster fields for the ACS-only and the
ACS+FORS2/GMOS colour selection schemes, respectively (averaged within the fit range and not corrected for magnification).
\end{justify}
\end{table*}

\subsection{Calibration of the source redshift distribution}
\label{se:redshiftdist}
The weak lensing shear $\gamma$ and convergence $\kappa$ \citep[see e.g.][]{schneider06}
 scale with the average geometric lensing efficiency
\begin{equation}
  \langle\beta\rangle=\frac{\sum  \beta(z_i) w_i}{\sum w_i}
\end{equation}
of the sources galaxies, where $w_i$ is the  shape weight\footnote{The shape weights are computed from the $\mathrm{log}_{10} (S/N_\mathrm{flux})$-dependent variance of bias-corrected ellipticity estimates of correspondingly selected CANDELS galaxies, see Appendix A5 in \citetalias{schrabback18}.} of  galaxy $i$,
and
\begin{equation}
\beta=\mathrm{max}\left[0,\frac{D_\mathrm{ls}}{D_\mathrm{s}}\right] \,
\end{equation}
is defined via the
angular diameter distances $D_\mathrm{s}$, $D_\mathrm{l}$, and $D_\mathrm{ls}$  to the source, to the lens, and between lens and source, respectively.
Since we have removed cluster members and other galaxies at or near the redshifts of the targeted clusters via the colour selection (see Sect.\thinspace\ref{se:photo_color}), there is no need to obtain individual photometric redshifts (photo-$z$s).
Instead, we can infer the redshift distribution and therefore $\langle\beta\rangle$ via observations of well-studied
reference fields, to which we apply a consistent source selection.
For this purpose, \citetalias{schrabback18} employed photo-$z$ catalogues computed by the 3D-HST team \citep[][\citetalias{skelton14} henceforth]{skelton14} for the CANDELS fields \citep{grogin2011}.
The five CANDELS fields have not only been observed by HST with at least four imaging filters \citep[including deep NIR, see][]{koekemoer11} plus slitless spectroscopy \citep{momcheva16}, but they also benefit from a wide range of additional imaging and spectroscopic observations obtained with other facilities (see \citetalias{skelton14}). Together with their significant sky coverage, which is needed to reduce the impact of sampling variance, this turns them into an outstanding reference sample to infer the redshift distribution for deep WL data (\citetalias{schrabback18}).

Through the comparison with even deeper  photometric and spectroscopic redshifts \citep{rafelski15,brammer12,brammer13} available in the overlapping
{\it Hubble} Ultra Deep Field, \citetalias{schrabback18} showed that the \citetalias{skelton14} photo-$z$s nevertheless suffer from systematic issues such as  catastrophic redshift outliers and redshift focusing effects \citep[e.g.][]{wolf09}, which would bias the
resulting cluster masses high by \mbox{$\sim  12\%$}
if unaccounted for.
In order to achieve an initial correction for this effect,  \citetalias{schrabback18} introduced an approximate empirical scheme to statistically correct the
\citetalias{skelton14} photo-$z$s for these effects.
Recently,
\citetalias{raihan20}
revisited this issue, also including new ultra-deep spectroscopic
data from MUSE \citep{inami17} in the comparison.
By varying both the inputs and the analysis scheme,  \citetalias{raihan20} show that the bias
in the inferred redshift distribution can be avoided by using \texttt{BPZ} \citep{benitez00} instead of \texttt{EAZY} \citep{brammer08}, for which in particular \texttt{BPZ}'s  template interpolation
 plays a crucial role.
 \citetalias{raihan20} compute  \texttt{BPZ} photo-$z$s for the five CANDELS fields based on the HST photometry and a subset of the ground-based photometric data provided by \citetalias{skelton14}.
 From their tests  \citetalias{raihan20} conclude that their catalogues are expected to provide accurate $\langle\beta\rangle$ estimates for observations similar to our data within a total systematic uncertainty of 3.0\%, which accounts for the impact of residual systematic photo-$z$ uncertainties and sampling variance.
 Recomputing the \citetalias{schrabback18} WL cluster mass constraints using their updated CANDELS catalogues for the redshift calibration, \citetalias{raihan20} find that the masses shift by \mbox{$\sim +1\%$} only compared to the \citetalias{schrabback18} results.
 This good agreement is an important confirmation of the robustness of the results, given that both approaches should provide
 unbiased $\langle\beta\rangle$ estimates within their systematic uncertainties.
 The joint uncertainty quoted by \citetalias{schrabback18} for  photo-$z$ uncertainties and sampling variance (2.4\%) is slightly smaller, but this ignores the
 impact depth variations between the different CANDELS fields have on the systematic biases and uncertainties.
 In contrast, this issue has been investigated by  \citetalias{raihan20} via the degradation of higher quality data and it is effectively accounted for in their analysis via their full photo-$z$ re-computation.
 We therefore use the  \citetalias{raihan20} CANDELS photo-$z$s as the redshift calibration reference sample for our analysis.

\begin{figure}
  \includegraphics[width=1\columnwidth]{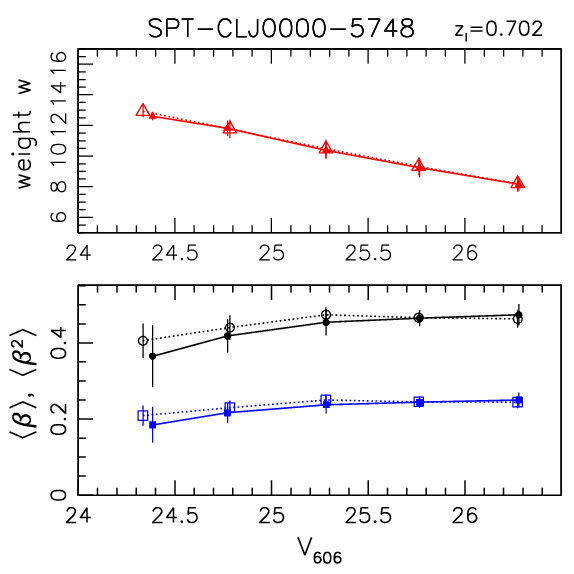}
\caption{Dependence of different parameters in the analysis of
SPT-CL{\thinspace}$J$0000$-$5748
 on $V_{606}$
  magnitude. Small solid (large open) symbols correspond to the analysis using ACS-only (ACS+FORS2) colours.
{\it Top:} Average weak lensing shape weight $w$, where the  error-bars show the
dispersion from all selected galaxies in the magnitude bin.
{\it Bottom:}
$\langle\beta \rangle$ (circles) and $\langle\beta^2 \rangle$ (squares), where the
error-bars correspond to the dispersion of their estimates between the cluster-field-sized CANDELS sub-patches.
\label{fig:beta_beta2_w}}
\end{figure}

\begin{figure}
  \includegraphics[width=0.99\columnwidth]{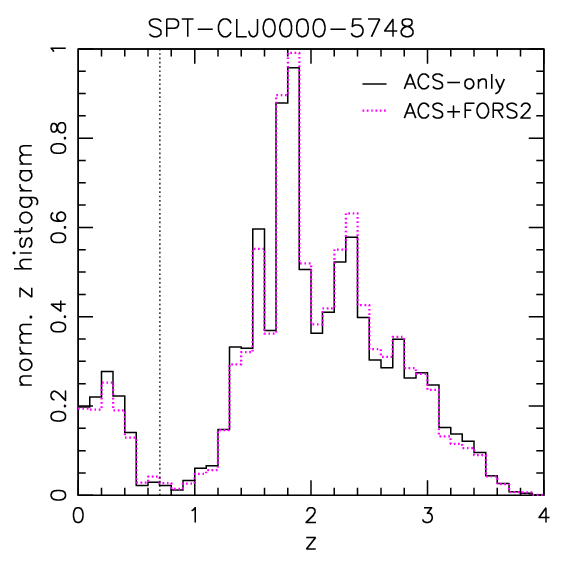}
   \caption{Inferred average redshift distribution of source galaxies using the ACS-only versus
     ACS+FORS2 colour selection for data with the noise properties of our observations of SPT-CL{\thinspace}$J$0000$-$5748, based on the CANDELS photometric redshift catalogues from \citetalias{raihan20}.
\label{fi:zdist}}
\end{figure}

In order to compute  $\langle\beta\rangle$ we first match the noise properties for the magnitude and $V_{606}-I_{814}$ colour selection between the corresponding cluster field and the CANDELS data as done by  \citetalias{schrabback18}, employing the photometric scatter distributions described in Sections \ref{se:color:acs_fors2} and \ref{se:color:acs_gmos} for the ACS+FORS2 and ACS+GMOS analyses, respectively.
Following the colour and magnitude selection we then compute $\langle\beta\rangle$ from the CANDELS catalogues in 0.5mag-wide $V_{606}$ magnitude bins (see Fig.\thinspace\ref{fig:beta_beta2_w}) to improve the weighting and tighten the overall constraints (see Sect.\thinspace\ref{sec:results:shearprofile} and
Table \ref{tab:beta} for  effective joint values).
We likewise compute  $\langle\beta^2\rangle(V_{606})$ to account for the impact of the broad width of the redshift distribution following \citet{seitz97,hoekstra00} and \citet{applegate14}.
In addition to obtaining global best estimates for the mean redshift distribution (see Fig.\thinspace\ref{fi:zdist} for an example) and  $\langle\beta\rangle(V_{606})$, we also
estimate the
line-of-sight
scatter $\sigma_{\langle
  \beta \rangle_j}$
by placing apertures $j$ of the size of our corresponding cluster-field observations into the CANDELS fields (see \citetalias{schrabback18}).

 The  total systematic uncertainty in the $\langle\beta\rangle$ estimates comprises the 3.0\% uncertainty estimate from
 \citetalias{raihan20},
 and in addition minor contributions from deblending differences and potential residual contamination of the source sample by very blue cluster members. For the latter, we use the estimates from \citetalias{schrabback18} of 0.5\% and 0.9\%,  respectively,
 yielding a joint uncertainty of 3.2\% (added in quadrature).

\section{Weak lensing results}
\label{sec:results}

\begin{table*}
\caption{Locations (\mbox{$\alpha, \delta$}) of the peaks in the mass reconstruction signal-to-noise ratio maps,
  their positional uncertainty (\mbox{$\Delta\alpha, \Delta\delta$}) as estimated by bootstrapping the galaxy
  catalogue, and their peak signal-to-noise ratio \mbox{$(S/N)_\mathrm{peak}$}.  Here we only include clusters with new observations and \mbox{$(S/N)_\mathrm{peak}>1.5$}. The top set of rows includes clusters with ACS mosaics, while the bottom set includes  clusters from the ACS+GMOS sample.
\label{tab:masspeaklocations}}
\begin{center}
\begin{tabular}{lccccccc}
\hline\hline
Cluster & $\alpha$ & $\delta$ & $\Delta\alpha$ & $\Delta\delta$ & $\Delta\alpha$ & $\Delta\delta$ &\mbox{$(S/N)_\mathrm{peak}$}   \\
 &  [deg J2000] &  [deg J2000] & [arcsec] & [arcsec] & [kpc] & [kpc] &\\
  \hline
  SPT-CL{\thinspace}$J$0000$-$5748  &   0.25607  & $ -57.80996 $ & 2.7  &  2.4  &  20  &  17  &  5.4 \\
SPT-CL{\thinspace}$J$0533$-$5005  &   83.39302  & $ -50.10844 $ & 7.8  &  7.1  &  61  &  55  &  3.3 \\
SPT-CL{\thinspace}$J$2040$-$5725  &   310.05696  & $ -57.42120 $ & 4.7  &  7.0  &  37  &  55  &  3.4 \\
SPT-CL{\thinspace}$J$2043$-$5035  &   310.81687  & $ -50.59325 $ & 4.3  &  7.8  &  31  &  56  &  3.3 \\
SPT-CL{\thinspace}$J$2337$-$5942  &   354.35873  & $ -59.70801 $ & 1.1  &  1.3  &  8  &  9  &  7.0 \\
SPT-CL{\thinspace}$J$2341$-$5119  &   355.30057  & $ -51.32996 $ & 2.1  &  3.4  &  17  &  27  &  3.8 \\
SPT-CL{\thinspace}$J$2359$-$5009  &   359.93212  & $ -50.16927 $ & 3.6  &  5.1  &  27  &  38  &  4.8 \\
  \hline
SPT-CL{\thinspace}$J$0058$-$6145  &   14.58664  & $ -61.76796 $ & 2.7  &  2.1  &  20  &  16  &  4.3 \\
SPT-CL{\thinspace}$J$0258$-$5355  &   44.52738  & $ -53.92520 $ & 3.5  &  3.3  &  28  &  27  &  4.0 \\
SPT-CL{\thinspace}$J$0339$-$4545  &   54.87871  & $ -45.75065 $ & 11.0  &  5.7  &  84  &  44  &  2.2 \\
SPT-CL{\thinspace}$J$0345$-$6419  &   56.25103  & $ -64.33496 $ & 9.9  &  6.4  &  78  &  51  &  2.5 \\
SPT-CL{\thinspace}$J$0346$-$5839  &   56.57704  & $ -58.65087 $ & 4.6  &  4.2  &  33  &  30  &  3.5 \\
SPT-CL{\thinspace}$J$0356$-$5337  &   59.09500  & $ -53.63168 $ & 11.4  &  10.5  &  92  &  85  &  1.6 \\
SPT-CL{\thinspace}$J$0422$-$4608  &   65.73875  & $ -46.14217 $ & 2.6  &  3.2  &  18  &  23  &  4.3 \\
SPT-CL{\thinspace}$J$0444$-$5603  &   71.10803  & $ -56.05631 $ & 6.4  &  5.3  &  51  &  42  &  3.0 \\
SPT-CL{\thinspace}$J$0516$-$5755  &   79.25988$^*$  & $ -57.89916^* $ & 4.2  &  6.6  &  33  &  52  &  3.4 \\
SPT-CL{\thinspace}$J$0530$-$4139  &   82.67820  & $ -41.65160 $ & 5.6  &  2.9  &  41  &  21  &  3.6 \\
SPT-CL{\thinspace}$J$0540$-$5744  &   84.99319  & $ -57.74324 $ & 6.2  &  7.8  &  45  &  57  &  3.3 \\
SPT-CL{\thinspace}$J$0617$-$5507  &   94.27795  & $ -55.13300 $ & 7.8  &  7.9  &  62  &  62  &  2.7 \\
SPT-CL{\thinspace}$J$2228$-$5828  &   337.17934  & $ -58.47028 $ & 7.7  &  12.8  &  56  &  93  &  2.5 \\
SPT-CL{\thinspace}$J$2311$-$5820  &   347.99784$^*$  & $ -58.36331^* $ & 10.4  &  10.2  &  81  &  81  &  2.5 \\
\hline
\end{tabular}
\end{center}
{\flushleft
  Note. --- $^*$: Indicates a less reliable  peak close to the edge of the field of view.
}
\end{table*}

\subsection{Mass reconstructions}
\label{sec:results:massrecon}
The weak lensing shear
$\gamma$
and convergence
$\kappa$,
which are linked to the reduced shear as
\begin{equation}
g=\frac{\gamma}{1-\kappa}\, ,
\end{equation}
are
both second-order derivatives of the lensing potential \citep[e.g.][]{bartelmann01}.
Therefore, it is possible
to reconstruct the convergence field from the shear field up to a constant, which is  the mass-sheet degeneracy \citep{kaiser93,schneider95}.
Following  \citetalias{schrabback18} we
employ a Wiener-filtered reconstruction algorithm  \citep{mcinnes09,simon09},
which also has the advantage of properly accounting for the spatially varying source
densities in our ACS+FORS2 data sets.
We fix the mass-sheet degeneracy by setting the average convergence inside each cluster field to zero. While this generally leads to an underestimation of $\kappa$,
this is a relatively minor effect for the clusters with ACS mosaics.
The impact is bigger for the clusters in the ACS+GMOS sample given the smaller field-of-view, but note that we only use the mass  reconstructions for illustrative purposes and not for quantitative mass constraints.

The left panels of Figs.\thinspace\ref{fi:wl_results_spt2043} to \ref{fi:wl_results_snap-6} show mass signal-to-noise ratio ($S/N$) contours overlaid on colour images for all clusters in
our sample with new observations.
To compute the $S/N$ maps we generate 500 noise shear fields for each cluster  by randomising the ellipticity phases, reconstruct the $\kappa$ field for each noise shear field, and then divide the actual  $\kappa$ reconstruction\footnote{We approximate the shear with the reduced shear when computing  $S/N$ maps.
See e.g.~\citet{schrabback18b} for the application of an iterative scheme to correct for the difference, which is more important when constraining  $\kappa$ (rather than $S/N$) for very massive clusters.}   by the r.m.s. image of the  noise field reconstructions.
For all clusters with ACS mosaics the mass  $S/N$ contours show a clear detection,
with peak ratios \mbox{$S/N_\mathrm{peak}\ge 3$} (see Table \ref{tab:masspeaklocations}).

Among the clusters in the ACS+GMOS sample, we obtain
detections with  \mbox{$S/N_\mathrm{peak}\ge 3$} for
SPT-CL\thinspace$J$0058$-$6145, SPT-CL\thinspace$J$0258$-$5355, SPT-CL\thinspace$J$0346$-$5839, SPT-CL\thinspace$J$0422$-$4608, SPT-CL\thinspace$J$0444$-$5603, SPT-CL\thinspace$J$0516$-$5755\footnote{\label{footnote:SPT-CLJ0516-5755}This cluster shows a very elongated reconstructed mass distribution, where the strongest peak in the $S/N$ mass map is located close to the edge of the field of view, making it less reliable.},
SPT-CL\thinspace$J$0530$-$4139, and SPT-CL\thinspace$J$0540$-$5744 (see Table \ref{tab:masspeaklocations}),
as well as tentative  detections
(\mbox{$S/N_\mathrm{peak}\ge 2$}) for
SPT-CL\thinspace$J$0339$-$4545, SPT-CL\thinspace$J$0345$-$6419, SPT-CL\thinspace$J$0617$-$5507, SPT-CL\thinspace$J$2228$-$5828\footnote{\label{footnote:SPT-CLJ2228-5828}The main peak in the  $S/N$  mass map of  SPT-CL\thinspace$J$2228$-$5828
  is located close to the Western edge of the field of view (see the top-left panel of Fig.\thinspace\ref{fi:wl_results_snap-6}), coinciding approximately with the position of the  candidate brightest cluster galaxy (BCG) from \citet{zenteno20}.
  The $S/N$ mass map of this cluster also shows a weak ($1.6\sigma$) secondary peak,
  located close to a second  concentration in the galaxy distribution, which surrounds a second bright candidate cluster galaxy.
These observations suggest that SPT-CL\thinspace$J$2228$-$5828 could be a
merger in the plane of the sky.}
and
SPT-CL\thinspace$J$2311$-$5820\textsuperscript{\ref{footnote:SPT-CLJ0516-5755}}.
Furthermore,
SPT-CL\thinspace$J$0356$-$5337, which is a potential dissociative merger based on
strong lensing features \citep{mahler20}, shows a weak peak (\mbox{$S/N_\mathrm{peak}=1.6$}, see
the bottom-left panel of Fig.\thinspace\ref{fi:wl_results_snap-3}) close to the BCG candidate from \citet{mahler20}.
We suspect that
the main reasons for
the poorer detection rate in the WL mass reconstructions
of the
ACS+GMOS sample are given by the smaller field covered by these observations and the (on average)
expected
lower masses of the clusters.

\subsection{NFW fits to reduced shear profiles}
\label{sec:results:shearprofile}

\begin{table*}
\caption{Weak lensing mass constraints from the NFW fits to the reduced shear
  profiles around the X-ray centres for the clusters with ACS mosaics
  for two different over-densities
  \mbox{$\Delta \in \{200\mathrm{c}, 500\mathrm{c}\}$}.
  The top (bottom) set of rows corresponds to clusters with (without) new observations.
$M_{\Delta}^\mathrm{biased,ML}$ are maximum likelihood mass estimates
in $10^{14}\mathrm{M}_\odot$ {\it without} corrections for mass modelling bias  applied.
The listed errors are statistical 68\%
uncertainties, including the contributions from shape noise (asymmetric errors),
uncorrelated
large-scale, and  line-of-sight variations in the redshift
distribution.
Systematic uncertainties are listed in Table \ref{tab:sys}.
$\hat{b}_\mathrm{\Delta,WL}=\text{exp}\left[\langle \text{ln}\,b_{\Delta,\text{WL}}\rangle\right]$ relates to the
mean
of the
 estimated
mass bias
distribution,
whose width is characterised by $\sigma(\mathrm{ln}\,b_\mathrm{\Delta,WL})$.
\label{tab:mass}}
\begin{center}

\begin{tabular}{crccrcc}
\hline\hline
Cluster&  \multicolumn{1}{c}{$M_{200\mathrm{c}}^\mathrm{biased,ML}\,[10^{14}\mathrm{M}_\odot]$} &
                                                                       $\hat{b}_{200\mathrm{c,WL}}$&
  $\sigma(\mathrm{ln}\, b_\mathrm{\mathrm{200c,WL}})$   &                                                                                                                                                             \multicolumn{1}{c}{$M_{500\mathrm{c}}^\mathrm{biased,ML}\,[10^{14}\mathrm{M}_\odot]$} &
                                                            $\hat{b}_{500\mathrm{c,WL}}$ &
   $\sigma(\mathrm{ln}\, b_\mathrm{\mathrm{500c,WL}})$ \\
    \hline
  SPT-CL{\thinspace}$J$0000$-$5748 & $6.0_{-2.2}^{+2.4}\pm 1.1\pm 0.3 $& $0.89\pm0.01$ &  $0.35\pm0.01$ & $4.1_{-1.5}^{+1.7} \pm 0.8\pm 0.2$ & $0.91\pm0.01$ &  $0.32\pm0.01$\\
SPT-CL{\thinspace}$J$0533$-$5005 & $4.0_{-2.0}^{+2.3}\pm 1.0\pm 0.2 $& $0.91\pm0.02$ &  $0.32\pm0.02$ & $2.7_{-1.4}^{+1.6} \pm 0.7\pm 0.2$ & $0.91\pm0.01$ &  $0.28\pm0.02$\\
SPT-CL{\thinspace}$J$2040$-$5726 & $3.5_{-2.1}^{+2.6}\pm 0.9\pm 0.2 $& $0.91\pm0.02$ &  $0.32\pm0.02$ & $2.4_{-1.5}^{+1.8} \pm 0.6\pm 0.2$ & $0.90\pm0.01$ &  $0.26\pm0.03$\\
SPT-CL{\thinspace}$J$2043$-$5035 & $4.3_{-1.9}^{+2.1}\pm 0.9\pm 0.3 $& $0.92\pm0.01$ &  $0.32\pm0.01$ & $2.9_{-1.3}^{+1.5} \pm 0.7\pm 0.2$ & $0.93\pm0.01$ &  $0.28\pm0.02$\\
SPT-CL{\thinspace}$J$2337$-$5942 & $10.9_{-2.6}^{+2.6}\pm 1.3\pm 0.6 $& $0.88\pm0.02$ &  $0.37\pm0.02$ & $7.6_{-1.8}^{+1.9} \pm 0.9\pm 0.4$ & $0.89\pm0.01$ &  $0.33\pm0.02$\\
SPT-CL{\thinspace}$J$2341$-$5119 & $3.5_{-2.1}^{+2.5}\pm 1.1\pm 0.2 $& $0.90\pm0.01$ &  $0.31\pm0.01$ & $2.4_{-1.5}^{+1.8} \pm 0.8\pm 0.2$ & $0.89\pm0.01$ &  $0.30\pm0.01$\\
SPT-CL{\thinspace}$J$2359$-$5009 & $6.2_{-2.3}^{+2.5}\pm 1.1\pm 0.3 $& $0.92\pm0.01$ &  $0.31\pm0.02$ & $4.3_{-1.6}^{+1.8} \pm 0.8\pm 0.2$ & $0.93\pm0.01$ &  $0.26\pm0.03$\\
\hline
SPT-CL{\thinspace}$J$0102$-$4915 & $11.9_{-2.9}^{+3.0}\pm 1.2\pm 0.9 $& $0.87\pm0.06$ &  $0.37\pm0.05$ & $8.6_{-2.2}^{+2.2} \pm 0.9\pm 0.6$ & $0.84\pm0.02$ &  $0.38\pm0.02$\\
SPT-CL{\thinspace}$J$0546$-$5345 & $5.1_{-3.2}^{+3.7}\pm 1.1\pm 0.5 $& $0.84\pm0.02$ &  $0.37\pm0.02$ & $3.5_{-2.2}^{+2.7} \pm 0.8\pm 0.3$ & $0.85\pm0.02$ &  $0.35\pm0.02$\\
SPT-CL{\thinspace}$J$0559$-$5249 & $8.1_{-3.0}^{+3.2}\pm 1.1\pm 0.5 $& $0.84\pm0.01$ &  $0.42\pm0.02$ & $5.5_{-2.1}^{+2.3} \pm 0.7\pm 0.4$ & $0.86\pm0.01$ &  $0.39\pm0.01$\\
SPT-CL{\thinspace}$J$0615$-$5746 & $8.1_{-2.7}^{+2.9}\pm 1.2\pm 0.7 $& $0.84\pm0.02$ &  $0.34\pm0.02$ & $5.6_{-1.9}^{+2.1} \pm 0.9\pm 0.5$ & $0.85\pm0.02$ &  $0.32\pm0.02$\\
SPT-CL{\thinspace}$J$2106$-$5844 & $8.4_{-4.5}^{+5.0}\pm 1.5\pm 0.9 $& $0.80\pm0.03$ &  $0.43\pm0.04$ & $5.9_{-3.3}^{+3.7} \pm 1.1\pm 0.6$ & $0.80\pm0.03$ &  $0.40\pm0.04$\\
SPT-CL{\thinspace}$J$2331$-$5051 & $4.8_{-2.3}^{+2.7}\pm 1.0\pm 0.3 $& $0.86\pm0.01$ &  $0.39\pm0.01$ & $3.3_{-1.6}^{+1.9} \pm 0.7\pm 0.2$ & $0.89\pm0.01$ &  $0.35\pm0.01$\\
SPT-CL{\thinspace}$J$2342$-$5411 & $8.8_{-3.7}^{+4.0}\pm 1.3\pm 0.9 $& $0.87\pm0.02$ &  $0.37\pm0.02$ & $6.2_{-2.6}^{+2.9} \pm 0.9\pm 0.6$ & $0.88\pm0.02$ &  $0.38\pm0.03$\\
  \hline
\end{tabular}
\end{center}
\end{table*}

\begin{table*}
\caption{As Table \ref{tab:mass}, but for the analysis centring on the SZ peak locations.
\label{tab:mass:SZ}}
\begin{center}

\begin{tabular}{crccrcc}
\hline\hline
Cluster&  \multicolumn{1}{c}{$M_{200\mathrm{c}}^\mathrm{biased,ML}\,[10^{14}\mathrm{M}_\odot]$} &
                                                                       $\hat{b}_{200\mathrm{c,WL}}$&
  $\sigma(\mathrm{ln}\, b_\mathrm{\mathrm{200c,WL}})$   &                                                                                                                                                             \multicolumn{1}{c}{$M_{500\mathrm{c}}^\mathrm{biased,ML}\,[10^{14}\mathrm{M}_\odot]$} &
                                                                                                                                                                                                                                                                                 $\hat{b}_{500\mathrm{c,WL}}$ &
   $\sigma(\mathrm{ln}\, b_\mathrm{\mathrm{500c,WL}})$ \\
  \hline
  SPT-CL{\thinspace}$J$0000$-$5748 & $6.5_{-2.2}^{+2.4}\pm 1.1\pm 0.3 $& $0.85\pm0.01$ &  $0.32\pm0.01$ & $4.4_{-1.5}^{+1.7} \pm 0.8\pm 0.2$ & $0.86\pm0.01$ &  $0.29\pm0.01$\\
SPT-CL{\thinspace}$J$0533$-$5005 & $2.0_{-1.5}^{+2.0}\pm 0.8\pm 0.1 $& $0.87\pm0.01$ &  $0.31\pm0.01$ & $1.3_{-1.0}^{+1.4} \pm 0.5\pm 0.1$ & $0.85\pm0.01$ &  $0.29\pm0.01$\\
SPT-CL{\thinspace}$J$2040$-$5726 & $4.3_{-2.3}^{+2.7}\pm 1.0\pm 0.3 $& $0.84\pm0.02$ &  $0.30\pm0.03$ & $2.9_{-1.6}^{+1.9} \pm 0.7\pm 0.2$ & $0.81\pm0.02$ &  $0.31\pm0.03$\\
SPT-CL{\thinspace}$J$2043$-$5035 & $3.6_{-1.8}^{+2.1}\pm 1.0\pm 0.3 $& $0.87\pm0.01$ &  $0.31\pm0.01$ & $2.4_{-1.2}^{+1.4} \pm 0.7\pm 0.2$ & $0.89\pm0.01$ &  $0.29\pm0.01$\\
SPT-CL{\thinspace}$J$2337$-$5942 & $10.7_{-2.6}^{+2.6}\pm 1.3\pm 0.6 $& $0.86\pm0.01$ &  $0.32\pm0.02$ & $7.5_{-1.8}^{+1.9} \pm 0.9\pm 0.4$ & $0.87\pm0.01$ &  $0.28\pm0.01$\\
SPT-CL{\thinspace}$J$2341$-$5119 & $3.2_{-2.1}^{+2.5}\pm 1.0\pm 0.2 $& $0.83\pm0.02$ &  $0.32\pm0.02$ & $2.2_{-1.4}^{+1.7} \pm 0.7\pm 0.2$ & $0.83\pm0.01$ &  $0.29\pm0.01$\\
SPT-CL{\thinspace}$J$2359$-$5009 & $5.4_{-2.2}^{+2.4}\pm 1.1\pm 0.3 $& $0.87\pm0.01$ &  $0.30\pm0.02$ & $3.7_{-1.5}^{+1.7} \pm 0.8\pm 0.2$ & $0.87\pm0.01$ &  $0.26\pm0.02$\\
  \hline
  SPT-CL{\thinspace}$J$0102$-$4915 & $15.2_{-2.8}^{+2.8}\pm 1.1\pm 1.1 $& $0.77\pm0.05$ &  $0.37\pm0.06$ & $11.1_{-2.1}^{+2.2} \pm 0.8\pm 0.8$ & $0.79\pm0.02$ &  $0.33\pm0.03$\\
SPT-CL{\thinspace}$J$0546$-$5345 & $2.5_{-2.3}^{+3.5}\pm 1.0\pm 0.2 $& $0.75\pm0.02$ &  $0.37\pm0.02$ & $1.7_{-1.6}^{+2.4} \pm 0.7\pm 0.2$ & $0.75\pm0.01$ &  $0.34\pm0.03$\\
SPT-CL{\thinspace}$J$0559$-$5249 & $4.2_{-2.4}^{+2.9}\pm 0.9\pm 0.3 $& $0.79\pm0.01$ &  $0.39\pm0.02$ & $2.8_{-1.6}^{+2.0} \pm 0.6\pm 0.2$ & $0.83\pm0.01$ &  $0.36\pm0.01$\\
SPT-CL{\thinspace}$J$0615$-$5746 & $7.1_{-2.6}^{+2.8}\pm 1.2\pm 0.6 $& $0.84\pm0.02$ &  $0.27\pm0.03$ & $4.9_{-1.8}^{+2.1} \pm 0.8\pm 0.4$ & $0.84\pm0.01$ &  $0.25\pm0.02$\\
SPT-CL{\thinspace}$J$2106$-$5844 & $8.2_{-4.4}^{+4.9}\pm 1.4\pm 0.8 $& $0.73\pm0.03$ &  $0.38\pm0.05$ & $5.7_{-3.2}^{+3.7} \pm 1.0\pm 0.6$ & $0.77\pm0.03$ &  $0.29\pm0.07$\\
SPT-CL{\thinspace}$J$2331$-$5051 & $5.2_{-2.4}^{+2.7}\pm 1.0\pm 0.3 $& $0.86\pm0.01$ &  $0.34\pm0.01$ & $3.5_{-1.6}^{+1.9} \pm 0.7\pm 0.2$ & $0.88\pm0.01$ &  $0.30\pm0.01$\\
SPT-CL{\thinspace}$J$2342$-$5411 & $7.4_{-3.5}^{+3.9}\pm 1.1\pm 0.7 $& $0.80\pm0.02$ &  $0.38\pm0.03$ & $5.1_{-2.5}^{+2.8} \pm 0.8\pm 0.5$ & $0.81\pm0.02$ &  $0.33\pm0.04$\\
\hline
\end{tabular}
\end{center}
\end{table*}

In order to constrain the cluster masses we estimate the binned
profiles of the tangential component of the reduced shear $g$
with respect to the corresponding cluster centre
\begin{equation}
  g_\mathrm{t}  =  - g_1 \cos{2 \phi} - g_2 \sin{2\phi}\, ,\label{eq:gt}
\end{equation}
where $\phi$ indicates the azimuthal angle with respect to the centre. Here, the reduced shear \mbox{$g=g_1+\mathrm{i} g_2$} is written in terms of its component along the coordinate grid $g_1$ and the 45deg-rotated component $g_2$.
Following \citetalias{schrabback18} we estimate the reduced shear profile
$\langle g_\mathrm{t} \rangle(r_k, V_j)=\sum (w_i
\epsilon_{\mathrm{t},i}) / \sum w_i$
for each cluster in bins of radius $r_k$ and magnitude $V_j$, where $w_i$ indicates the shape weight and the sum is computed over all galaxies $i$ falling into the corresponding radius and magnitude bin combination.
Accounting for the magnitude dependence increases the sensitivity of the analysis given the dependence of $\langle\beta\rangle$ on $V_{606}$ (see Fig.\thinspace\ref{fig:beta_beta2_w}).
For each cluster we then jointly fit the $\langle g_\mathrm{t} \rangle(r_k, V_j)$ profiles
with
predictions for  spherical NFW \citep{navarro97}
density profiles according to \citet{brainerd00}, assuming the concentration--mass ($c(M)$) relation from \citet{diemer19}.
When computing model predictions we also correct for the impact of
weak lensing magnification on  $\langle\beta\rangle$ following \citetalias{schrabback18},
as well as
the finite width of the redshift distribution
following  \citet{seitz97,hoekstra00} and \citet{applegate14}.
For the clusters with ACS mosaics we compute shear profiles both around the SZ peak locations and the X-ray centroids\footnote{We do not employ brightest cluster galaxies (BCGs) as centre proxies, because \citetalias{schrabback18} found that in their analysis BCG centres resulted in  a larger r.m.s.~offset  with respect to the peak in the weak lensing mass reconstruction than the X-ray and SZ centres.} (see Table \ref{tab:clusters_mos}).
Since high-resolution X-ray observations are presently unavailable for most clusters in the ACS+GMOS sample, we employ the SZ peak locations as centres when computing shear profiles for these clusters.
Both centre proxies typically deviate from the location of the halo centre, which would be used in simulation analyses to define over-density cluster masses.
We describe in Sect.\thinspace\ref{sec:massmodelling} how we account for the
mass  modelling bias that results from this and other effects, but to limit their impact we only include scale \mbox{$r>500$kpc} in the fit, as done by \citetalias{schrabback18}.
Following them we also limit the fit to scales  \mbox{$r<1.5$Mpc} for the clusters with ACS mosaics.
The right panels of Figs.\thinspace\ref{fi:wl_results_spt2043} to \ref{fi:wl_results_snap-6} show the resulting reduced tangential shear profiles (scaled to the average $\langle\beta\rangle$ and combined as done by \citetalias{schrabback18}), best-fit NFW models, and profiles of the 45deg-rotated reduced shear cross component
\begin{equation}
g_\times  =  g_1 \sin{2\phi} - g_2 \cos{2 \phi}  \,,
\label{eq:gx}
\end{equation}
which should be consistent with zero in the absence of PSF-related systematics.

We list the constraints on the best-fitting mass within a sphere that has an average density of 200 times the critical density of the Universe at the cluster redshift ($M_\mathrm{200c}$) and the corresponding $M_\mathrm{500c}$ estimates  \citep[assuming the $c(M)$ relation from][]{diemer19} in Tables \ref{tab:mass}, \ref{tab:mass:SZ}, and  \ref{tab:mass:snap}.
There we not only list the statistical uncertainties
from the NFW shear profile fit and shape noise, but also contributions from uncorrelated large-scale structure projections \citep[computed using Gaussian cosmic shear field realisations following][see also \citetalias{schrabback18}]{simon12} and line-of-sight variations in the source redshift distribution (see Sect.\thinspace\ref{se:redshiftdist}).
For the clusters in the ACS+GMOS sample
(see Table \ref{tab:clusters_snap}) we also list the mass uncertainty resulting from the photometric cluster redshift uncertainties.
Note that the maximum likelihood mass estimates reported in Tables \ref{tab:mass} to \ref{tab:mass:snap} have not yet been corrected for mass modelling biases. Our procedure to correct for these biases is described in Sect.\thinspace\ref{sec:massmodelling} and applied in the scaling relation analysis is Sect.\thinspace\ref{sec:scaling_relations}.

Comparing entries in Tables \ref{tab:mass} and \ref{tab:mass:SZ} versus Table  \ref{tab:mass:snap} it is evident that the observations using \mbox{$2\times 2$} ACS mosaics
yield much tighter mass constraints given their better radial coverage. E.g., comparing the results for SPT-CL{\thinspace}J2337$-$5942 and SPT-CL{\thinspace}J0530$-$4139, which have similar cluster redshifts and best-fit WL mass estimates, we find that the relative statistical mass errors are larger by a factor 1.8 for the single-pointing ACS data.
We expect that these large fit uncertainties, together with a larger intrinsic scatter (see Sect.\thinspace\ref{sec:massmodelling}) are primarily responsible for the large spread in best-fitting mass estimates reported in Table \ref{tab:mass:snap} for the ACS+GMOS sample,
for which we would expect a relatively low scatter in halo mass based on their SZ signature (Table \ref{tab:clusters_snap}).

\begin{table*}
\caption{Weak lensing mass constraints from the NFW fits to the reduced shear
  profiles around the SZ peaks of the clusters in the ACS+GMOS sample
  for two different over-densities
  \mbox{$\Delta \in \{200\mathrm{c}, 500\mathrm{c}\}$}.
$M_{\Delta}^\mathrm{biased,ML}$ are maximum likelihood mass estimates
in $10^{14}\mathrm{M}_\odot$ \emph{without} corrections for mass modelling bias  applied.
The listed errors are statistical 68\%
uncertainties, including the contributions from shape noise (asymmetric errors),
uncorrelated
large-scale,  line-of-sight variations in the redshift
distribution, and the uncertainty in the (photometric) cluster redshift.
Systematic uncertainties are listed in Table \ref{tab:sys}.
$\hat{b}_\mathrm{\Delta,WL}=\text{exp}\left[\langle \text{ln}\,b_{\Delta,\text{WL}}\rangle\right]$
relates to the
mean
of the estimated mass bias
distribution,
whose width is characterised by $\sigma(\mathrm{ln}\,b_\mathrm{\Delta,WL})$.
\label{tab:mass:snap}}
\addtolength{\tabcolsep}{-2pt}
\begin{center}
\begin{tabular}{crccrcc}
\hline\hline
Cluster&  \multicolumn{1}{c}{$M_{200\mathrm{c}}^\mathrm{biased,ML}\,[10^{14}\mathrm{M}_\odot]$} &
                                                                       $\hat{b}_{200\mathrm{c,WL}}$&
  $\sigma(\mathrm{ln}\, b_\mathrm{\mathrm{200c,WL}})$   &                                                                                                                                                             \multicolumn{1}{c}{$M_{500\mathrm{c}}^\mathrm{biased,ML}\,[10^{14}\mathrm{M}_\odot]$} &
                                                                                                                                                                                                                                                                                 $\hat{b}_{500\mathrm{c,WL}}$ &
   $\sigma(\mathrm{ln}\, b_\mathrm{\mathrm{500c,WL}})$ \\
  \hline
  SPT-CL{\thinspace}$J$0044$-$4037 & $-0.2_{-2.9}^{+2.1}\pm 0.3\pm 0.0\pm 0.4 $& $0.76\pm0.05$ &  $0.40\pm0.08$ & $-0.1_{-2.0}^{+1.4} \pm 0.2\pm 0.0\pm 0.2$ & $0.74\pm0.03$ &  $0.39\pm0.06$\\
SPT-CL{\thinspace}$J$0058$-$6145 & $7.7_{-4.1}^{+4.4}\pm 0.9\pm 0.8\pm 0.8 $& $0.76\pm0.02$ &  $0.41\pm0.03$ & $5.3_{-2.9}^{+3.2} \pm 0.6\pm 0.6\pm 0.5$ & $0.78\pm0.02$ &  $0.33\pm0.04$\\
SPT-CL{\thinspace}$J$0258$-$5355 & $13.9_{-4.5}^{+4.2}\pm 1.0\pm 1.5\pm 1.9 $& $0.64\pm0.04$ &  $0.53\pm0.08$ & $10.1_{-3.4}^{+3.2} \pm 0.8\pm 1.1\pm 1.6$ & $0.68\pm0.04$ &  $0.39\pm0.07$\\
SPT-CL{\thinspace}$J$0339$-$4545 & $2.5_{-2.5}^{+4.0}\pm 0.7\pm 0.3\pm 0.5 $& $0.75\pm0.04$ &  $0.43\pm0.06$ & $1.7_{-1.7}^{+2.8} \pm 0.5\pm 0.2\pm 0.4$ & $0.73\pm0.03$ &  $0.46\pm0.05$\\
SPT-CL{\thinspace}$J$0344$-$5452 & $6.8_{-5.4}^{+6.2}\pm 1.1\pm 0.7\pm 1.7 $& $0.71\pm0.03$ &  $0.44\pm0.05$ & $4.8_{-3.8}^{+4.6} \pm 0.8\pm 0.5\pm 1.2$ & $0.68\pm0.03$ &  $0.49\pm0.06$\\
SPT-CL{\thinspace}$J$0345$-$6419 & $0.0_{-3.0}^{+2.9}\pm 0.3\pm 0.0\pm 0.2 $& $0.79\pm0.04$ &  $0.40\pm0.08$ & $0.0_{-2.1}^{+1.9} \pm 0.2\pm 0.0\pm 0.1$ & $0.80\pm0.03$ &  $0.32\pm0.06$\\
SPT-CL{\thinspace}$J$0346$-$5839 & $12.2_{-6.0}^{+6.0}\pm 1.0\pm 1.2\pm 1.5 $& $0.78\pm0.05$ &  $0.41\pm0.10$ & $8.5_{-4.2}^{+4.3} \pm 0.7\pm 0.8\pm 1.1$ & $0.80\pm0.05$ &  $0.36\pm0.07$\\
SPT-CL{\thinspace}$J$0356$-$5337 & $1.1_{-2.2}^{+4.0}\pm 0.7\pm 0.1\pm 0.0 $& $0.76\pm0.03$ &  $0.42\pm0.06$ & $0.8_{-1.5}^{+2.8} \pm 0.5\pm 0.1\pm 0.0$ & $0.77\pm0.03$ &  $0.35\pm0.05$\\
SPT-CL{\thinspace}$J$0422$-$4608 & $9.8_{-6.1}^{+6.6}\pm 1.0\pm 0.8\pm 1.6 $& $0.73\pm0.05$ &  $0.46\pm0.09$ & $6.7_{-4.3}^{+4.7} \pm 0.7\pm 0.6\pm 1.1$ & $0.79\pm0.04$ &  $0.39\pm0.09$\\
SPT-CL{\thinspace}$J$0444$-$5603 & $7.4_{-4.4}^{+4.8}\pm 0.9\pm 0.8\pm 0.6 $& $0.76\pm0.04$ &  $0.41\pm0.09$ & $5.1_{-3.1}^{+3.4} \pm 0.7\pm 0.5\pm 0.3$ & $0.76\pm0.04$ &  $0.41\pm0.06$\\
SPT-CL{\thinspace}$J$0516$-$5755 & $2.8_{-3.2}^{+5.3}\pm 0.9\pm 0.3\pm 0.1 $& $0.77\pm0.05$ &  $0.37\pm0.10$ & $1.9_{-2.2}^{+3.8} \pm 0.6\pm 0.2\pm 0.1$ & $0.78\pm0.04$ &  $0.32\pm0.06$\\
SPT-CL{\thinspace}$J$0530$-$4139 & $8.8_{-4.5}^{+4.6}\pm 0.9\pm 0.9\pm 0.7 $& $0.76\pm0.02$ &  $0.42\pm0.04$ & $6.1_{-3.1}^{+3.3} \pm 0.7\pm 0.6\pm 0.5$ & $0.78\pm0.02$ &  $0.39\pm0.04$\\
SPT-CL{\thinspace}$J$0540$-$5744 & $8.4_{-4.3}^{+4.7}\pm 0.9\pm 0.8\pm 0.0 $& $0.79\pm0.02$ &  $0.40\pm0.04$ & $5.8_{-3.0}^{+3.4} \pm 0.7\pm 0.5\pm 0.0$ & $0.81\pm0.03$ &  $0.39\pm0.03$\\
SPT-CL{\thinspace}$J$0617$-$5507 & $1.3_{-2.3}^{+4.1}\pm 0.6\pm 0.2\pm 0.6 $& $0.75\pm0.04$ &  $0.41\pm0.07$ & $0.9_{-1.5}^{+2.9} \pm 0.4\pm 0.1\pm 0.4$ & $0.72\pm0.03$ &  $0.40\pm0.06$\\
SPT-CL{\thinspace}$J$2228$-$5828 & $2.4_{-2.4}^{+3.7}\pm 0.7\pm 0.2\pm 0.4 $& $0.64\pm0.03$ &  $0.54\pm0.06$ & $1.6_{-1.6}^{+2.6} \pm 0.5\pm 0.1\pm 0.3$ & $0.64\pm0.03$ &  $0.47\pm0.05$\\
SPT-CL{\thinspace}$J$2311$-$5820 & $9.5_{-4.8}^{+5.0}\pm 0.9\pm 0.9\pm 0.8 $& $0.78\pm0.04$ &  $0.41\pm0.07$ & $6.6_{-3.4}^{+3.7} \pm 0.7\pm 0.7\pm 0.6$ & $0.74\pm0.03$ &  $0.37\pm0.06$\\
 \hline
\end{tabular}
\end{center}
{\flushleft
  Note. ---
  Because of noise  (from the intrinsic galaxy shapes and large-scale structure projections)  the tangential reduced shear profiles of individual clusters may become slightly negative on average, as is the case for SPT-CL{\thinspace}$J$0044$-$4037  (see Fig.\thinspace\ref{fi:wl_results_snap-1}).
  For the mass limits reported in this table we model such negative profiles by allowing for (unphysical) negative cluster masses. Here we  employ the  NFW reduced shear profile prediction of the corresponding positive mass, but switch the sign of the model.
\\
}

\addtolength{\tabcolsep}{-2pt}
\end{table*}

\subsection{Correction for mass modelling biases}
\label{sec:massmodelling}
Systematic deviations from the NFW model, uncertainties and scatter in the assumed $c(M)$ relation \citep[e.g.][]{child18}, triaxiality, correlated large-scale structure, and miscentring of the fitted profile can lead to  systematic biases in the measured masses. Here we describe our method of constraining the distribution of the net bias, excluding contributions from
uncorrelated large-scale structure (the latter effect is discussed in Sect.\thinspace\ref{sec:results:shearprofile}). Following \citet{becker11}, we define the bias for an individual target through
\begin{equation}
  \label{eq:MWL}
    M_{\Delta,\text{WL}} = b_{\Delta,\text{WL}} M_{\Delta, \text{halo}},
\end{equation}
where $M_{\Delta,\text{WL}}$ is the mass at overdensity $\Delta$ measured from the reduced shear profile, $M_{\Delta, \text{halo}}$ is the corresponding halo mass and $b_{\Delta,\text{WL}}$ is the bias factor.

We use simulations to estimate the bias distribution for each of our targets, including a mass dependence. In particular, following \citetalias{schrabback18} and \citetalias{dietrich19} we use the \mbox{$z=0.25$} and \mbox{$z=1.0$} snapshots of the Millennium-XXL simulations \citep{angulo12}, from which reduced shear fields of massive halos are derived, using the lensing efficiencies of the individual targets.
After choosing a centre that either corresponds to the 3D halo centre or a miscentred position (explained below), we
bin the tangential reduced shear profile, to which we add shape noise that matches the uncertainties of the actual cluster tangential reduced shear estimates in each corresponding radial bin.
We then fit the cluster masses from the noisy mock data as done for the real cluster observations.
Halo-related properties, such as the lens redshift, are scaled appropriately, while cosmology-related properties, such as the redshift dependence in the mass-concentration relation, are kept at the redshifts of the simulation in the respective snapshots.

In the presence of noise, it is difficult to model the bias distribution generally. Following recent work including \citetalias{schrabback18}, \citetalias{dietrich19}, and  \citetalias{bocquet19}
we therefore make the simplifying assumption of a log-normal (in halo mass) distribution.
The distribution is defined by the expectation value $\mu$ and the
dispersion $\sigma$ in the natural-log space of $b_{\Delta,\text{WL}}$,
such that \mbox{$\mu = \langle\text{ln}\,b_{\Delta,\text{WL}}\rangle$} and $\sigma^2$ is the
variance of $\text{ln}\,b_{\Delta,\rm{WL}}$. We further define
\begin{equation}
        \hat{b}_{\Delta,\text{WL}}=\text{exp}\left[
\langle\text{ln}\,b_{\Delta,\text{WL}}\rangle\right]\,
\end{equation}
as a measure of the bias in linear space (with the caveat that this
measure alone cannot be used in order to remove the mass bias).
The Bayesian framework for this analysis was already summarised in \citetalias{schrabback18} and \citetalias{dietrich19}.
It closely matches the approach employed in \cite{lee18}, to which we refer
the reader for a more detailed description.
For each target, mass bin, overdensity ($500c$ and $200c$) and simulation snapshot, we derive the mean and scatter of the log-normal, and interpolate linearly between the snapshots to the redshift of the target. We note that for any mass bin, the bias amplitudes inferred from the two simulation snapshots differ by at most $10\%$.

As a prior on the mass, we use the SZ-derived masses ($M_{500c,\text{SZ}}$ and $M_{200c,\text{SZ}}$) from
\citetalias{bocquet19}. We use the asymmetric distributions of these mass priors to marginalise over the mass dependence of the weak-lensing bias, to arrive at a final mean and dispersion for each target.

To additionally account for miscentring, we add a step to the procedure described above.
Prior to fitting masses, we offset the shear field on the sky in a random direction, where the  magnitude of the offset is drawn from a miscentring distribution.
In this
paper, we use the two miscentring distributions also used by \citetalias{schrabback18},
derived from the Magneticum Pathfinder Simulation \citep{dolag16} and based on using X-ray centroids and SZ peaks from the simulation as proxies (see Appendix \ref{app:miscentring} for details).
A shortcoming of this approach is that all clusters, mergers and relaxed systems alike, are treated in the same way\footnote{We note, however, that our current SZ-miscentring correction already depends on the cluster core radius $\theta_\mathrm{c}$ (see Appendix \ref{app:miscentring}), which has some dependence on cluster morphology.}, while we would expect the miscentring to be greater, on average, in merging systems \citep[e.g.][]{bleem20,zenteno20}. In a future work
(Sommer et al., \textit{in prep.}),
we plan to use hydrodynamical simulations to explore the bias magnitude due to miscentring for different dynamical states.

We list the estimates
for $\hat{b}_{\Delta,\text{WL}}$
and the scatter  $\sigma(\mathrm{ln} \,b_\mathrm{\Delta,\text{WL}})$
including their statistical uncertainties for the different clusters incorporating miscentring in Tables \ref{tab:mass}, \ref{tab:mass:SZ}  and \ref{tab:mass:snap}.
For clusters already studied in \citetalias{schrabback18},
slight to moderate shifts can occur in the reported bias values for two reasons: First, we now account for a mass dependence of the bias
\citep[see also][]{sommer21}.
Second, our modifications in the source selection (especially for the clusters with new VLT data) changes the relative contributions of scales at different radii, thereby affecting the mass modelling bias.

The average bias values are summarised in Table \ref{tab:bias},
showing that masses are expected to be biased  by \mbox{$-5\%$} to $-6\%$ when centred on the 3D halo centre.
Miscentring increases the bias by \mbox{$-7\%$} for ACS mosaics and X-ray centres, \mbox{$-12\%$}  in the case of  ACS mosaics and SZ centres,  and \mbox{$-19\%$} for the ACS+GMOS observations  and SZ centres, which are more strongly affected because of the smaller field of view.
Comparing Tables \ref{tab:mass:SZ} and \ref{tab:mass:snap} we see that the  limited radial coverage provided by the ACS+GMOS observations also leads to a substantial increase in the
 estimated intrinsic scatter  $\sigma(\mathrm{ln}\, b_\mathrm{\mathrm{500c,WL}})$.

The largest systematic uncertainty related to these bias estimates is given by the uncertainty in the miscentring correction.
As a conservative estimate \citetalias{schrabback18} assume that this uncertainty would at most be half of the actual correction.
Here we follow their conservative assumption, not only because of uncertainties in the assumed miscentring distributions, but also because our simulation analysis suggests  that the assumption of a log-normal
scatter is not strictly met when  miscentring is included
\citep[see][]{sommer21}.
 This constitutes the largest contribution to our systematic error budget for the analysis using SZ centres (see Sect.\thinspace\ref{se:sys_budget}), highlighting the importance  of reducing this uncertainty
  in future WL studies
  of larger samples.

\begin{figure}
    \includegraphics[width=\columnwidth]{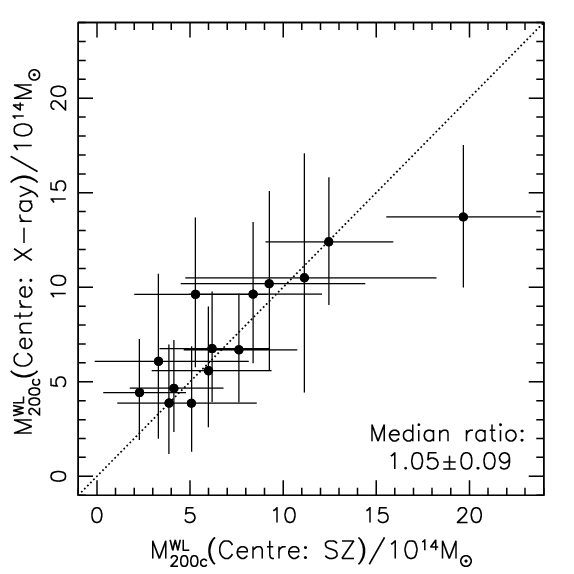}
    \caption{Comparison of the best-fitting WL mass estimates obtained for the clusters with
      ACS mosaics  using the X-ray centres versus the SZ centres,
      approximately corrected for mass modelling bias as \mbox{$M_{200\mathrm{c}}^\mathrm{WL}=M_{200\mathrm{c}}^\mathrm{biased,ML}/\hat{b}_{200\mathrm{c,WL}}$}. The error-bars correspond to the combined statistical errors given in Tables \ref{tab:mass} and \ref{tab:mass:SZ}, respectively, but do not include the additional
      scatter inferred from the simulations.}
    \label{fig:mass_mass_centre}
\end{figure}

  As a cross-check for the miscentring correction we compare the
   WL mass estimates  obtained for the clusters with ACS mosaics
   using the X-ray centres versus
using
   the SZ centres in Fig.\thinspace\ref{fig:mass_mass_centre},
applying  approximate corrections for the corresponding mass modelling biases.
The median  ratio \mbox{$1.05\pm 0.09$}
of the corrected estimates
  is consistent with unity, as one would expect
  in  case of
  an accurate correction.
  Given the limited sample size the statistical uncertainty of this ratio (estimated by bootstrapping the clusters) is still substantial, exceeding the estimated systematic uncertainty of the miscentring correction (compare Table \ref{tab:sys}).
  Future studies using larger samples should, however, be able to use similar
  cross-checks to test their miscentring corrections
  at a useful precision.

An additional source of systematic uncertainty
is given by the impact of baryons, which may systematically shift the distributions of cluster concentrations compared to the N-body simulations we are using to calibrate mass modelling biases.
\citetalias{schrabback18}
estimate that this could lead to a mass bias uncertainty of 2--4\%, where we conservatively assume a 4\% uncertainty in our systematic error budget.

\begin{table}
  \caption{Simulation-derived estimates of $\hat{b}_\mathrm{\Delta,WL}$
      for different miscentring distributions and overdensities,  averaged over different cluster samples.
    The individual bias estimates and their statistical errors are listed in Tables  \ref{tab:mass}, \ref{tab:mass:SZ} and \ref{tab:mass:snap}.
\label{tab:bias}}
\begin{center}
\begin{tabular}{cccc}
\hline\hline
  Miscentring & Setup & \mbox{$\langle \hat{b}_\mathrm{200c,WL} \rangle$} & \mbox{$\langle \hat{b}_\mathrm{500c,WL} \rangle$} \\

\hline
None &  ACS mosaics & 0.95 & 0.95 \\
None & ACS+GMOS & 0.94 & 0.95 \\
X-ray &  ACS mosaics & 0.87 & 0.88 \\
X-ray & ACS+GMOS & 0.83 & 0.83 \\
SZ &  ACS mosaics & 0.82 & 0.83 \\
SZ & ACS+GMOS & 0.75 & 0.75 \\
\hline
\end{tabular}
\end{center}
\end{table}

\subsection{Systematic error summary}
\label{se:sys_budget}

We summarise the systematic error contributions described in Sections \ref{se:shapes}, \ref{se:redshiftdist}, and \ref{sec:massmodelling}
in Table \ref{tab:sys}.
For the clusters with ACS mosaics the total systematic uncertainty amounts to 7.5\% when using X-ray centres and 9.0\% for SZ centres.
The systematic uncertainty increases to 11.6\% for the analysis using SZ centres and ACS+GMOS observations due to their smaller field of view.

\begin{table}
  \caption{Systematic error budget for our current study.
\label{tab:sys}}
\begin{center}
\addtolength{\tabcolsep}{-4pt}
\begin{tabular}{lccc}
\hline
  \hline
Source & rel.~error   &  \multicolumn{2}{c}{rel.~error $M_\mathrm{500c}$} \\
& signal & & \\
\hline
{\bf Shape measurements:} &&\\
  Shear calibration & 1.5\% & \multicolumn{2}{c}{2.3\%} \\
\hline
{\bf Redshift distribution:} &&\\
  Photo-$z$~sys. + sampling variance  & 3.0\% & \multicolumn{2}{c}{4.5\%} \\
  Deblending & 0.5\% & \multicolumn{2}{c}{0.8\%}\\
  Blue member contamination & 0.9\% & \multicolumn{2}{c}{1.4\%} \\
\hline
{\bf Mass model:}  &&\\
  $c(M)$ relation &  & \multicolumn{2}{c}{4\%} \\
  Miscentring for &&\\
  $\quad$ ACS mosaics + X-ray  centres  &  & \multicolumn{2}{c}{3.5\%}\\
  $\quad$ / ACS mosaics + SZ  centres &  & \multicolumn{2}{c}{6\%}\\
  $\quad$ / ACS+GMOS + SZ  centres  &  & \multicolumn{2}{c}{9.5\%}\\
\hline
{\bf Total}: &  & \multicolumn{2}{c}{7.5\% / 9.0\% / 11.6\%} \\

\hline
\end{tabular}
\addtolength{\tabcolsep}{+4pt}
\end{center}
\end{table}

\section{Constraints on the SPT observable--mass relation}
\label{sec:scaling_relations}

We use the extended and updated HST WL data-set (HST-30) to constrain the SPT observable--mass relation. As in other recent SPT work, we also use the set of 19 weak-lensing observations of SPT clusters from Magellan/Megacam presented in \citetalias{dietrich19} (Megacam-19). Our full sample of SPT clusters with WL data then contains 49 objects.
In some comparisons conducted below we alternatively employ the previous HST weak lensing data set of 13 clusters (HST-13) from \citetalias{schrabback18} (not applying our updated calibrations and source selections).

\subsection{Observable--mass relation model and likelihood function}

Following previous SPT work \citep[e.g.,][]{vanderlinde10}, we describe the \emph{unbiased detection significance} $\zeta$ as a power law in mass and
the dimensionless Hubble parameter \mbox{$E(z)\equiv H(z)/H_0$}
\begin{equation}
\langle\ln\zeta\rangle = \ln\left[\gamma_\mathrm{field}\asz \left( \frac{M_{500c}}{3 \times 10^{14} \mathrm{M}_{\odot}/h} \right)^\bsz \left(\frac{E(z)}{E(0.6)}\right)^\csz\right],
\end{equation}
where \asz, \bsz, and \csz\ are the scaling relation parameters\footnote{In practice, we sample the parameter $\ln\asz$ instead of \asz.} and $\gamma_\mathrm{field}$ describes the effective depth of each of the SPT fields \citep[e.g.,][]{dehaan16}.
The unbiased significance $\zeta$ is related to the detection significance $\xi$ via
\begin{equation}
\label{eq:zetaxi}
P(\xi|\zeta) = \mathcal N(\sqrt{\zeta^2+3}, 1).
\end{equation}
The relationship between the lensing mass $M_\mathrm{WL}$ and the halo mass was defined earlier in Eq.~\ref{eq:MWL} (in the current section we always use \mbox{$\Delta=500\mathrm{c}$} and therefore suppress this index for better readability). The following covariance matrix describes the correlated intrinsic scatter between the logarithms of the two observables $\zeta$ and $M_\mathrm{WL}$
\begin{equation}
    \mathbf\Sigma_{\zeta-M_\mathrm{WL}} = \begin{pmatrix} \sigma_{\ln\zeta}^2 & \rho_{\mathrm{SZ}-\mathrm{WL}}\sigma_{\ln\zeta}\sigma_{\ln M_\mathrm{WL}} \\
    \rho_{\mathrm{SZ}-\mathrm{WL}}\sigma_{\ln\zeta}\sigma_{\ln M_\mathrm{WL}} & \sigma_{\ln M_\mathrm{WL}}^2
    \end{pmatrix}.
\end{equation}
The joint scaling relation then reads
\begin{equation}
    P\Bigl(\begin{bmatrix}\ln\zeta \\ \ln M_\mathrm{WL}\end{bmatrix}|M,z\Bigr) = \mathcal N\Bigl(\begin{bmatrix}\langle\ln\zeta\rangle(M,z) \\ \langle\ln M_\mathrm{WL}\rangle(M,z)\end{bmatrix}, \mathbf\Sigma_{\zeta-M_\mathrm{WL}}\Bigr).
\end{equation}
Following previous work (\citetalias{dietrich19}, \citetalias{bocquet19}), we compute the likelihood function for each cluster with weak-lensing data as
\begin{equation}
\label{eq:gt_given_xi}
\begin{split}
  P(g_\mathrm{t} |\xi, z, \boldsymbol p) = \iiint&
  \mathrm{d}M\, \mathrm{d}\zeta \,\mathrm{d}M_\mathrm{WL}
  \left[\right.\\
    &P(\xi|\zeta)\, P(g_\mathrm{t}|M_\mathrm{WL}, N_\mathrm{source}(z), \boldsymbol p)\\
    &P(\zeta, M_\mathrm{WL}|M,z,\boldsymbol p)\, P(M|z, \boldsymbol p) \left.\right]\,,
\end{split}
\end{equation}
with the lensing source redshift distribution $N_\mathrm{source}(z)$, and where $\boldsymbol p$ is the vector of astrophysical and cosmological modelling parameters and $(M|z, \boldsymbol p)$ is the halo mass function \citep{tinker08}. The total log-likelihood is then obtained by summing the logarithms of the individual cluster likelihoods.

\subsection{Priors and Sampling}

Our WL data-set is not able to provide useful constraints on the mass-slope \bsz\ and the intrinsic scatter $\sigma_{\ln\zeta}$. We therefore apply Gaussian priors motivated by our latest cosmological analysis $\bsz\sim \mathcal{N} (1.53, 0.1^2)$ (\citetalias{bocquet19}) and a simulation-based prior $\sigma_{\ln\zeta}\sim \mathcal{N}(0.13, 0.13^2)$ \citep{dehaan16}.
{The intrinsic scatter in the WL mass \mbox{$\sigma_{\ln M_\mathrm{WL}}=\sigma(\mathrm{ln} \,b_\mathrm{500c,\text{WL}})$} and the employed correction for mass modelling bias are estimated from simulations as described in Sect.\thinspace\ref{sec:massmodelling}.}
The correlation coefficient $\rho_{\mathrm{SZ}-\mathrm{WL}}$ is allowed to vary in the range $[-1,1]$; our analysis prefers a positive correlation but this preference is not statistically significant.

We update the cosmology and scaling relation pipeline\footnote{\url{https://github.com/SebastianBocquet/SPT_SZ_cluster_likelihood}} used, e.g., for the latest cosmological analysis of SPT clusters (\citetalias{bocquet19}), to include the HST data presented in this work. The pipeline is embedded in the \textsc{cosmosis} framework \citep{zuntz15}. We explore the likelihood using the \textsc{multinest} sampler \citep{feroz09}, employing 500~\textsc{live\_points}, an \textsc{efficiency} of 0.1, and a \textsc{tolerance} of 0.01.

\subsection{The $\zeta$--mass relation}

With the likelihood machinery in place, we determine the parameters of the $\zeta$--mass relation by exploring the likelihood described in Eq.~\ref{eq:gt_given_xi}. The results are summarised in Table~\ref{tab:zeta_mass}.

\begin{table*}
	\centering
	\caption{The parameters of the $\zeta$--mass relation. The constraints from the  HST-30 + Megacam-19 data-set constitute a key result of this work. The SPTcl ($\nu\Lambda$CDM) results are obtained from the SPT cluster counts together with the weak-lensing and X-ray mass calibration from \citetalias{bocquet19}. The {\it Planck} + SPTcl ($\nu\Lambda$CDM) results are obtained using {\it Planck} (TT,TE,EE+lowE) and SPT cluster counts (without weak-lensing mass calibration).}
	\label{tab:zeta_mass}
	\begin{tabular}{lccccc}
		\hline
		\hline
		Parameter & Prior & \multicolumn{2}{c}{HST-30 + Megacam-19} & SPTcl ($\nu\Lambda$CDM) & {\it Planck} + SPTcl ($\nu\Lambda$CDM)\\
		&& Fiducial & Binned &(\citetalias{bocquet19})& (SPTcl abundance only)\\
		\hline
		$\ln\asz$ & flat & $1.63\pm0.19$ & -- & $1.67\pm0.16$ & $1.27^{+0.08}_{-0.15}$\\
		$\ln A_\mathrm{SZ}(0.25<z<0.5)$ & flat &  -- & $1.69\pm0.21$ & -- & -- \\
		$\ln A_\mathrm{SZ}(0.5<z<0.88)$ & flat &  -- & $1.51\pm0.27$ & -- & -- \\
		$\ln A_\mathrm{SZ}(0.88<z<1.2)$ & flat &  -- & $1.95^{+0.40}_{-0.56}$ & -- & -- \\
		\csz & flat/fixed & $1.78\pm1.11$ & $1.78$ & $0.63^{+0.48}_{-0.30}$ & $0.73^{+0.17}_{-0.19}$\\
		\hline
		\multicolumn{6}{l}{Parameters that are prior-dominated in our analysis:}\\
		\bsz & $\mathcal{N}(1.53, 0.1^2)^a$ & $1.56\pm0.09$ & $1.56\pm0.09$ & $1.53\pm0.09$ & $1.68\pm0.08$\\
		$\sigma_{\ln\zeta}$ & $\mathcal{N}(0.13, 0.13^2)$ & $0.17^{+0.06}_{-0.14}$ & $0.17^{+0.07}_{-0.13}$ & $0.17\pm0.08$ & $0.16^{+0.05}_{-0.14}$\\
		\hline
		\multicolumn{6}{l}{$^a$ The Gaussian prior on \bsz\ is only applied for the HST + Megacam analyses.}\\
	\end{tabular}
\end{table*}

\begin{figure}
    \includegraphics[width=\columnwidth]{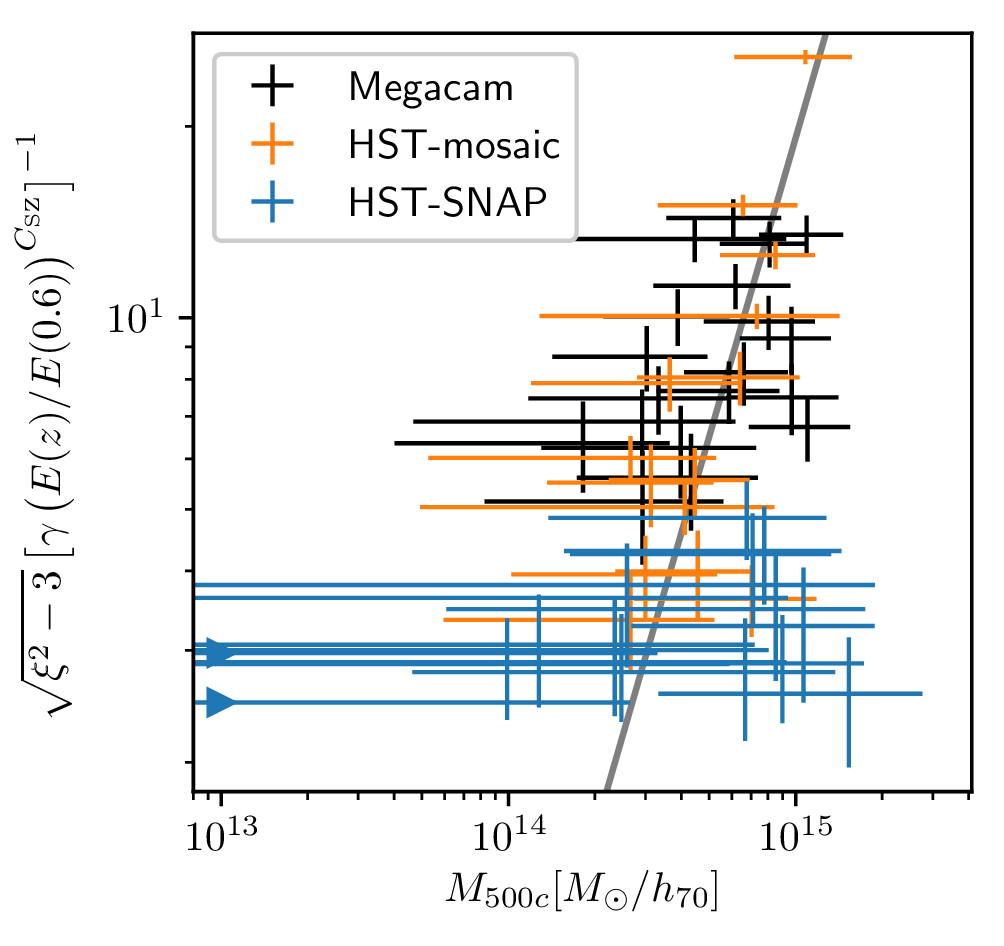}
   \caption{Normalised, debiased, and redshift-evolution-corrected SPT detection significance $\xi$ versus mass. The WL mass estimates are plotted at the best-fitting mass estimate, corrected for mass modelling bias. The error-bars include both shape noise and the  scatter in the mass estimates inferred from the simulations. The data points are colour-coded according to the source of the WL data. The solid line shows the best-fit $\zeta$--mass relation. The blue triangles mark clusters from the Snapshot programme which have a best-fitting mass $M_\mathrm{500c}<10^{13}\mathrm{M}_\odot$.}
    \label{fig:zeta_mass}
\end{figure}

In Figure~\ref{fig:zeta_mass}, we show the relationship between the normalised, debiased, and redshift-evolution-corrected SPT detection significance and the WL-based halo mass estimate $M_{500c}$. For each cluster, the best-fit WL mass estimate corresponds to the minimum $\chi^2$ between the measured and the modelled shear profiles, taking only the shape noise into account. The mass uncertainty is computed via $\Delta\chi^2=1$.
For the purpose of this figure, the WL mass estimates and the respective uncertainties are scaled with the WL mass bias (see Eq.~\ref{eq:MWL}) and the uncertainties are inflated with the intrinsic WL scatter.
We remind the reader that our scaling relation pipeline does not fit for a lensing mass; instead, it evaluates the likelihood of the measured shear profile $g_\mathrm{t}$, see Equation~\ref{eq:gt_given_xi}.

\subsection{The redshift evolution of the $\zeta$--mass relation}

\begin{figure*}
    \includegraphics[width=\textwidth]{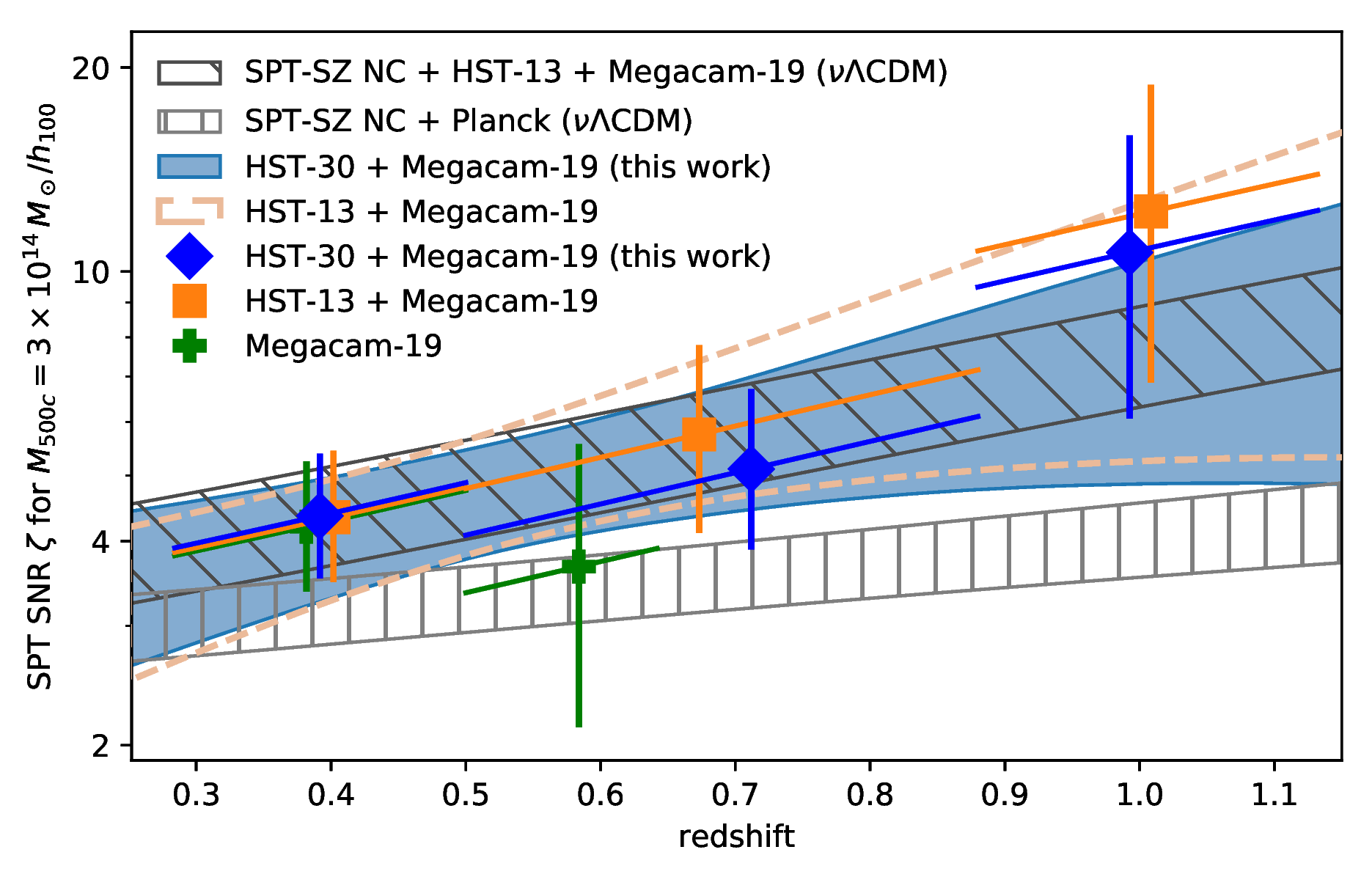}
    \caption{{The redshift evolution of the unbiased SPT detection significance $\zeta$ at the pivot mass $3\times10^{14}\mathrm{M}_\odot/h_{100}$. The bands and error bars show the 68\% credible interval for the overall relation and the redshift-binned analysis, respectively.
Our main analysis is shown in blue, while a corresponding analysis using the WL data employed by \citetalias{bocquet19} is shown in orange.
The data points are placed at the mean cluster redshift within each bin. The low-redshift data points are slightly shifted in redshift for better readability. The redshift evolution within each bin is set by $\csz=1.78$.
The hatched regions correspond to the scaling relations derived from the SPT-SZ cluster counts for a {\it Planck} $\nu\Lambda$CDM cosmology and the WL-informed SPT cluster cosmology analysis by \citetalias{bocquet19}, respectively.}
    }
    \label{fig:zeta_z}
\end{figure*}

An important result from the previous subsection is that, with our WL data-set, we are able to place a constraint on the redshift evolution $\csz=1.78\pm1.11$, albeit weak.
We show the evolution of $\zeta$ with redshift in Fig.\thinspace\ref{fig:zeta_z}.
Coloured bands show the results for the fiducial scaling relation:
the predecessor HST-13 + Megacam-19 data-set in orange and the updated data-set from this work in blue.
The diagonally-hatched band shows the constraint obtained from a simultaneous analysis of the predecessor HST-13 + Megacam-19 cluster weak-lensing data, X-ray data, and cluster abundance measurements (\citetalias{bocquet19}),\footnote{The MCMC chain can be downloaded at \url{https://pole.uchicago.edu/public/data/sptsz-clusters/}.} marginalising over cosmological parameters for a flat $\nu\Lambda$CDM cosmology.
Finally, the vertically-hatched band shows the result from a joint analysis of {\it Planck} primary CMB anisotropies \citep[TT,TE,EE+lowE,][]{planck19-5} and the SPT-SZ cluster abundance, also marginalising over cosmological parameters for a flat $\nu\Lambda$CDM cosmology, but without any weak-lensing mass calibration.
In this case, the cosmology is essentially set by {\it Planck}, and mass calibration is achieved through the cluster abundance likelihood.

We observe an offset between the mass calibration required to match the {\it Planck} $\nu\Lambda$CDM cosmology and the mass calibration preferred by our weak-lensing data-set (compare the vertically-hatched band with the blue band in Fig.~\ref{fig:zeta_z} and the constraints in Table~\ref{tab:zeta_mass}).
The recovered parameters suggest that, at our pivot redshift \mbox{$z=0.6$}, the WL-preferred mass scale is lower than the mass scale required to match the {\it Planck} $\nu\Lambda$CDM cosmology by a factor
$0.76^{+0.10}_{-0.14}$.
This observation is equivalent to the observation that the parameter constraints on $\Omega_\mathrm{m}$ and $\sigma_8$ obtained from SPT clusters with WL mass calibration are somewhat lower than the constraints favoured by {\it Planck} (see e.g., \citealt{bocquet15a}, \citealt{dehaan16}, \citetalias{bocquet19}).

Because our set of WL clusters spans a rather wide range in redshift, we wish to investigate whether the simple scaling relation model adopted is able to provide a good description of the data. We split our WL clusters into separate redshift bins, limited by $z=0.25, 0.5, 0.88, 1.2$. The bin limits are chosen such that the full sample of 49~clusters has (almost) equal numbers of objects in each of the three bins. We then repeat the scaling relation analysis as discussed above, with the difference that each redshift bin now has its own normalisation parameter $\ln\asz(z)$. The redshift evolution within each bin is modelled as usual and we fix \csz\ to $1.78$, the best-fit result from the full analysis\footnote{Fixing \csz\ to $0.5$ -- a value that is close to the one recovered from the joint analysis of SPT number counts and WL mass calibration -- has negligible impact on the binned test.}.
The parameter constraints are also listed in Table~\ref{tab:zeta_mass}. We compare the recovered constraints on $\ln\asz(z)$ with the result obtained in the fiducial analysis. For each of the three redshift bins, the probability that the recovered amplitude $\ln\asz(z)$ and the fiducial $\ln\asz$ are consistent with 0 difference is larger than $p=0.6\,(\text{agreement within } 0.5\sigma)$\footnote{We use the code available at \url{https://github.com/SebastianBocquet/PosteriorAgreement}.}.

In Fig.~\ref{fig:zeta_z}, the data points with error bars show the results from the binned approach we just described.
We apply this binned analysis to three WL data combinations: ground-based Magellan/Megacam-19 data (green), the predecessor data-set HST-13 + Megacam-19 (orange), and the full data-set presented in this work (blue).
As discussed, we find no evidence that our simple description of the redshift evolution of the SPT observable--mass relation with a single parameter \csz\ is in disagreement with the data (compare the blue data points with the blue band in Fig.\thinspace\ref{fig:zeta_z}). Note that the slightly larger value of $\ln\asz$ in the highest-redshift bin would imply that a halo with a given SPT SZ signal would be less massive than implied by the fiducial scaling relation.
However, the highest-redshift data points above redshift $\sim0.9$ are still only weakly constrained and this test thus remains inconclusive.

\section{Summary, discussion, and conclusions}
\label{sec:conclusions}
In this work we presented weak lensing (WL) measurements for a total sample of 30 distant SPT-SZ clusters based on high-resolution galaxy shape measurements from HST.
This includes new observations for 16 clusters using
single-pointing ACS F606W images and one cluster with ACS mosaics, as well as a reanalysis of 13 clusters with ACS mosaics.
In order to remove cluster galaxies and preferentially select background sources we complemented the single-pointing ACS observations with new Gemini-South GMOS $i$-band imaging (ACS+GMOS sample).
For six of the 13 previously studied clusters with ACS mosaics (updated ACS+FORS2 sample) we included new FORS2 $I$-band imaging for the source selection, allowing us to significantly boost the WL source density compared to earlier work. This is  not only due to the longer integration times, but also benefited from the excellent image quality of these observations.
Studying the source density profiles we confirmed the success of the employed colour selection scheme to remove contaminating cluster galaxies from the source sample.
For all targets we employed new calibrations for the source redshift distribution \citep{raihan20} and shear recovery \citep{hernandez20}, which also allowed us to include galaxies with slightly lower signal-to-noise ratios in the analysis.

Based on the WL shear measurements we reconstructed the projected mass distributions, yielding clear cluster detections with peak signal-to-noise ratios \mbox{$S/N_\mathrm{peak}>3$} for all clusters with ACS mosaics and eight out of 16 clusters with  single-pointing ACS data.
In order to constrain the cluster masses we fitted NFW model predictions to the tangential reduced shear profiles, applying corrections for the impact of weak lensing magnification and the finite width of the source redshift distribution.
These mass constraints are expected to be biased because of miscentring and variations in cluster density profile.
We estimated and corrected for these mass modelling biases using simulated data sets based on the Millennium-XXL simulations \citep{angulo12}.

We have used our measurements in combination with  earlier WL constraints for lower-redshift clusters from Magellan (\citetalias{dietrich19}) to derive refined constraints on
the scaling relation between the debiased SPT cluster detection significance $\zeta$ and the cluster mass.
In particular,
we obtained constraints on the redshift evolution
of the scaling relation, which do not rely on information from the cluster counts.
While  yielding a steeper best-fit power-law index $C_\mathrm{SZ}$ for the redshift evolution, our analysis is still consistent with the scaling relation derived from the combination of the SPT clusters counts with earlier WL data (\citetalias{dietrich19}, \citetalias{schrabback18}) by \citetalias{bocquet19}.
As a cross-check for the scaling relation analysis we split the clusters into three redshift bins,  finding reasonable agreement between the redshift-binned analysis and the overall relation.

We have not yet used our expanded high-$z$ WL data set to derive improved cosmological constraints from SPT clusters, but postpone this to future work, which will also incorporate additional WL data for clusters at lower redshifts.
However,
we have compared our WL-derived scaling relation constraints to the scaling relation that would be expected from the SPT cluster counts in a flat {\it Planck} $\nu\Lambda$CDM cosmology (compare Fig.\thinspace\ref{fig:zeta_z}).
In all redshift bins the WL-based analysis yields higher   $\zeta$ at a given reference mass,
consistent with the previously reported offset in the best-fit $\sigma_8$ estimates between {\it Planck} and  SPT clusters (\citetalias{bocquet19}).
However, the overall significance of the discrepancy is still low, which is why larger WL data sets will be needed to sensitively test the level of agreement between SPT clusters and {\it Planck} CMB constraints.

Compared to the earlier work from \citetalias{schrabback18} we were able to reduce the total
systematic uncertainty for the analysis of clusters with ACS mosaics, for which we can use X-ray centroids to centre the WL reduced shear profiles, from 9.2\% to 7.5\%,
mostly due to our smaller shear calibration uncertainty.
Now the largest contribution to the systematic error budget comes from residual uncertainties in the mass modelling correction.
This is even more severe for the clusters in our ACS+GMOS sample for two reasons.
First, their smaller field of view (single ACS pointing) limits the constraints to  scales
\mbox{$r\lesssim 900\,$kpc}. Although we generally exclude the cluster cores (\mbox{$r<500\,$kpc}) from our analysis, this still amplifies the impact especially of miscentring uncertainties.
In addition, nearly all of the clusters currently lack high-resolution X-ray observations,
which would provide a tighter centre proxy than the SZ peak positions.
As a result, the analysis of these data is currently subject to a 11.6\%
total systematic uncertainty, which is dominated by mass modelling uncertainties.

While systematic errors do not yet dominate our total error budget, it will be crucial to reduce them for future WL analyses of larger samples of massive high-$z$ clusters.
As one step to reduce mass modelling uncertainties, X-ray centres should become available for large samples of massive clusters in the near future from eROSITA \citep{merloni12}.
In addition,
it will be important
to
reduce uncertainties in our understanding of miscentring distributions.
One route for this is given by the comparison of different centre proxies.
E.g.,
\citet{zhang19} compare the centres derived
from the  redMaPPer cluster finding algorithm to X-ray centres.
This was also done by \citet{bleem20}, who furthermore compared redMaPPer and SZ centres.
However, even X-ray centres do not exactly correspond to the 3D halo centres.
As argued by \citetalias{schrabback18}, a possible solution could be provided
by studying offset distributions between centre proxies (from X-ray, SZ, or optical data)
and weak lensing mass peaks \citep[which provide noisy tracers for the 3D halo centre,][]{dietrich12}, and comparing these distributions between the real data and mock data from hydrodynamical simulations with matched noise properties.
The two noisy distributions should agree
if the  hydrodynamical simulations accurately describe the true miscentring.

As a further
approach to reduce mass modelling uncertainties
we recommend to
generally obtain observations with a larger field of view (e.g.~the \mbox{$2\times 2$} ACS mosaics studied here) when obtaining pointed follow-up for massive high-$z$ clusters.
In addition to reducing systematic uncertainties this also reduces the weak lensing fit uncertainties  and the intrinsic scatter (compare Tables \ref{tab:mass:SZ} and \ref{tab:mass:snap}).

For clusters at redshifts \mbox{$0.7\lesssim z\lesssim 1.0$} a more cost-effective alternative
to
HST mosaics
may be provided by deep good-seeing ground-based $K_\mathrm{s}$ imaging.
This also has the benefit of reducing systematic uncertainties related to the calibration of the redshift distribution compared to the source selection scheme applied in this paper \citep{schrabback18b}.
We however stress that the depth and the resolution of HST observations (including NIR imaging for the source selection) are still critically needed for weak lensing measurements of massive clusters at \mbox{$z>1$}.

Samples of massive, well-selected clusters that extend out to high redshifts have been increasing rapidly in recent years \citep[e.g.][]{hilton18,hilton21,bleem20, huang20}, and will continue to do so thanks to the latest surveys, including the one conducted by eROSITA  \citep{merloni12}.
In order to exploit their full potential for constraints on dark energy properties and other cosmological parameters it will be crucial to further tighten the cluster mass calibration by   reducing systematic uncertainties and adding new WL data. This includes for example the observations conducted by \textit{Euclid} \citep[][]{laureijs11} and the Vera C.~Rubin Observatory \citep{lsst09},
especially for the calibration of more common intermediate-mass clusters,  as well as further deep high-resolution follow-up for rare high-mass, high-$z$ clusters.

\vspace{-0.5cm}
\section*{Acknowledgements}

This work is based on observations made with the NASA/ESA {\it Hubble Space
  Telescope}, using imaging data from the SPT follow-up GO programmes
12246 (PI: C.~Stubbs), 12477  (PI: F.~W.~High),  14352 (PI: J.~Hlavacek-Larrondo), and
13412 (PI:
Schrabback), as well as archival data from
GO programmes 9425, 9500, 9583, 10134, 12064, 12440, and 12757, obtained via the data archive at the Space
Telescope Science Institute,
and catalogues based on observations taken by the 3D-HST Treasury Program
(GO 12177 and 12328) and the UVUDF Project (GO 12534, also based on data
from GO programmes 9978,
10086, 11563, 12498).
STScI is operated by the Association of Universities for Research in Astronomy, Inc. under NASA contract NAS 5-26555.
It is also based on observations made with ESO Telescopes at the La Silla Paranal Observatory under programmes
086.A-0741 (PI: Bazin), 088.A-0796 (PI: Bazin), 088.A-0889 (PI: Mohr), 089.A-0824 (PI: Mohr),
0100.A-0217 (PI: Hern\'andez-Mart\'in), 0101.A-0694 (PI: Zohren), and 0102.A-0189 (PI: Zohren).
It is also based on observations obtained at the Gemini Observatory, which is operated by the Association of Universities for Research in Astronomy, Inc., under a cooperative agreement with the NSF on behalf of the Gemini partnership: the National Science Foundation (United States), National Research Council (Canada), CONICYT (Chile), Ministerio de Ciencia, Tecnolog\'{i}a e Innovaci\'{o}n Productiva (Argentina), Minist\'{e}rio da Ci\^{e}ncia, Tecnologia e Inova\c{c}\~{a}o (Brazil), and Korea Astronomy and Space Science Institute (Republic of Korea),
under programmes  2014B-0338
and 	2016B-0176 (PI: B.~Benson).

The scientific results reported in this article are based in part on
observations made by the {\it Chandra} X-ray Observatory
(ObsIDs 9332, 9333, 9334, 9335, 9336, 9345, 10851, 10864, 11738, 11739, 11741, 11742, 11748, 11799, 11859, 11864, 11870, 11997, 12001, 12002, 12014, 12091, 12180, 12189, 12258, 12264, 13116, 13117, 13478, 14017, 14018, 14022, 14023, 14349, 14350, 14351, 14437, 15572, 15574, 15579, 15582, 15588, 15589, 18240, 18241).

We  thank
Raul Angulo for providing
data from the Millennium XXL Simulation, and Klaus Dolag for providing access to data from the Magneticum Pathfinder Simulation.

The Bonn group acknowledges support from  the German Federal Ministry for
Economic Affairs and Energy (BMWi) provided through DLR under projects  50OR1407,  50OR1610,  50OR1803, 50OR2002, 50QE1103, and 50QE2002, as well as support provided  by the Deutsche Forschungsgemeinschaft (DFG, German Research Foundation) under grant 415537506.
HHo acknowledges support from Vici grant 639.043.512
from the Netherlands Organisation for Scientific Research (NWO).
AAS acknowledges support from U.S. NSF grant AST-1814719.
AS is supported by the ERC-StG 'ClustersXCosmo' grant agreement 716762, and by the FARE-MIUR grant 'ClustersXEuclid' R165SBKTMA.
BB is supported by the Fermi Research Alliance LLC under contract no. De-AC02-07CH11359 with the U.S. Department of Energy.
JLvdB is supported by the European Research Council (Grant No. 770935).
HZ and SFR are members of and received financial support
from
the International Max Planck Research School (IMPRS) for Astronomy and
Astrophysics at the Universities of Bonn and Cologne.

This work was performed in the context of the South-Pole Telescope scientific programme. SPT is supported by the National Science Foundation through grants PLR-1248097 and OPP-1852617. Partial support is also provided by the NSF Physics Frontier Center grant PHY-0114422 to the Kavli Institute of Cosmological Physics at the University of Chicago, the Kavli Foundation and the Gordon and Betty Moore Foundation grant GBMF 947 to the University of Chicago. This work is also supported by the U.S. Department of Energy. PISCO observations are supported by NSF AST-1814719.
Argonne National Laboratory's work was supported by the U.S. Department of Energy, Office of High Energy Physics, under contract DE-AC02-06CH11357.

\vspace{-0.3cm}
\section*{Data availability}
SPT cluster data products are available at \url{https://pole.uchicago.edu/public/data/sptsz-clusters/}.
The scaling relation pipeline used in our analysis is available at \url{https://github.com/SebastianBocquet/SPT_SZ_cluster_likelihood}. It will be updated to include the tangential reduced shear profiles and source redshift distributions derived in our analysis once this paper is accepted.

\vspace{-0.3cm}
\bibliographystyle{mn2e}

\bibliography{oir}

\section*{Affiliations}
\altaffilmark{\Bonn} Argelander-Institut f\"{u}r Astronomie, Universit\"{a}t Bonn, Auf dem
H\"{u}gel 71, 53121, Bonn, Germany\\
\altaffilmark{\Munich} Faculty of Physics, Ludwig-Maximilians University, Scheinerstr.\ 1, 81679 M\"{u}nchen, Germany\\
\altaffilmark{\ExcellenceCluster} Excellence Cluster ORIGINS,
Boltzmannstr. 2, 85748 Garching, Germany\\
\altaffilmark{\Bochum} Ruhr-University Bochum, Astronomical Institute, German Centre for Cosmological Lensing, Universit\"atsstr. 150, 44801 Bochum, Germany\\
\altaffilmark{\Leiden} Leiden Observatory, Leiden University, Niels Bohrweg
2, NL-2300 CA Leiden, The Netherlands\\
\altaffilmark{\MPIA} Max-Planck-Institut f\"ur Astronomie, K\"onigstuhl 17, D-69117 Heidelberg, Germany\\
\altaffilmark{\KICPChicago} Kavli Institute for Cosmological Physics, University of Chicago, 5640 South Ellis Avenue, Chicago, IL 60637 \\
\altaffilmark{\Cincinnati} Department of Physics, University of Cincinnati, Cincinnati, OH 45221, USA \\
\altaffilmark{\FNAL} Fermi National Accelerator Laboratory, Batavia, IL 60510-0500, USA\\
\altaffilmark{\AAUChicago} Department of Astronomy and Astrophysics, University of Chicago, 5640 South Ellis Avenue, Chicago, IL 60637 \\
\altaffilmark{\PhysicsUChicago} Department of Physics, University of Chicago,
  5640 South Ellis Avenue, Chicago, IL 60637 \\
\altaffilmark{\ANL} Argonne National Laboratory, 9700 S. Cass Avenue,
Argonne, IL, USA 60439\\
\altaffilmark{\Missouri} Department of Physics and Astronomy, University of Missouri--Kansas City, 5110 Rockhill Road, Kansas City, MO 64110, USA\\
\altaffilmark{\Montreal} D\'epartement de Physique, Universit\'e de Montr\'eal, Montr\'eal, QC, Canada\\
\altaffilmark{\MIT} MIT Kavli Institute for Astrophysics and Space
Research, Massachusetts Institute of Technology, 77 Massachusetts
Avenue, Cambridge, MA 02139 \\
\altaffilmark{\Trieste}  Astronomy Unit, Department of Physics, University of Trieste, via Tiepolo 11, I-34131 Trieste, Italy\\
\altaffilmark{\IFPU} IFPU - Institute for Fundamental Physics of the Universe, Via Beirut 2, 34014 Trieste, Italy \\
\altaffilmark{\INAF} INAF - Osservatorio Astronomico di Trieste, via G. B. Tiepolo 11, I-34143 Trieste, Italy \\
\altaffilmark{\INFN} INFN -  National Institute for Nuclear Physics, Via Valerio 2, I-34127 Trieste, Italy \\
\altaffilmark{\CfA} Center for Astrophysics $|$ Harvard \& Smithsonian
60 Garden Street $|$ MS 42 $|$ Cambridge, MA 02138 \\

\appendix

\section{Dependence of the average geometric lensing efficiency on galaxy size}
\label{se:beta_of_rf}
\begin{figure}
  \includegraphics[width=0.99\columnwidth]{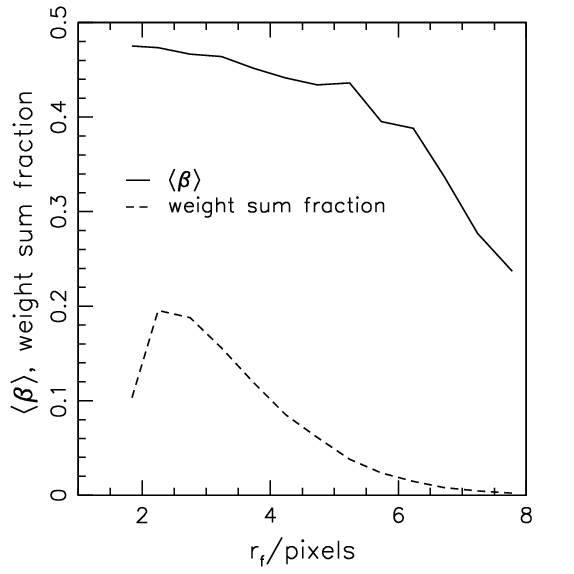}
   \caption{Dependence of the estimated $\langle\beta\rangle$ for  SPT-CL{\thinspace}$J$0000$-$5748 on the galaxy flux radius $r_\mathrm{f}$, including all galaxies with \mbox{$24<V<26.5$} that pass the ACS-only colour selection, averaged over the five CANDELS fields (solid curve). The dashed curve indicates the fraction  of the summed shape weights of the galaxies located within the corresponding bin (width \mbox{$\Delta r_\mathrm{f}=0.5$} pixels).
\label{fi:beta_rf}}
\end{figure}
Following the ACS-only colour selection and   including all galaxies with \mbox{$24<V<26.5$},
Fig.\thinspace\ref{fi:beta_rf} shows the dependence of the estimated $\langle\beta\rangle$ for  SPT-CL{\thinspace}J0000$-$5748
on the galaxy flux radius $r_\mathrm{f}$. The dashed curve in the figure represents the fraction  of the summed shape weights of the galaxies that are located within the corresponding $r_\mathrm{f}$ bin.
This shows that most galaxies are located in the regime \mbox{$r_\mathrm{f}\lesssim 5$} pixels, where $\langle\beta\rangle$  depends only weakly on $r_\mathrm{f}$.

\section{Cross-check from overlapping shear estimates}
\begin{figure}
 \includegraphics[width=0.99\columnwidth]{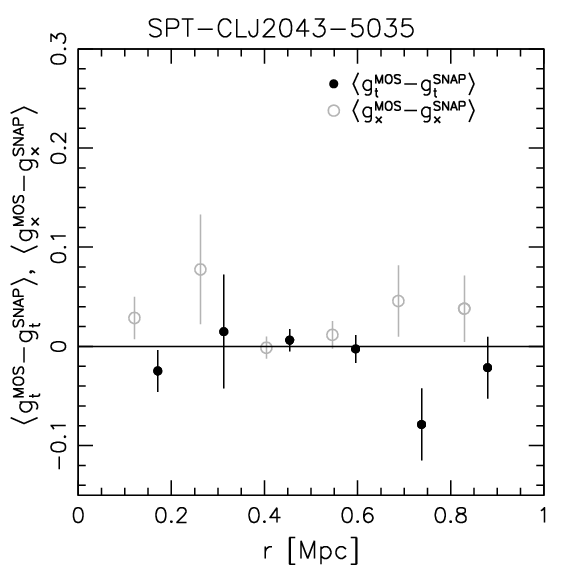}
 \vspace{-0.1cm}
 \caption{Difference in the reduced tangential shear (black solid points) and the cross shear (grey open points, shifted along the $x$-axis for clarity) estimates from the ACS mosaic versus the single-pointing SNAP observation
   of SPT-CL\thinspace$J$2043$-$5035, computed using only galaxies that are present in both catalogues  and plotted as a function of distance from the SZ centre. The outlier at \mbox{$r=0.75$ Mpc} is dominated by a single noisy galaxy with complex morphology. For this figure error-bars have been computed by randomising the phases of the ellipticity differences. The actual reduced tangential and cross shear profiles of the cluster are shown in Fig.\thinspace\ref{fi:wl_results_spt2043}.
\label{fi:spt2043_shear_comparison}}
\end{figure}

 The cluster SPT-CL\thinspace$J$2043$-$5035
 has been observed by two separate HST programmes (see Sect.\thinspace\ref{se:data:acs:highmass}).
 For this target we have therefore computed ACS F606W shape measurements from both the $2\times 2$ ACS mosaic and the central single pointing observation.
 This provides us with an opportunity to cross-check our shape measurements in the overlap region.
The difference profiles between the two reduced shear estimates is shown in
Fig.\thinspace\ref{fi:spt2043_shear_comparison},
decomposed into tangential and cross-components with respect to the cluster centre.
For the tangential component (which is used to constrain the mass models, see Sect.\thinspace\ref{sec:results}), the difference profile is well consistent with zero, as expected.
For the cross-component the difference is slightly positive on average, but combining the different radial bins the deviation from zero is significant at the \mbox{$\sim 1.6\sigma$} level only, which is therefore not a concern.
Note that for galaxies with two successful shear estimates we use the average of the two estimates in the actual cluster WL analysis  (see Sect.\thinspace\ref{sec:results}).

\section{Miscentring distributions}
\label{app:miscentring}
\begin{figure}
 \includegraphics[width=0.99\columnwidth]{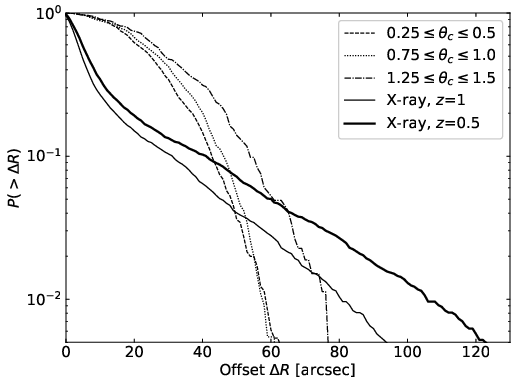}
 \vspace{-0.1cm}
 \caption{Cumulative offset distribution measured between the deepest point in the halo potential and
   centre proxies in the SPT-SZ mock observations.
   The dashed, dotted, and dash-dotted curves correspond to SZ centres for simulated clusters with
   core sizes in the ranges \mbox{$0\farcm25\le \theta_\mathrm{c} \le 0\farcm5$}, \mbox{$0\farcm75\le \theta_\mathrm{c} \le 1\farcm0$}, and \mbox{$1\farcm25\le \theta_\mathrm{c} \le 1\farcm5$}, which
have
r.m.s. offsets of $29^{\prime\prime}$, $31^{\prime\prime}$, and $36^{\prime\prime}$, respectively.
The solid curves correspond to  X-ray centres, whose r.m.s. offset of 159 kpc
equals
$20^{\prime\prime}$
 at \mbox{$z=1$} (thin curve) and
   $26^{\prime\prime}$ at \mbox{$z=0.5$} (thick curve).
\label{fi:miscentring}}
\end{figure}

To correct our weak lensing mass estimates for the impact of miscentring (see Sect.\thinspace\ref{sec:massmodelling})
 we employ miscentring distributions for the X-ray
 centroids and SZ peaks which  are based on the Magneticum Pathfinder Simulation
 \citep{dolag16}.
 These distributions were already employed in \citetalias{schrabback18} and are shown in Fig.\thinspace\ref{fi:miscentring}. In this Appendix we briefly summarise how these
 distributions were derived.

The large-volume, high-resolution cosmological hydrodynamical Magneticum
Pathfinder simulations were carried out with P-GADGET3, a modification of
P-GADGET-2 \citep{springel05b},
using an entropy-conserving formulation of Smoothed Particle Hydrodynamics
\citep[SPH,][]{springel02}
and including treatment of radiative cooling,
heating by a UV background, star formation and feedback processes
from supernovae explosions and active galactic nuclei
\citep{springel03,fabjan10}.
From this set of simulations
\citet{saro14}
and
\citet{gupta17}
employed
a high-resolution  simulation, which is based on a  WMAP7 cosmology
\citep{komatsu11}
and contains
$1512^3$ dark matter particles and as many gas particles
in a comoving box of $896 h^{-1}$ Mpc per side,
to create thermal Sunyaev-Zel'dovich Effect (SZE) light cones and SPT mock observations.
To replicate the observing conditions of the SPT-SZ survey they built five thermal
SZE light-cones up to \mbox{$z \sim 2$}, each of size $13\deg \times 13\deg$,
from which mock SPT observations at 95 GHz and 150 GHz were extracted.
In each mock, they accounted for three contributions: (1) primary CMB
anisotropies; (2) the SPT beam and transfer function
\citep{schaffer11}
at
the two frequencies; (3) instrumental noise consistent with the observed
SPT-SZ map depths of 18 and 44 $\mu$K-arcmin for the 150 GHz and
95 GHz bands, respectively. From these mocks, cluster candidates were
identified with the same approach adopted for real SPT clusters
\citep[e.g.][]{staniszewski09,reichardt13,bleem20}.
In particular,  a $\beta$-profile \citep{cavaliere76} with $\beta=1$ was used as a cluster template, and
different cluster core sizes $\theta_\mathrm{c}$ were adopted, also in line with
the SPT data analysis, with twelve discrete values evenly spaced in the
range
$0\farcm25$--$3\farcm0$ (only values $\le 1\farcm5$ are relevant for the present
analysis).
The resulting sample of SPT-SZ-like clusters (selected to have detection significances  \mbox{$\xi > 4.5$}) was used to characterise
both the SZE and X-ray miscentring distributions.
These were computed with respect
to the deepest
point in the potential of the halo, which is the centre  typically used for computations of the halo mass function \citep[e.g.][]{bocquet20}.
The SZ centres directly correspond to the SZ peak location, as employed for the real survey data.
X-ray centres were defined as the peak\footnote{For the analysis of the real clusters
  we employ X-ray centroids instead of peaks, since they are less affected by noise.
We expect that this has a minimal impact on our analysis.
As a consistency check we repeated the weak lensing analysis using the reduced shear profiles around the X-ray peaks instead of the centroids, leading to a change in the median weak lensing-estimated $M_{200\mathrm{c}}^\mathrm{biased,ML}$ of the clusters with {\it Chandra} X-ray centres by $-3\pm 2\%$ only (uncertainty estimated by bootstrapping the sample). This difference is within the estimated systematic uncertainty of our miscentring correction and negligible compared to our overall systematic uncertainty (see Table \ref{tab:sys}).}
in the
 X-ray surface brightness maps
within
a radius of  $2r_\mathrm{500c}$ around
 the deepest
point of the potential.

The resulting offset distributions are shown in Fig.\thinspace\ref{fi:miscentring}.
We find that the SZ miscentring
 mostly depends on the  cluster core size $\theta_\mathrm{c}$.
We therefore
compute and employ separate offset distributions for each of the discrete $\theta_\mathrm{c}$ values.
The X-ray offset distribution is measured in transverse physical
separation, leading to a redshift dependence of the offset distribution in angular separation
(see Fig.\thinspace\ref{fi:miscentring}).
In most cases the simulated X-ray centres trace the halo centres with smaller offsets  compared to the SZ centres, leading to smaller mass modelling bias corrections.
However, there is a slightly increased tail towards large offsets for the X-ray centres, especially at lower cluster redshifts.
For parametric fits to the SZ miscentring distribution in this simulation see \citet{gupta17}.

\vspace{-0.3cm}
\section{Weak lensing results for individual clusters}
In this Appendix we present the mass reconstructions and shear profiles of the individual clusters as described in Sections
\ref{sec:results:massrecon}
and
\ref{sec:results:shearprofile}.
These figures
are available as supplementary material.

\clearpage

\begin{figure*}
 \includegraphics[width=1.05\columnwidth]{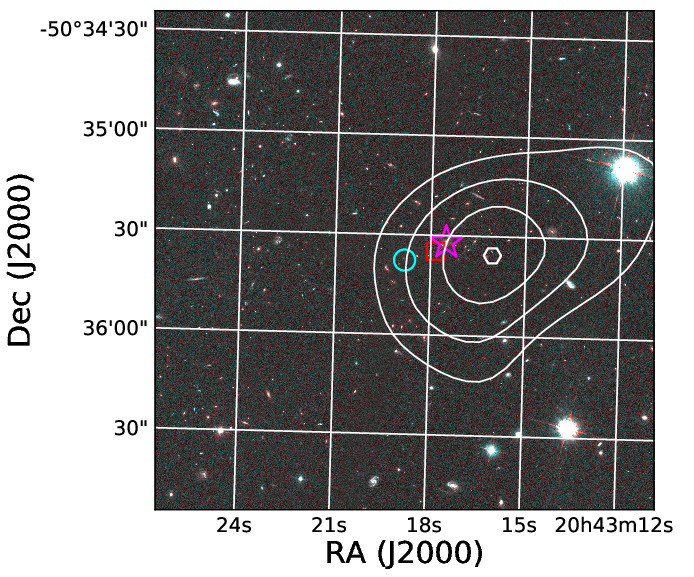}
 \includegraphics[width=0.9\columnwidth]{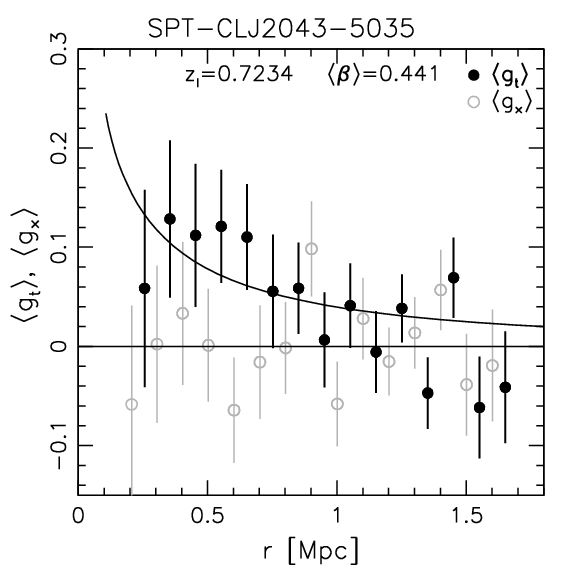}
 \caption{WL results for SPT-CL{\thinspace}$J$2043$-$5035. {\it Left:} Signal-to-noise ratio map of the WL mass reconstruction (starting at $2\sigma$ in steps of $0.5\sigma$, with the peak indicated by the hexagon), overlaid on the ACS F606W/F814W mosaic image (\mbox{$2\farcm5 \times 2\farcm5$} cutout, using
   F814W for the red channel and F606W for the blue and green channels).
 The BCG \citep[from][]{mcdonald19}, SZ peak, and X-ray centroid are shown by the magenta star, cyan circle, and red square, respectively.   {\it Right:} Combined reduced shear profile around the X-ray centre, showing the tangential (black solid circles with best-fitting NFW model) and cross (grey open circles, shifted along the $x$-axis for clarity) components. \label{fi:wl_results_spt2043}}
\end{figure*}

\begin{figure*}
 \includegraphics[width=1.05\columnwidth]{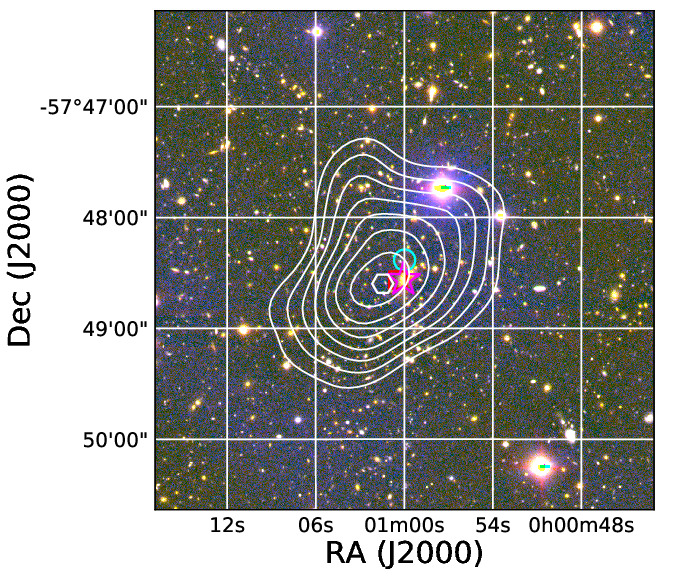}
 \includegraphics[width=0.9\columnwidth]{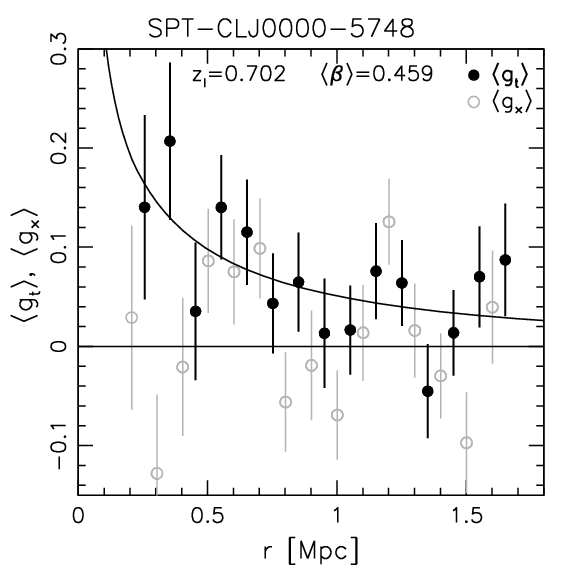}
 \vspace{-0.2cm}
 \includegraphics[width=1.05\columnwidth]{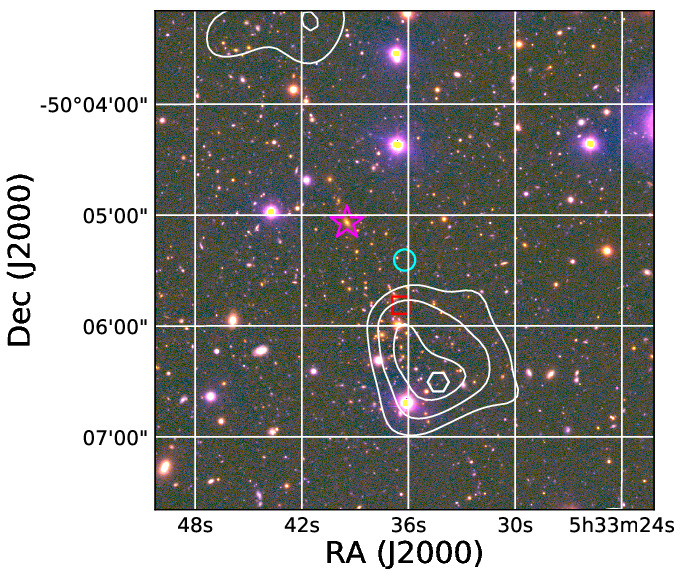}
 \includegraphics[width=0.9\columnwidth]{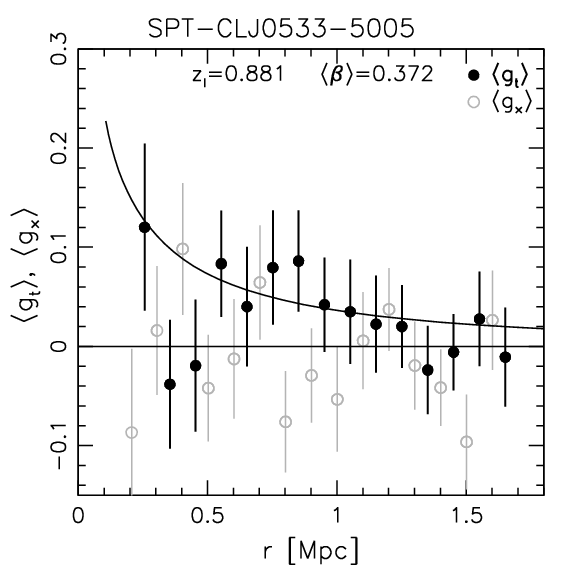}
 \vspace{-0.2cm}
 \includegraphics[width=1.05\columnwidth]{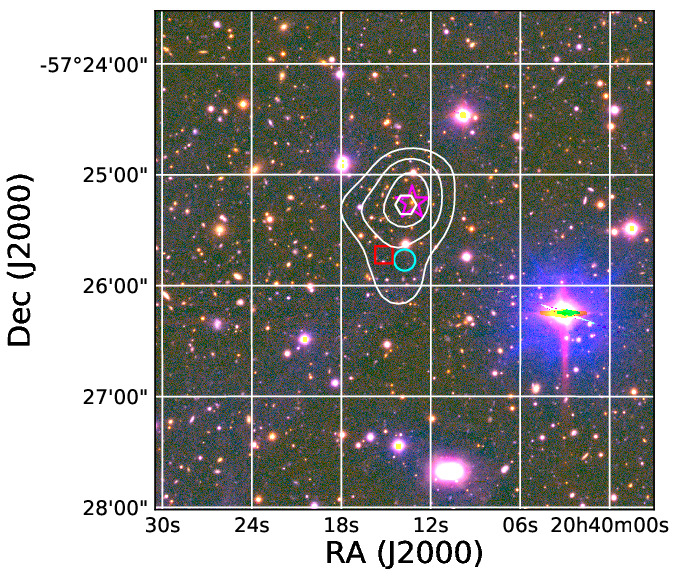}
 \includegraphics[width=0.9\columnwidth]{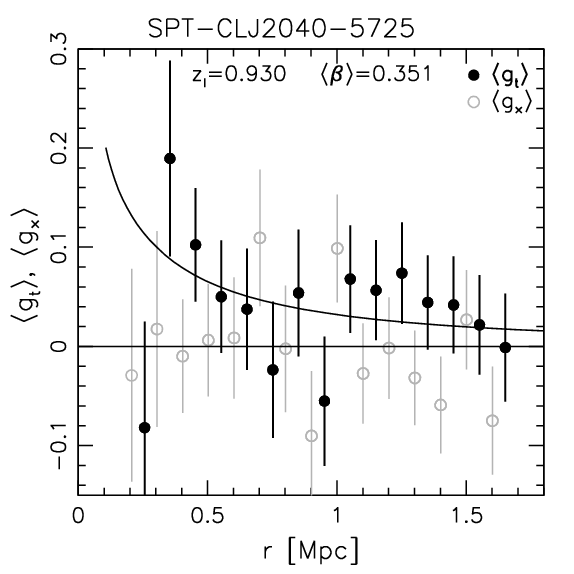}
 \caption{WL results for clusters from the updated ACS+FORS2 sample. {\it Left:} Signal-to-noise ratio map of the WL mass reconstruction (starting at $2\sigma$ in steps of $0.5\sigma$, with the peak indicated by the hexagon), overlaid on a VLT/FORS2 $zIB$ colour image \citep[\mbox{$4\farcm5 \times 4\farcm5$} cutout, using $z$- and $B$-band images from][]{chiu16}. The BCG \citep[from][]{chiu16}, SZ peak, and X-ray centroid are shown by the magenta star, cyan circle, and red square, respectively.   {\it Right:} Combined reduced shear profile around the X-ray centre, showing the tangential (black solid circles with best-fitting NFW model) and cross (grey open circles, shifted along the $x$-axis for clarity) components.
   \label{fi:wl_results_hst13-2019-1}}
\end{figure*}

\begin{figure*}
 \includegraphics[width=1.05\columnwidth]{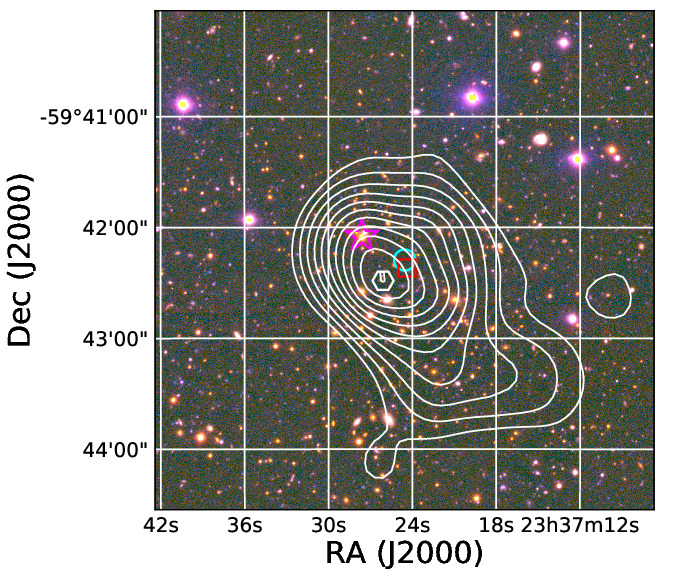}
 \includegraphics[width=0.9\columnwidth]{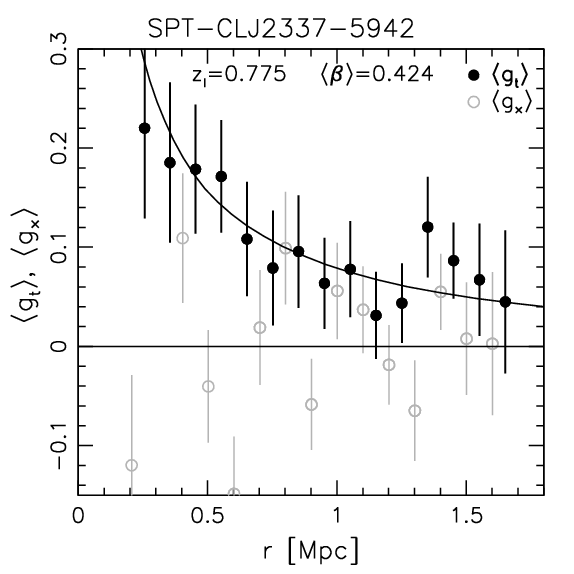}
 \includegraphics[width=1.05\columnwidth]{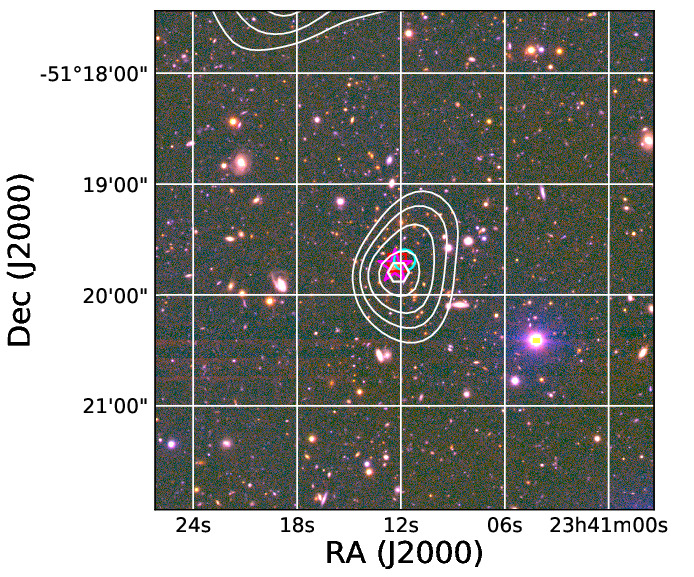}
 \includegraphics[width=0.9\columnwidth]{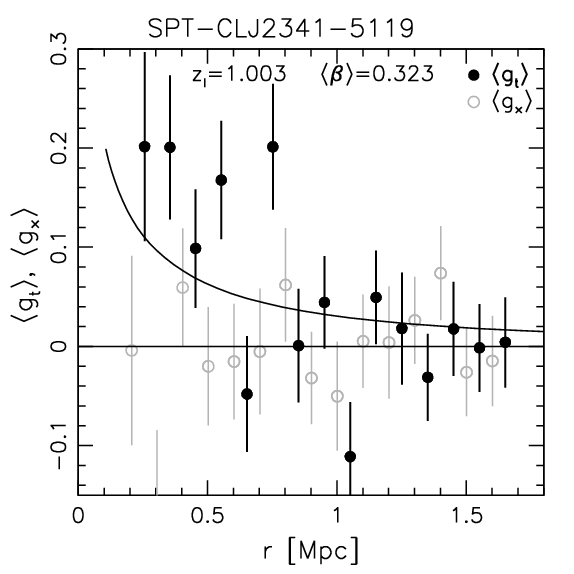}
 \includegraphics[width=1.05\columnwidth]{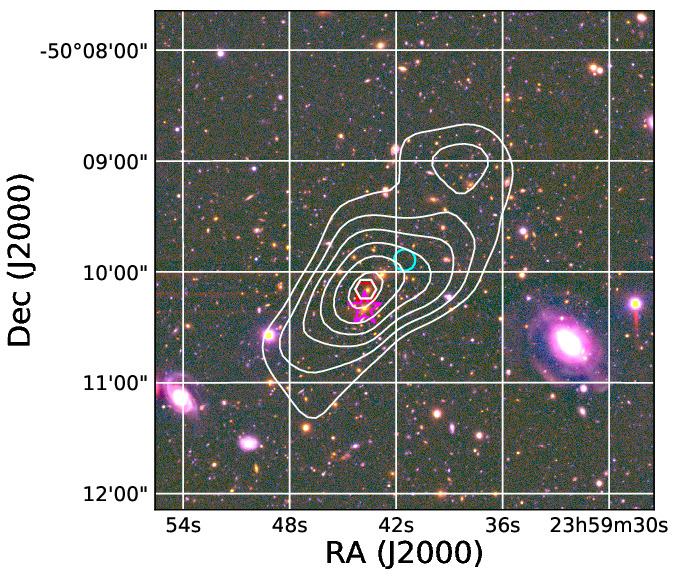}
 \includegraphics[width=0.9\columnwidth]{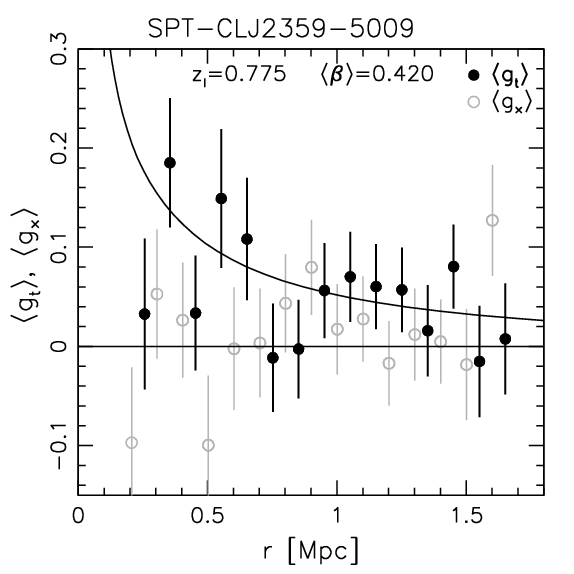}
   \caption{WL results for clusters from the updated ACS+FORS2 sample (continued, see Fig.\thinspace\ref{fi:wl_results_hst13-2019-1} for details). \label{fi:wl_results_hst13-2019-2}}
\end{figure*}

\begin{figure*}
 \includegraphics[width=1.05\columnwidth]{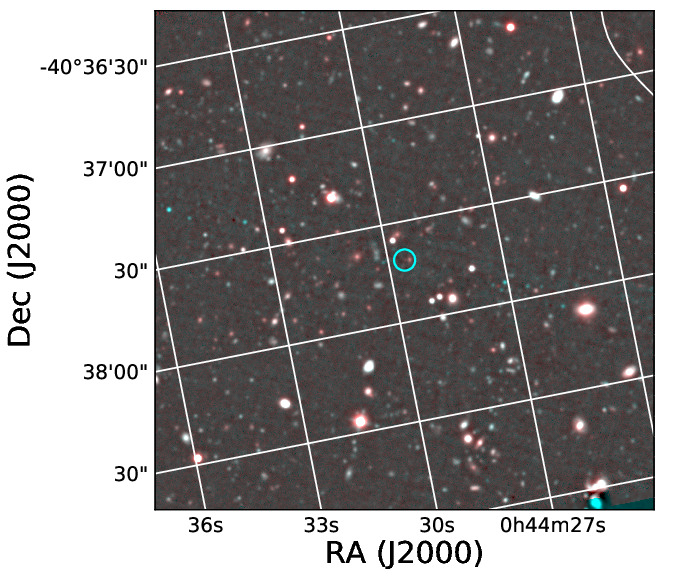}
 \includegraphics[width=0.9\columnwidth]{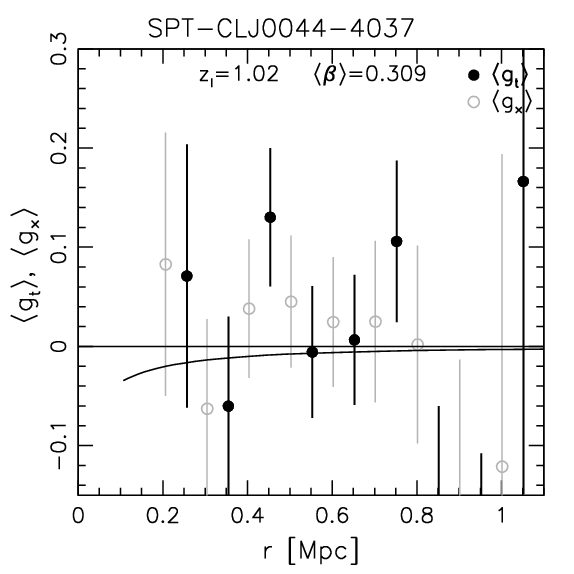}
 \vspace{-0.2cm}

 \includegraphics[width=1.05\columnwidth]{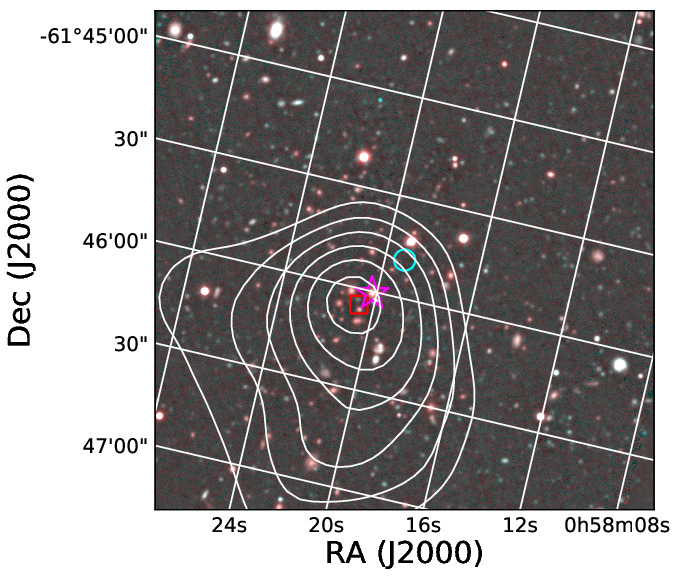}
 \includegraphics[width=0.9\columnwidth]{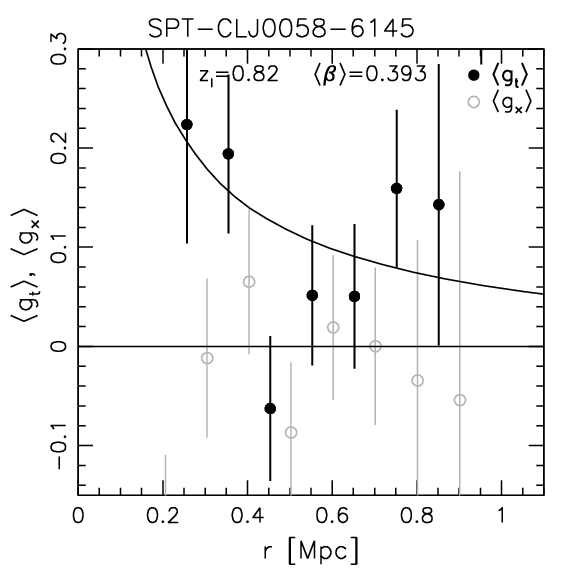}

 \caption{WL results for clusters from the ACS+GMOS sample. {\it Left:} Signal-to-noise ratio map of the WL mass reconstruction (starting at
   $1.5\sigma$
   in steps of $0.5\sigma$, with the peak indicated by the hexagon), overlaid on a
   colour image using the PSF-homogenised ACS/F606W image for the blue and green channels, and the GMOS $i$-band image for the red channel   (\mbox{$2\farcm5 \times 2\farcm5$} cutout).
   Differing from Figs.\thinspace\ref{fi:wl_results_spt2043} to \ref{fi:wl_results_hst13-2019-2} we show the extra contour at $1.5\sigma$ since the mass reconstructions  are more strongly affected by the mass-sheet degeneracy for the ACS+GMOS sample (see Sect.\thinspace\ref{sec:results:massrecon}), leading to a stronger underestimation of the true $S/N$.
The SZ peak
is shown by the
cyan circle.
For SPT-CL{\thinspace}$J$0058$-$6145 we additionally mark the X-ray centre from \citet[][red square]{mcdonald13} and the BCG candidate from \citet[][magenta star]{zenteno20}.
{\it Right:} Combined reduced shear profile around the SZ centre, showing the tangential (black solid circles with best-fitting NFW model) and cross (grey open circles, shifted along the $x$-axis for clarity) components.
   \label{fi:wl_results_snap-1}}
\end{figure*}

\begin{figure*}
 \includegraphics[width=1.05\columnwidth]{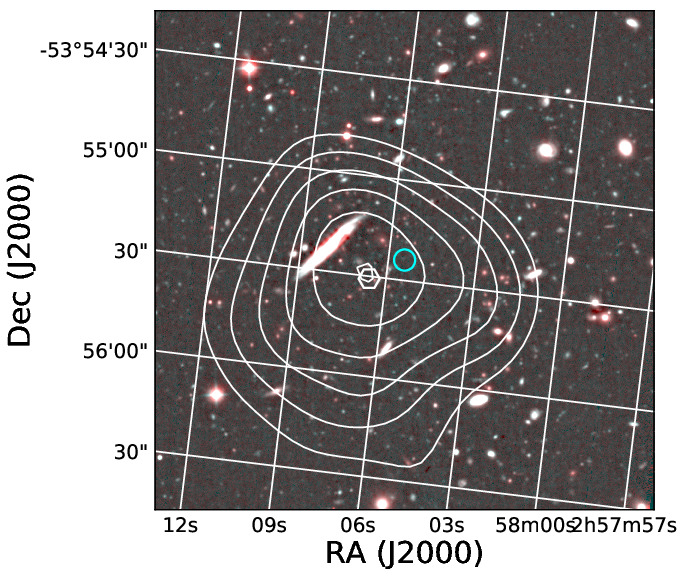}
 \includegraphics[width=0.9\columnwidth]{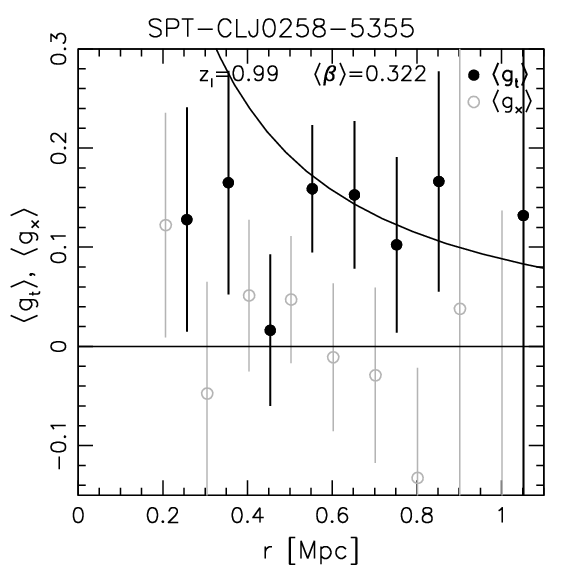}
 \includegraphics[width=1.05\columnwidth]{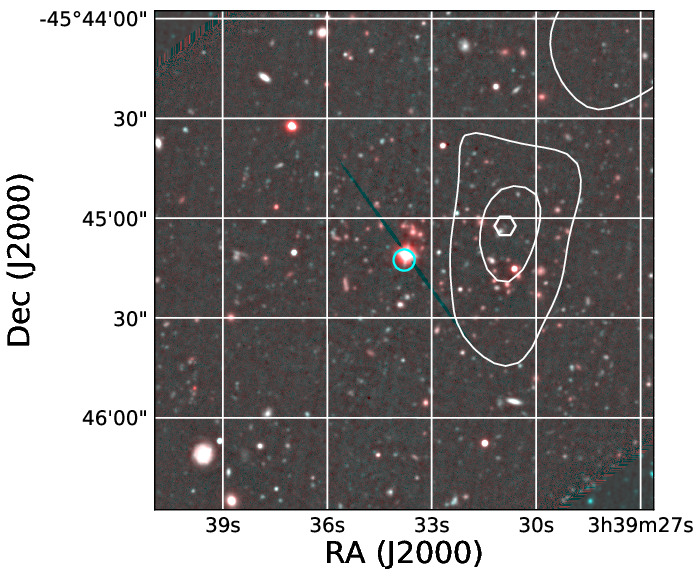}
 \includegraphics[width=0.9\columnwidth]{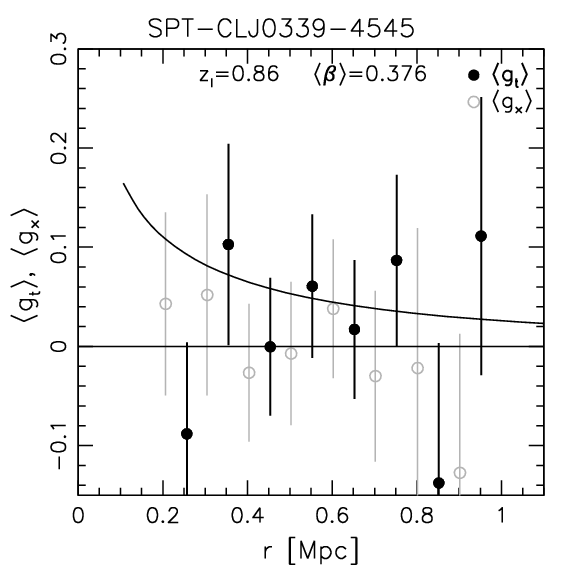}
 \includegraphics[width=1.05\columnwidth]{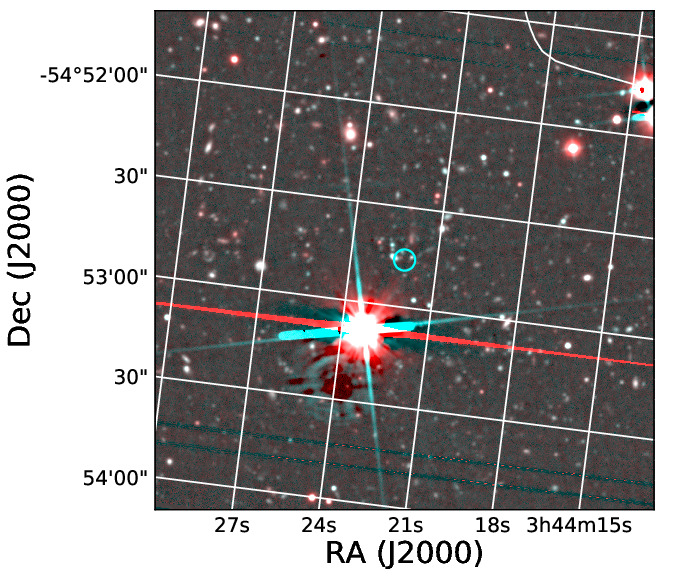}
 \includegraphics[width=0.9\columnwidth]{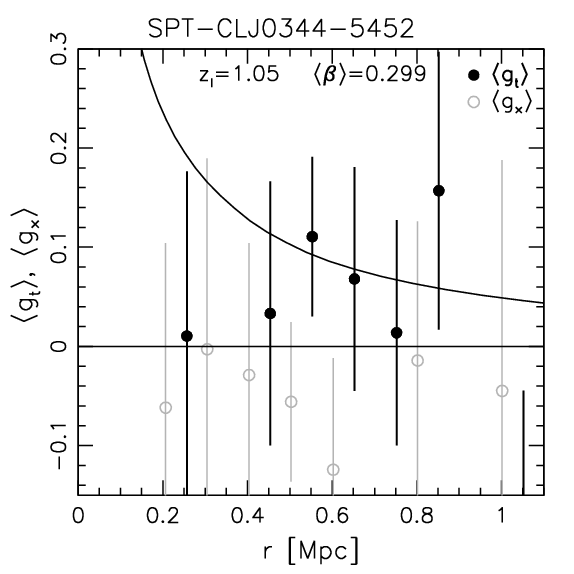}
 \caption{WL results for clusters from the ACS+GMOS sample (continued, see Fig.\thinspace\ref{fi:wl_results_snap-1} for details).
   \label{fi:wl_results_snap-2}}
\end{figure*}

\begin{figure*}
 \includegraphics[width=1.05\columnwidth]{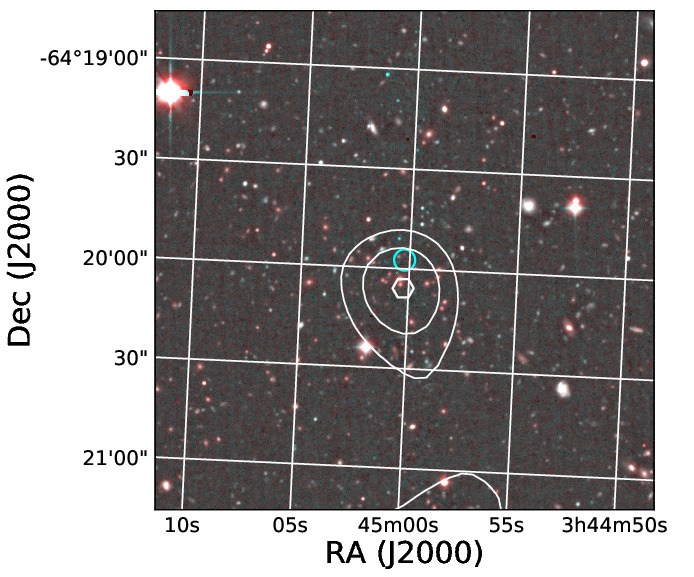}
 \includegraphics[width=0.9\columnwidth]{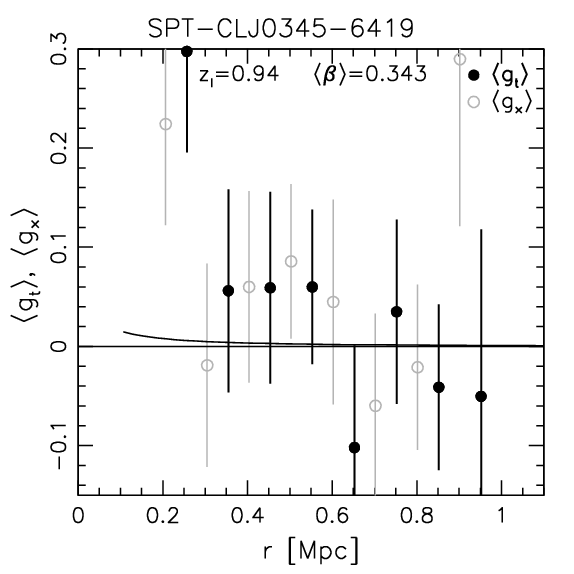}
 \includegraphics[width=1.05\columnwidth]{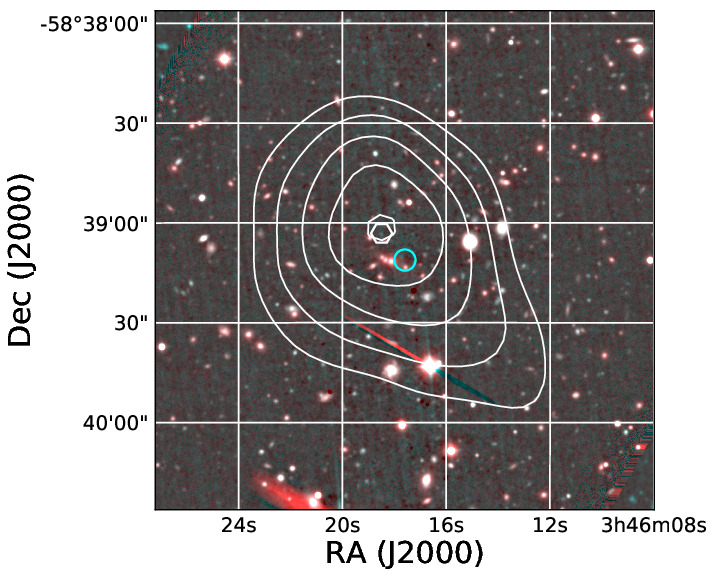}
 \includegraphics[width=0.9\columnwidth]{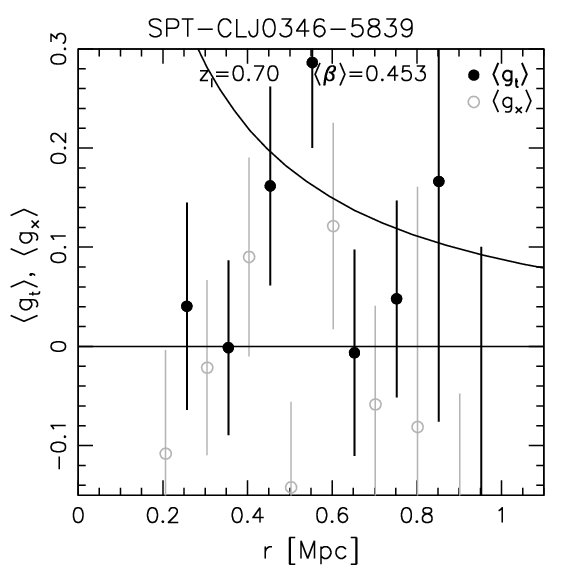}
 \includegraphics[width=1.05\columnwidth]{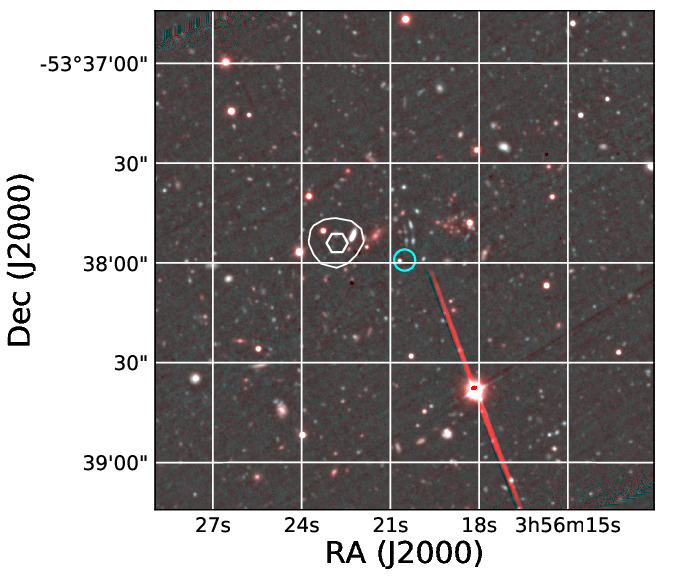}
 \includegraphics[width=0.9\columnwidth]{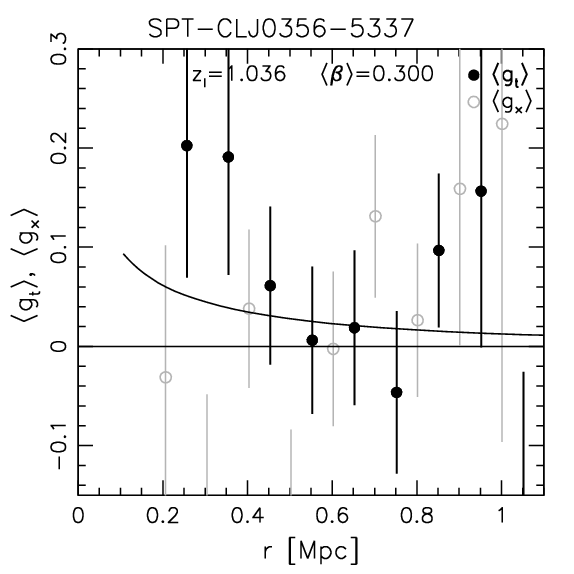}
 \caption{WL results for clusters from the ACS+GMOS sample (continued, see Fig.\thinspace\ref{fi:wl_results_snap-1} for details).
   \label{fi:wl_results_snap-3}}
\end{figure*}

\begin{figure*}
\includegraphics[width=1.05\columnwidth]{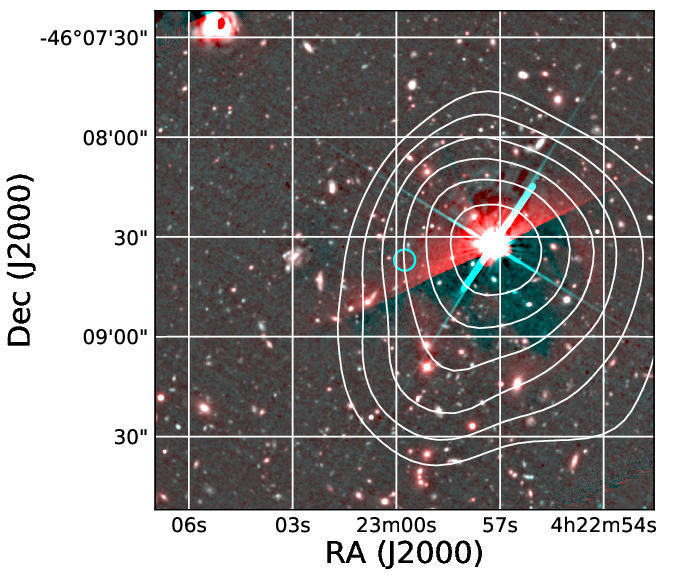}
 \includegraphics[width=0.9\columnwidth]{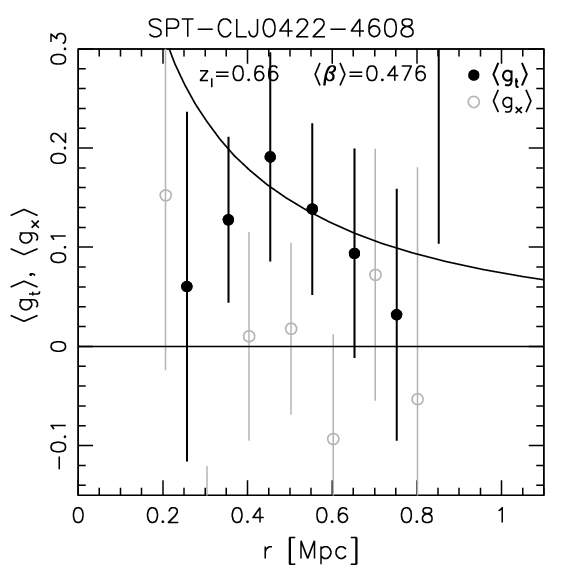}
 \includegraphics[width=1.05\columnwidth]{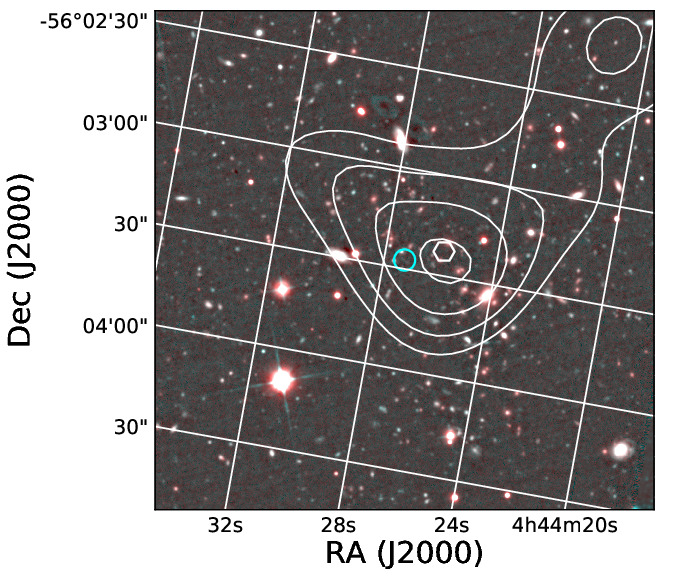}
 \includegraphics[width=0.9\columnwidth]{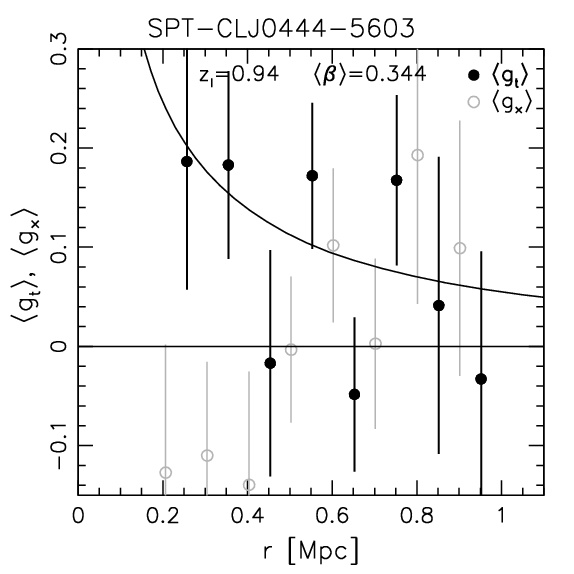}
 \includegraphics[width=1.05\columnwidth]{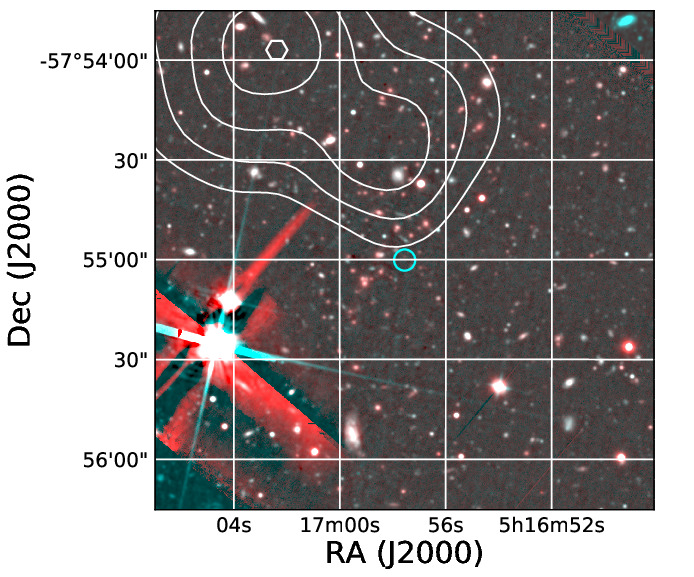}
 \includegraphics[width=0.9\columnwidth]{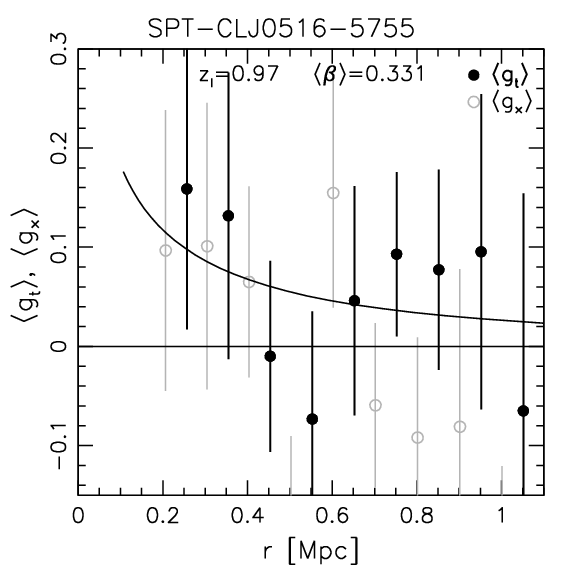}
 \caption{WL results for clusters from the ACS+GMOS sample (continued, see Fig.\thinspace\ref{fi:wl_results_snap-1} for details).
   We stress that the regions affected by bright stars (especially in the GMOS images) are well masked in our analysis. We therefore expect that the apparent alignment of a bright star with the
peak in the signal-to-noise ratio map of the WL mass reconstruction for
SPT-CL{\thinspace}$J$0422$-$4608 is purely by chance.
   \label{fi:wl_results_snap-4}}
\end{figure*}

\begin{figure*}
 \includegraphics[width=1.05\columnwidth]{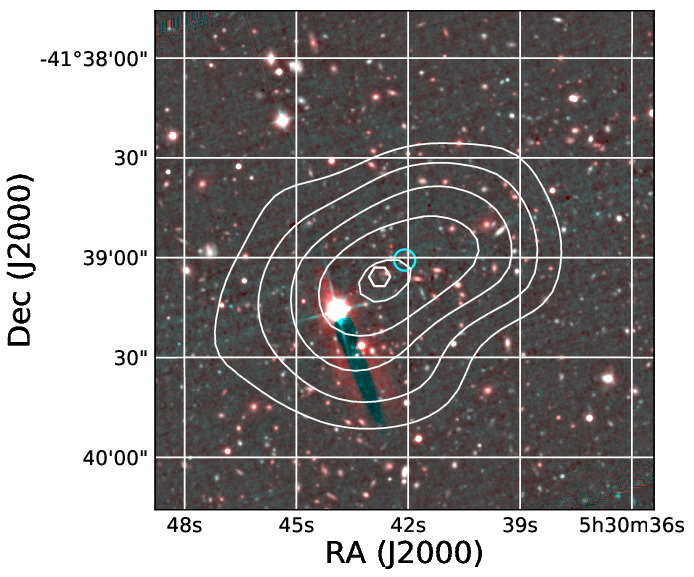}
 \includegraphics[width=0.9\columnwidth]{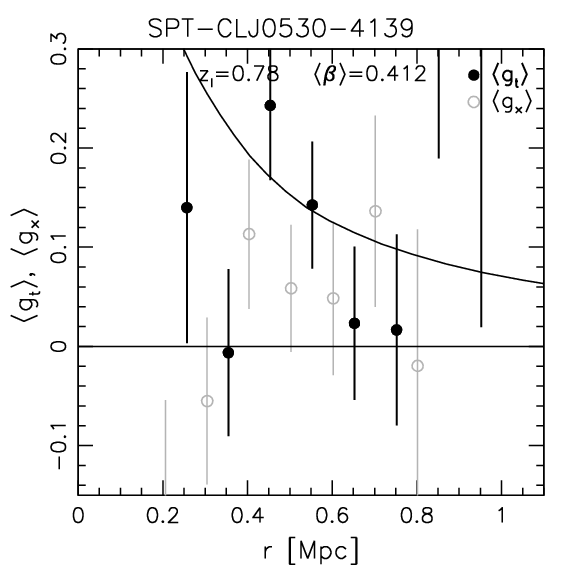}
\includegraphics[width=1.05\columnwidth]{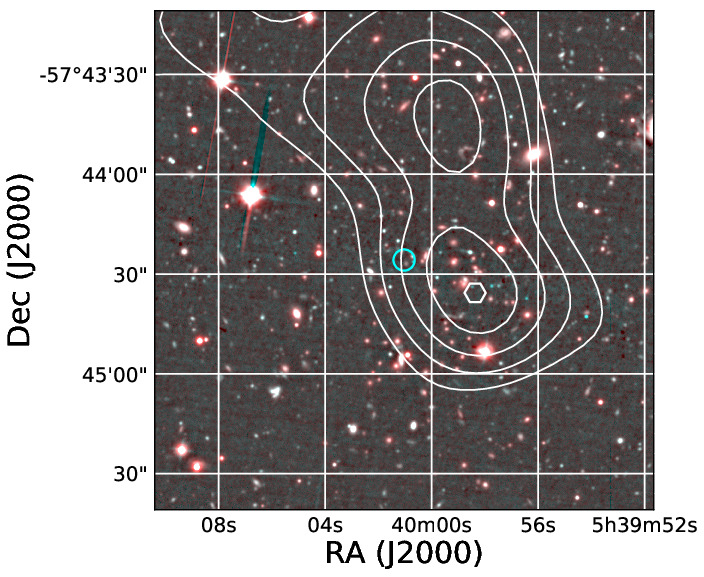}
 \includegraphics[width=0.9\columnwidth]{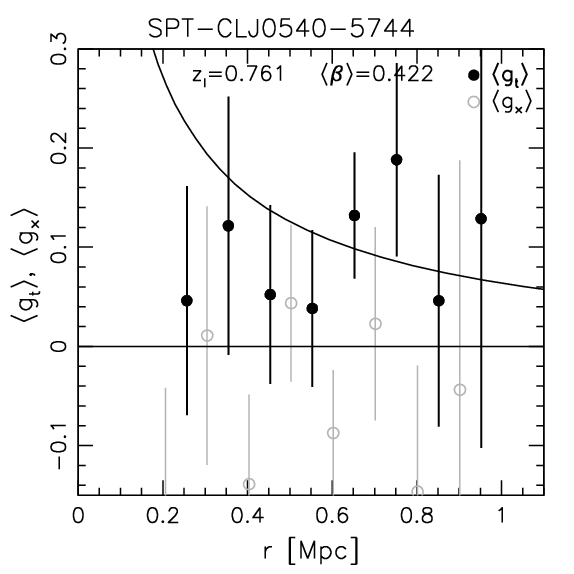}
 \includegraphics[width=1.05\columnwidth]{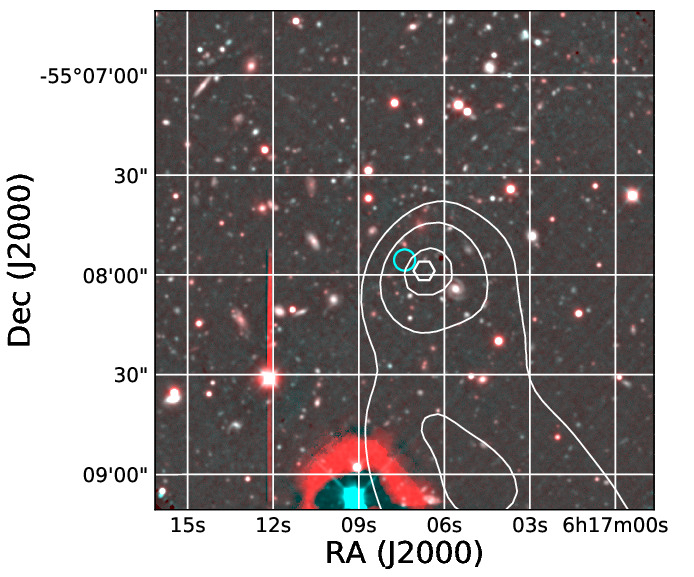}
 \includegraphics[width=0.9\columnwidth]{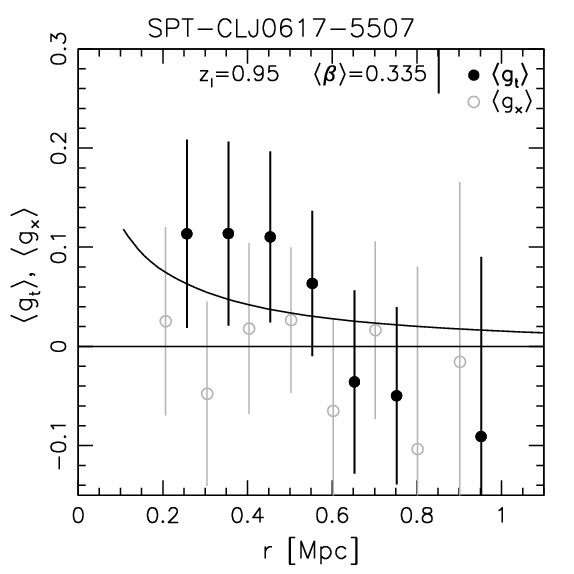}
 \caption{WL results for clusters from the ACS+GMOS sample (continued, see Fig.\thinspace\ref{fi:wl_results_snap-1} for details).
   \label{fi:wl_results_snap-5}}
\end{figure*}

\begin{figure*}
\includegraphics[width=1.05\columnwidth]{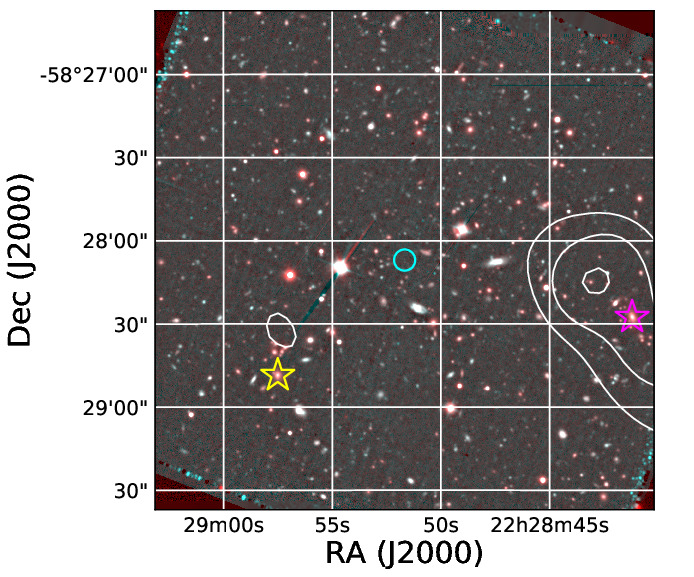}
 \includegraphics[width=0.9\columnwidth]{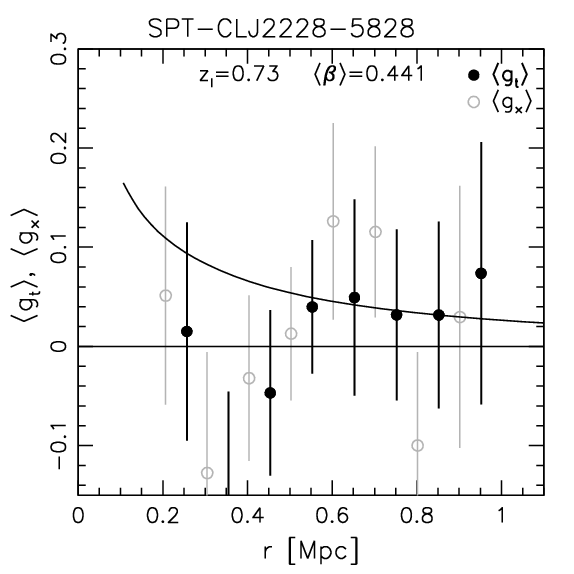}
 \includegraphics[width=1.05\columnwidth]{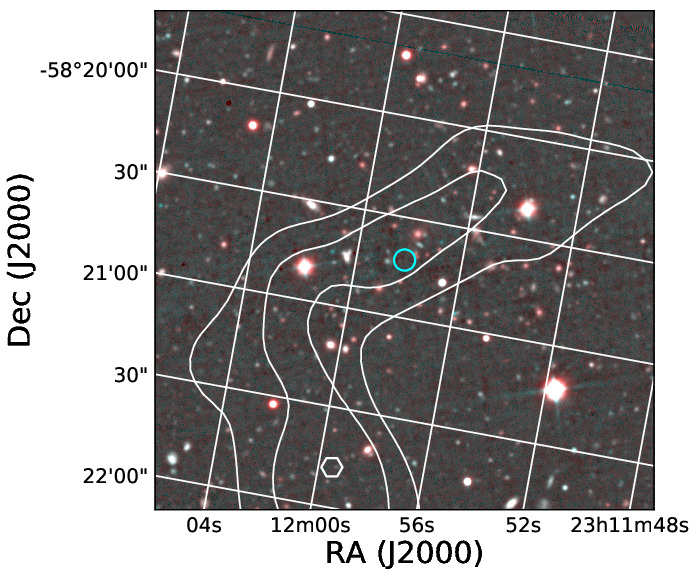}
 \includegraphics[width=0.9\columnwidth]{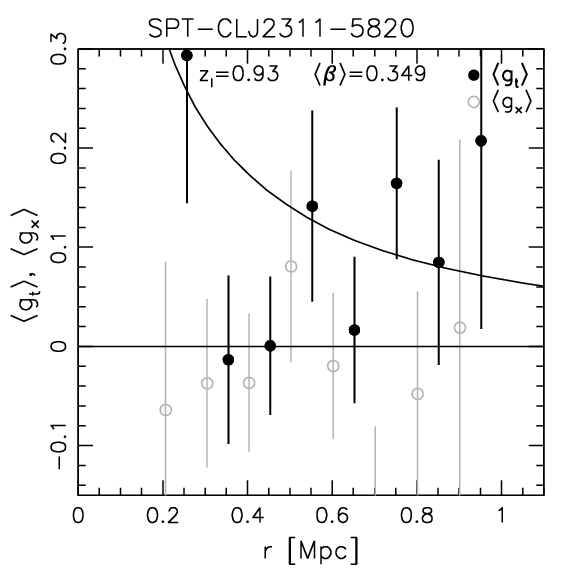}
 \caption{WL results for clusters from the ACS+GMOS sample (continued, see Fig.\thinspace\ref{fi:wl_results_snap-1} for details).
   As an exception, we show a slightly larger (\mbox{$3^\prime \times 3^\prime$}) cut-out for SPT-CL{\thinspace}$J$2228$-$5828 (top left),
   where we also mark  the BCG candidate
   from \citet{zenteno20} and the brightest galaxy in a secondary concentration of candidate cluster galaxies
   (\mbox{$\alpha=337.2396\thinspace$}deg, \mbox{$\delta=-58.4801\thinspace$}deg) with the magenta and yellow stars, respectively
   (see also footnote \ref{footnote:SPT-CLJ2228-5828}).
   The cyan dots visible near the corners of this image are residual cosmic rays and other image reduction artefacts, which occur close to the edge of the ACS field-of-view, but are masked in the science analysis.
   \label{fi:wl_results_snap-6}}
\end{figure*}

\end{document}